\documentclass[12pt]{article}
\usepackage[utf8]{inputenc}
\usepackage{setspace}
\usepackage[textwidth=7.5in,top=1.2in,bottom=1.2in]{geometry}
\usepackage[margin=1cm]{caption}
\title{\vspace{-3cm} Proximal Learning for Individualized Treatment Regimes Under Unmeasured Confounding}

\author{Zhengling Qi\thanks{Co-first Authors}\,\,\thanks{Department of Decision Sciences, The George Washington University}, Rui Miao\footnotemark[1]\,\,\thanks{Department of Statistics, University of California, Irvine}, Xiaoke Zhang\thanks{Department of Statistics, The George Washington University}}

\date{}

\usepackage{amsthm,amsmath,amssymb,bbm,bm}
\usepackage{natbib}
\usepackage{longtable}
\usepackage{multirow}
\usepackage{setspace}
\usepackage[dvipsnames]{xcolor}
\definecolor{d1(L,Z)}{HTML}{0173b2}
\definecolor{d2(L,W)}{HTML}{cc78bc}
\definecolor{d1(L)}{HTML}{56b4e9}
\definecolor{d2(L)}{HTML}{fbafe4}
\definecolor{d3DR(L)}{HTML}{029e73}
\definecolor{d4}{HTML}{949494}
\definecolor{dEARL(L)}{HTML}{ece133}
\definecolor{dEARL(L,W)}{HTML}{d55e00}
\definecolor{dEARL(L,Z)}{HTML}{de8f05}
\definecolor{dEARL(L,W,Z)}{HTML}{ca9161}
\definecolor{dPESS}{HTML}{ff0000}
\definecolor{NUC}{HTML}{ffffff}
\newcommand{\thiscolor}[1]{\raisebox{.6\height}{\fcolorbox{black}{#1}{\hspace{2mm}}} \texttt{#1}}
\usepackage{wrapfig}
\usepackage{tikz}
\usepackage{centernot}
\usepackage{array}
\usepackage{url}
\usepackage{algorithmic}
\usepackage[linesnumbered, ruled, vlined]{algorithm2e}
\usepackage{mathrsfs}
\usepackage{dsfont}
\usepackage{titling}
\usepackage{relsize}
\usepackage{rotating}
\usepackage{enumitem}
\usepackage{titling}
\usepackage{float}
\usepackage{subcaption}
\usepackage{multirow}

\usepackage{xr-hyper}
\usepackage{hyperref}
\hypersetup{
	colorlinks=true,
	linkcolor=red,
	urlcolor=blue,
	citecolor=blue
}

\usepackage{comment}

\newenvironment{smashedalign*}
 {\par$\!\aligned}
 {\endaligned$\par}

\newtheorem{ex}{Example}
\newtheorem{remark}{Remark}
\newtheorem{assumption}{Assumption}

\newtheorem{theorem}{Theorem}[section]
\newtheorem{proposition}{Proposition}[section]

{\textwidth=6.5in}

\newcommand{\bI}{{\boldsymbol{I}}}
\newcommand{\bK}{{\boldsymbol{K}}}

\newcommand{\bH}{{\boldsymbol{H}}}
\newcommand{\bM}{{\boldsymbol{M}}}

\newcommand{\bSigma}{\mbox{\boldmath $\Sigma$}}

\newcommand{\expit}{{\text{expit}}}

\newcommand{\sign}{{\text{sign}}}

\def\calA{{\cal A}}

\def\calD{{\cal D}}

\def\calF{{\cal F}}
\def\calG{{\cal G}}
\def\calH{{\cal H}}

\def\calL{{\cal L}}

\def\calQ{{\cal Q}}
\def\calR{{\cal R}}

\def\calU{{\cal U}}

\def\calW{{\cal W}}
\def\calX{{\cal X}}

\def\calZ{{\cal Z}}
\newcommand{\indep}{\rotatebox[origin=c]{90}{$\models$}}
\newcommand{\E}{\mathbb{E}}

\newcommand{\tr}{\mbox{{\rm tr}}}
\newcommand{\diag}{\mbox{{\rm diag}}}

\DeclareMathOperator*{\argmin}{arg\,min}
\DeclareMathOperator*{\argmax}{arg\,max}

\newcommand\independent{\protect\mathpalette{\protect\independenT}{\perp}}
\def\independenT#1#2{\mathrel{\rlap{$#1#2$}\mkern2mu{#1#2}}}

\newcommand{\numit}{\stepcounter{equation}\tag{\theequation}}

\newcommand{\rui}[1]{{\color{blue}\bf [Rui: #1]}}

\begin{document}
\maketitle
\vspace{-1cm}
\abstract{
Data-driven individualized decision making has recently received increasing research interest. 
However, most existing 
methods 
rely on the assumption of no unmeasured confounding, which 
cannot be ensured in practice  especially in observational studies. Motivated by the recently proposed proximal causal inference,  
we develop several proximal learning methods to estimate optimal individualized treatment regimes (ITRs) in the presence of unmeasured confounding. Explicitly, in terms of two types of proxy variables, 
we are able to establish several identification results for different classes of ITRs respectively, exhibiting the trade-off between the risk of making untestable assumptions and the potential improvement of the value function in decision making. Based on these identification results, we propose several classification-based approaches to finding a variety of restricted in-class optimal ITRs and establish their theoretical properties. 
The appealing numerical performance of our proposed methods is demonstrated via  extensive simulation experiments and a real data application.}


\newpage

\section{Introduction}\label{sec:intro}

In recent years, there is a surge of interest in studying data-driven individualized decision making in various scientific fields. For example, in precision medicine, clinicians leverage biomedical data to discover the best personalized treatments for heterogeneous patients \citep[e.g.,][]{rashid2020high}. In mobile health, due to recent advances in smart devices and sensing technology, real time information can be collected and used to learn the most effective interventions for patients to promote healthy behaviors \citep[e.g.,][]{klasnja2015microrandomized}. In robotics, tremendous amounts of simulated data are generated to train robots for making optimal decisions to complete human tasks 
 \citep[e.g.,][]{kober2013reinforcement}. In operations management, learning the optimal resource allocation based on current conditions, logistics and costs, etc, is necessary to improve the efficiency of operations \citep[e.g.,][]{seong2006optimal}.  
Apparently a common goal of the aforementioned applications is to find an optimal individualized treatment regime (ITR) that can optimize the utility of each instance. 



Recently, many statistical learning methods 
have been developed for learning the optimal ITR.
For example, \cite{qian2011performance} 
proposed to learn the optimal ITR by 
first fitting a high-dimensional regression model for the so-called Q-function, which is the conditional expectation of the outcome given the treatment and covariates 
\citep{watkins1992q},  
 and then assigning the optimal treatment to each individual corresponding to the largest Q-function value.
In the binary treatment setting,
this method is equivalent to estimating the conditional average treatment effect. Methods of such type are 
usually referred to as model-based methods 
\citep[e.g.,][]{zhao2009reinforcement,shi2018high}. 
Alternatively, one may obtain the optimal ITR by directly maximizing the value function, as defined in 
\eqref{eq:value fun} below. In the literature, such method is called a direct method.
For example, \cite{dudik2011doubly} and \cite{zhao2012estimating} applied inverse probability weighting (IPW)  to estimate the value function for each ITR, and then leveraged modern classification techniques to learn the optimal one. 
Along this line of research, various extensions have been proposed, such as \cite{zhao2014doubly} for censored outcomes, \cite{chen2016personalized} 
for ordinal outcomes, and \cite{wang2017quantile} for quantile ITRs. 
To alleviate potential model misspecifications of 
the 
Q-function or the propensity score, 
augmented IPW has been used
so that the value function estimator enjoys 
the doubly robust property
\citep[e.g.,][]{zhang2012robust}. 
Recently, 
borrowed from semiparametric statistics \citep{bickel1982adaptive}, cross-fitting techniques have been incorporated in ITR learning \citep{athey2017efficient,zhao2019efficient} so that flexible black-box machine learning methods can be used for estimating the Q-function and the propensity score without sacrificing the efficiency of the resulting estimated ITR. 
Finally, a review of various ITR learning methods can be found in \cite{kosorok2019precision} and references therein. 



Most existing methods for learning the optimal ITR rely on the unconfoundedness assumption so that the value function can be identified nonparametrically using the observed data. However, 
it is difficult, or even impossible, to verify this assumption in practice, especially in observational studies or randomized trials with non-compliance. 
Therefore, to remove confounding effects and thus identify optimal ITRs, 
practitioners often collect and adjust for as many variables as possible. 
While this might be the best approach in practice,
it is often very costly and sometimes unethical. To address this problem, instrumental variables (IVs) have been used in the literature to find an optimal ITR
in the presence of unmeasured confounding.
For example, motivated by \cite{wang2018bounded}, \cite{cui2020IV} and \cite{qiu2020optimal} 
independently established similar identification results on the value function 
using an IV and proposed different optimal ITR learning methods, one for deterministic ITRs and the other for stochastic ITRs. 
While these two methods are particularly useful in randomized trials with non-compliance, their restrictions on the setting of binary treatments and IVs 
are very restrictive, which may limit their 
applicability. 
Recently, instead of aiming to exactly identify the value function under unmeasured confounding, \cite{han2019optimal} and \cite{pu2020estimating} considered partial identification in terms of an IV to provide robustness in estimating the optimal ITR. 
In addition, \cite{kallus2018confounding} leveraged a sensitivity analysis in causal inference where the value function is partially identified and developed a confounding-robust policy improvement method. 
Although partial identification can still lead to valuable ITRs to policy makers, they are likely to be sub-optimal.



In this paper, we propose an alternative remedy to estimate optimal ITRs under endogeneity. Our approach is built upon 
\emph{proximal causal inference} %
recently developed 
by \cite{miao2018confounding} and \cite{tchetgen2020introduction}.
The salient idea behind proximal causal inference 
is to identify the causal effect under unmeasured confounding via either treatment-inducing or outcome-inducing confounding proxies, which connects existing identification results on the causal effect 
based on IVs and negative controls. 
The applicability of proximal causal inference is very promising since the existence of such 
proxies is common in many applications. See \citet{miao2018confounding} and \citet{tchetgen2020introduction} for examples. Moreover, in contrast with the aforementioned IV-based ITR learning approaches, there is no restriction on the data type of these proxy variables.
Due to these merits, in this paper, we adapt the idea of proximal causal inference to establish identification results for various classes of
ITRs under unmeasured confounding and accordingly propose several classification-based methods to estimate corresponding in-class optimal ITRs. 


The contribution of this paper can be mainly summarized into three folds. 
First, we establish several new identification results for various classes of ITRs under unmeasured confounding in terms of treatment-inducing and/or outcome-inducing confounding proxies. 
All these results can show an interesting trade-off between the risk of making untestable assumptions and the potential gain of value function in decision making. Note that since covariates are involved in ITRs, our identification results are focused on conditional treatment effect identification, which are different from those for average treatment effect estimation in the proximal causal inference literature. 
Second, based on these identification results, we propose several classification-based methods to estimate optimal ITRs. 
For nuisance functions 
involved in these 
methods, which are characterized by conditional moment restrictions,
we 
apply the min-max learning approach in \cite{dikkala2020minimax}
to estimate them nonparametrically.
Similar approaches to estimating nuisance functions have also been used in
\cite{kallus2021causal} and \cite{ghassami2021minimax}, which were posted on \url{arxiv.org} very recently, and are independent works from this paper. 
Under one specific setting where the value function can be identified via either a treatment-inducing confounding bridge function or an outcome-inducing confounding bridge function (see Section \ref{sec:identification under U} below), motivated by \cite{cui2020semiparametric}, we develop a doubly robust optimal ITR learning method with cross-fitting (see Section \ref{sec:DRlearning} below). 
 Third, we establish a theoretical guarantee for the proposed doubly robust proximal learning method. Specifically, we provide a finite sample bound for the value function difference between the optimal ITR and our estimated in-class optimal ITR. 
The bound can be decomposed into four components: 
an irreducible error due to unmeasured confounding, an approximation error due to 
the restricted treatment regime class, and two estimation errors caused by the finite sample, illustrating different sources of errors in finding the optimal ITR under unmeasured confounding. Theoretical results for other proposed methods can be similarly derived.


The rest of our paper is organized as follows. In Section \ref{sec:NoConfound}, we briefly introduce the framework of learning optimal ITRs without unmeasured confounding. In Section \ref{sec:identification under U}, adapting the idea from the proximal causal inference, we establish nonparametric value function identification results for various classes of ITRs under unmeasured confounding. In Section \ref{sec:policy}, we develop several corresponding proximal learning methods based on the identification conditions established in Section \ref{sec:identification under U}. Theoretical guarantees of the proposed methods are presented in Section \ref{sec:theory}. In Sections \ref{sec:simu} and \ref{sec: real data} respectively, we demonstrate the superior performance of our methods via extensive simulation studies and one real data application. 
Discussions and future research directions in Section \ref{sec:disc} conclude the paper. 


\section{Optimal ITR 
Without Unmeasured Confounding} \label{sec:NoConfound}





In this section, we give a brief introduction to optimal ITR learning under no unmeasured confounding. 
Let $A$ be a binary treatment which takes values in the space $\calA = \left\{-1, 1\right\}$. Let $Y(1)$ and $Y(-1)$ be the potential outcomes when $A=1$ and $A=-1$ respectively, but in practice $Y(1)$ and $Y(-1)$ are not both observable.  
Under the consistency assumption that $Y=Y(A)$, we can write $Y= Y(1) \mathbb{I}(A = 1) + Y(-1)\mathbb{I}(A = - 1)$, where $\mathbb{I}(\bullet)$ denotes the indicator function. Moreover, let $X$ be the observed $p$-dimensional covariate that belongs to a covariate space $\calX \subset \mathbb{R}^p$. Without loss of generality, we assume that a large outcome $Y$ is always preferred.


For an ITR $d$, which is a measurable function mapping from the covariate space $\calX$ into the treatment space $\calA$, 
the potential outcome under $d$ 
is defined by $
	Y(d) = Y(1) \mathbb{I}(d(X) = 1) + Y(-1) \mathbb{I}(d(X) = -1)$,
and then the value function of $d$  \citep{manski2004statistical,qian2011performance} can be defined as
\begin{align}\label{eq:value fun}
	V(d) = \E\left\{Y(d)\right\}.
\end{align}

Under the following three standard assumptions in the potential outcome framework \citep{robins1986new}: (i) Unconfoundedness: $\{Y(1), Y(-1)\} \independent A \mid X$ where $\independent$ represents independence, (ii) Positivity: $\Pr(A = a \mid X) >0$ for every $a \in \calA$ almost surely, and (iii) Consistency: $Y = Y(A)$,
we can nonparametrically identify the value function $V(d)$ using $(X, A, Y)$ via
\begin{align}\label{eq:value fun under no unmeasured confounders}
	V(d) = \E\left\{\frac{Y \cdot\mathbb{I}(A = d(X))}{\Pr(A \mid X)}\right\}.
\end{align} 
Then by maximizing $V(d)$ in \eqref{eq:value fun under no unmeasured confounders} over $\calD$, the class of all ITRs, the \textit{global} optimal ITR is 
$$
d^\ast(X) = \sign\left\{\E\left(Y \mid X, A = 1\right) - \E\left(Y \mid X, A = -1\right) \right\},
$$
almost surely. 
See \cite{qian2011performance} and \cite{zhao2012estimating} for more details. Note that the optimal ITR remains same if we use other coding schemes, e.g., $1/0$, for the binary $A$. Moreover, the optimal ITR learning method under no unmeasured confounding above is also applicable to multiple treatments and so are our proposed methods below.


\section{Optimal ITR With Unmeasured Confounding}\label{sec:identification under U}


The optimal ITR learning without unmeasured confounding in Section \ref{sec:NoConfound} relies on the unconfoundedness assumption, i.e., $\{Y(1), Y(-1)\} \independent A \mid X$. 
In practice, however, one typically cannot ensure the unconfounded assumption to hold. 
If there exist unmeasured confounders $U$ 
that affect both the treatment $A$ and the outcome $Y$, 
unless we make 
proper assumptions, 
we are unable to identify the value function based on the observed data $(X, A, Y)$ as in (\ref{eq:value fun under no unmeasured confounders}) or further to find optimal ITRs \citep{pearl2009causality}. 
Inspired by 
proximal casual inference that was recently proposed by \citet{miao2018confounding} and \citet{tchetgen2020introduction}, 
we propose to adapt its idea to optimal ITR learning in the presence of unmeasured confounding. 

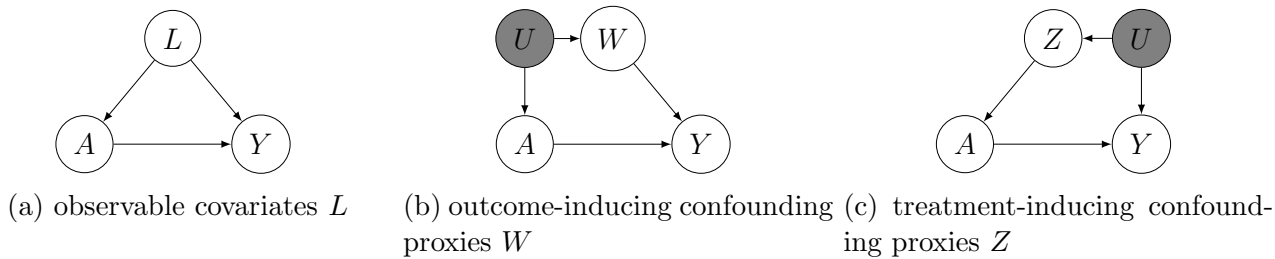
\begin{figure}[H]
	\centering
    \begin{subfigure}[b]{0.3\textwidth}
      \centering
	    \resizebox{100pt}{!}{
		\begin{tikzpicture}[state/.style={circle, draw, minimum size=0.7cm}]
  \def\Ax{0}
  \def\Ay{0}
  \def\offset{2.5}
  \def\Bx{\Ax+5}
  \def\By{\Ay}

  \node[state,shape=circle,draw=black] (Y) at (\Bx,\By) {$Y$};
  \node[state,shape=circle,draw=black] (A) at (\Bx-2.5,\By) {$A$};
  \node[state,shape=circle,draw=black] (L) at (\Bx-1.25,\By+1.5) {$L$};

  \draw [-latex] (L) to [bend left=0] (A);
  \draw [-latex] (L) to [bend left=0] (Y);
  \draw [-latex] (A) to [bend left=0] (Y);

\end{tikzpicture}
        }
        \caption{observable covariates $L$\\ ~}
    \end{subfigure}
    \begin{subfigure}[b]{0.3\textwidth}
      \centering
	    \resizebox{100pt}{!}{
      \begin{tikzpicture}[state/.style={circle, draw, minimum size=0.7cm}]
  \def\Ax{0}
  \def\Ay{0}
  \def\offset{2.5}
  \def\Bx{\Ax+5}
  \def\By{\Ay}
  
  \node[state,shape=circle,draw=black, fill=gray] (U) at (\Bx-2.5,\By+1.5) {$U$};
  \node[state,shape=circle,draw=black] (Y) at (\Bx,\By) {$Y$};
  \node[state,shape=circle,draw=black] (A) at (\Bx-2.5,\By) {$A$};
  \node[state,shape=circle,draw=black] (L) at (\Bx-1.25,\By+1.5) {$W$};

  \draw [-latex] (L) to [bend left=0] (Y);
  \draw [-latex] (U) to [bend left=0] (L);
  \draw [-latex] (A) to [bend left=0] (Y);
\draw [-latex] (U) to [bend left=0] (A);

\end{tikzpicture}
      }
      \caption{outcome-inducing confounding proxies $W$}
    \end{subfigure}
    \begin{subfigure}[b]{0.3\textwidth}
      \centering
	    \resizebox{100pt}{!}{
        \begin{tikzpicture}[state/.style={circle, draw, minimum size=0.7cm}]
  \def\Ax{0}
  \def\Ay{0}
  \def\offset{2.5}
  \def\Bx{\Ax+5}
  \def\By{\Ay}
  
  \node[state,shape=circle,draw=black, fill= gray] (U) at (\Bx,\By+1.5) {$U$};
  \node[state,shape=circle,draw=black] (Y) at (\Bx,\By) {$Y$};
  \node[state,shape=circle,draw=black] (A) at (\Bx-2.5,\By) {$A$};
  \node[state,shape=circle,draw=black] (L) at (\Bx-1.25,\By+1.5) {$Z$};

  \draw [-latex] (L) to [bend left=0] (A);
  \draw [-latex] (U) to [bend left=0] (L);
  \draw [-latex] (A) to [bend left=0] (Y);
\draw [-latex] (U) to [bend left=0] (Y);

\end{tikzpicture}
        }
        \caption{treatment-inducing confounding proxies $Z$}
     \end{subfigure}
	\caption{Directed acyclic graph representations when no unmeasured confounding holds.} 
	\label{fig:sra dag0}
  \end{figure}
\begin{figure}[H]
	\centering
	\vfill
	\resizebox{130pt}{!}{%
		\begin{tikzpicture}[state/.style={circle, draw, minimum size=0.6cm}]
  \def\Ax{0}
  \def\Ay{0}
  \def\offset{2.5}
  \def\Bx{\Ax+5}
  \def\By{\Ay}
  \node[state,shape=circle,draw=black] (Z) at (\Bx-4,\Ay+1.5) {$Z$};
  \node[state,shape=circle,draw=black, fill=gray] (U1) at (\Bx-2.5,\Ay+1.5) {$U_1$};
  \node[state,shape=circle,draw=black, fill=gray] (U2) at (\Bx,\By+1.5) {$U_2$};
  \node[state,shape=circle,draw=black] (Y) at (\Bx,\By) {$Y$};
  \node[state,shape=circle,draw=black] (A) at (\Bx-2.5,\By) {$A$};
  \node[state,shape=circle,draw=black] (X) at (\Bx-1.25,\By+1.5) {$L$};
    \node[state,shape=circle,draw=black, fill=gray] (U3) at (\Bx-1.25,\By+3) {$U_3$};
  \node[state,shape=circle,draw=black] (W) at (\Bx+1.5,\Ay+1.5) {$W$};

  \draw [-latex] (A) to [bend left=0] (Y);
  \draw [-latex] (X) to [bend left=0] (A);
  \draw [-latex] (X) to [bend left=0] (Y);
  \draw [-latex] (Z) to [bend left=0] (A);
  \draw [-latex] (W) to [bend left=0] (Y);

  \draw [-latex] (U3) to [bend left=0] (U2);
  \draw [-latex] (U3) to [bend left=0] (U1);
  \draw [-latex] (U1) to [bend left=0] (Y);
  \draw [-latex] (U2) to [bend left=0] (A);
    \draw [-latex] (U1) to [bend left=0] (Z);
  \draw [-latex] (U2) to [bend left=0] (W);

\end{tikzpicture}
	}
	\vfill
	\caption{An unidentifiable causal effect when $L$, $W$, $Z$ and $U=(U_1, U_2, U_3)$ coexist.}
	\label{fig:sra dag2}
\end{figure}
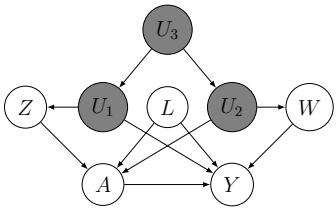

Following \citet{tchetgen2020introduction}, suppose that we can decompose $X$ into three types 
$X = (L, W, Z)$, where $L$ 
are observable convariates that affect both $A$ and $Y$, $W$ are 
\emph{outcome-inducing confounding proxies} that are only related to $A$ through $(L, U)$, and $Z$ are 
\emph{treatment-inducing confounding proxies} that are only related to $Y$ through $(L, U)$. The terminologies we adopt here for $Z$ and $W$ follow those in \citet{tchetgen2020introduction},  
but they may also be called negative control exposures and negative control outcomes respectively in other literature
\citep[e.g.,][]{miao2018confounding}.  Denote the spaces $L$, $U$, $W$, and $Z$ belong to as $\calL$, $\calU$, $\calW$ and $\calZ$ respectively.
Figure \ref{fig:sra dag0} illustrates some of their relationships with $A$ and $Y$. 
In general, such decomposition does not guarantee the identifiability of the causal effect. 
Figure \ref{fig:sra dag2} 
shows an example 
where $L$, $W$, $Z$ and $U=(U_1, U_2, U_3)$ coexist, 
but the path 
$A$-$U_2$-$U_3$-$U_1$-$Y$ prevents us from identifying the causal effect of $A$ on $Y$. 
However, there exist scenarios where no unmeasured confounding
still holds despite the presence of $U$, $W$, and/or $Z$ to make the causal effect of $A$ on $Y$ identifiable. 
See Figure \ref{fig:sra dag0}~(b)-(c) and Figure 1 (d) of \cite{tchetgen2020introduction} for examples.
As a result, given the decomposition $X = (L, W, Z)$, it is possible to relax the no unmeasured confounding assumption and still identify the causal effect of $A$ on $Y$.


Compared to average treatment effect estimation in causal inference, 
we need to further take the class of treatment regimes into consideration in ITR learning since an ITR 
could depend on observed covariates. Such dependency may hinder us from identifying the value function. In the following, 
we first list several basic assumptions and then provide several different value function identification results based on various sets of additional assumptions, together with their corresponding optimal ITRs.

\subsection{Basic Assumptions} \label{sec:basicass}

Below we list several basic assumptions used for value function identification with unmeasured confounding. 
See \cite{tchetgen2020introduction} for more details.

\begin{assumption}[Consistency]\label{ass: consistency}
	$Y = Y(A)$ almost surely.
\end{assumption}

\begin{assumption}[Positivity]\label{ass: positivity}
	$\Pr(A = a \mid U, L) > 0$ for every $a\in \calA$ almost surely.
\end{assumption}

\begin{assumption}[Outcome-inducing confounding proxies]\label{ass: outcome proxy}
	$W(a, z) = W$ for all $a \in \calA$ and $z \in \calZ$ almost surely.
\end{assumption}

\begin{assumption}[Treatment-inducing confounding proxies]\label{ass: treament proxy}
	$Y(a, z) = Y(a)$ for all $a \in \calA$ and $z \in \calZ$ almost surely.
\end{assumption}

\begin{assumption}[Latent exchangeability]\label{ass: latent exchangeability}
	$(Z, A) \independent (Y(a), W) \mid (U, L)$ for every $a \in \calA$.
\end{assumption}

\begin{figure}[h]
	\centering
	\vfill
	\resizebox{170pt}{!}{%
		\begin{tikzpicture}[state/.style={circle, draw, minimum size=1.1cm}]
  \def\Ax{0}
  \def\Ay{0}
  \def\offset{2.5}
  \def\Bx{\Ax+5}
  \def\By{\Ay}
  \node[state,shape=circle,draw=black] (Z) at (\Bx-4,\Ay+1.5) {$Z$};
  \node[state,shape=circle,draw=black] (Y) at (\Bx,\By) {$Y$};
  \node[state,shape=circle,draw=black] (A) at (\Bx-2.5,\By) {$A$};
  \node[state,shape=circle,draw=black] (L) at (\Bx-1.25,\By+1.5) {$L$};
    \node[state,shape=circle,draw=black, fill=gray] (U) at (\Bx-1.25,\By+3.5) {$U$};
  \node[state,shape=circle,draw=black] (W) at (\Bx+1.5,\Ay+1.5) {$W$};

  \draw [-latex] (L) to [bend left=0] (W);
  \draw [-latex] (A) to [bend left=0] (Y);
  \draw [-latex] (L) to [bend left=0] (A);
  \draw [-latex] (L) to [bend left=0] (Z);
  \draw [-latex] (L) to [bend left=0] (Y);
  \draw [-latex] (Z) to [bend left=0] (A);
  \draw [-latex] (W) to [bend left=0] (Y);

  \draw [-latex] (U) to [bend left=0] (A);
  \draw [-latex] (U) to [bend left=0] (Z);
  \draw [-latex] (U) to [bend left=0] (Y);
  \draw [-latex] (U) to [bend left=0] (W);
  \draw [-latex] (U) to [bend left=0] (L);

\end{tikzpicture}
	}
	\vfill
	\caption{A causal DAG as a representation of Assumptions \ref{ass: outcome proxy}--\ref{ass: latent exchangeability}.
	}
	\label{fig:sra dag 1}
\end{figure}
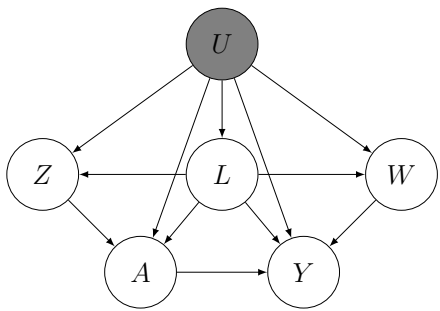

Assumptions \ref{ass: consistency} and \ref{ass: positivity} are standard in the literature of causal inference with no unmeasured confounding. Assumptions \ref{ass: consistency} 
links the potential outcome with the observed outcome while Assumption \ref{ass: positivity} states that each treatment has a positive probability of being assigned. Assumption \ref{ass: outcome proxy} essentially states that $A$ and $Z$ do not have a causal effect on $W$ while Assumption 
\ref{ass: treament proxy} indicates that 
there is no direct causal effect of $Z$ on $Y$ except intervening on $A$. 
In many practical applications, Assumptions \ref{ass: outcome proxy} and \ref{ass: treament proxy} can hold with the help of domain experts to identify valid $W$ and $Z$. See our real data application and also the air pollution example in \citep{miao2018confounding}.
Assumption \ref{ass: latent exchangeability} is also standard in the literature. It indicates that by adjusting for $(U, L)$, one is able to jointly identify the causal effect of $(Z, A)$ on $Y$ and $W$, which holds in principle  
as $U$ is not observed. 
A directed acyclic graph (DAG) of Assumptions \ref{ass: outcome proxy}--\ref{ass: latent exchangeability} is depicted in Figure \ref{fig:sra dag 1}.
Note that 
if $\calD$ is now defined as the class of all ITRs mapping from $(L, U)$ to $\calA$, then the \textit{global} optimal ITR within $\calD$ becomes $
d^\ast(L, U) = \sign\left\{\E\left(Y \mid L,U, A = 1\right) - \E\left(Y \mid L,U, A = -1\right) \right\},
$ almost surely, but it is unattainable in practice since $U$ is unobsevable. 
\vspace{0.5cm}
\vspace{0.5cm}

Under Assumptions \ref{ass: consistency}--\ref{ass: latent exchangeability} and a few different sets of additional assumptions to be listed below, we are able to establish several value function identifications under unmeasured confounding. 

\subsection{Optimal ITR via Outcome Confounding Bridge} \label{sec:identifyoutcome}


Our first value function identification requires the following technical assumptions, which were used by \citet{miao2018identifying} to 
identify the population average treatment effect unmeasured confounding.

\begin{assumption}[Completeness]\label{ass: completeness (1)}
	\begin{itemize}
		\item[(a)] For any $a \in \calA, l \in \calL$ and measurable function $g$ defined on $\calU$, if $\E\left\{g(U) \mid Z, A = a, L = l\right\} = 0$ almost surely, then $g(U) = 0$ almost surely.
		\item[(b)] For any $a \in \calA, l \in \calL$ and measurable function $g$ defined on $\calZ$, if $\E\left\{g(Z) \mid W, A = a, L = l\right\} = 0$ almost surely, then $g(Z) = 0$ almost surely.
	\end{itemize}
\end{assumption}

\begin{assumption}[Outcome confounding bridge]\label{ass: outcome bridge}
	There exists an outcome confounding bridge function $h_0$ defined on $(\calW, \calA, \calL)$ such that
		\begin{align}\label{eq: outcome bridge}
			\E\left(Y \mid Z, A, L\right) = \E\left\{h_0(W, A, L) \mid  Z, A, L\right\},
		\end{align}
		almost surely.
\end{assumption}


 
The completeness assumption is commonly seen in mathematical statistics and can be satisfied by many parametric or semiparametric models such as those for exponential families \citep{newey2003instrumental}. For more  examples including nonparametric models, we refer readers to \citet{d2011completeness} and \citet{chen2014local}. Assumption \ref{ass: completeness (1)}~(a) essentially requires that the variability of $U$ can be accounted for by $Z$ and
Assumption \ref{ass: completeness (1)}~(b) can be similarly interpreted. 
Unlike the requirement by \cite{cui2020IV} and \cite{qiu2020optimal} that the IV must be binary, 
Assumption \ref{ass: completeness (1)} does not require any specific type of $Z$, which is appealing. 
Assumption \ref{ass: outcome bridge} basically states that there exists a solution to \eqref{eq: outcome bridge}, 
 which is called 
 a linear integral equation of the first kind \citep{kress1989linear}.  Assumption \ref{ass: completeness (1)}~(b), together with some regularity conditions 
 given by  \cite{miao2018identifying}, 
can ensure the existence of $h_0$ satisfying \eqref{eq: outcome bridge}. 
 For more details and practical examples where these assumptions are satisfied, we refer readers to \cite{miao2018identifying}, \cite{shi2020multiply} and \cite{miao2018confounding}. 

 Based on the outcome confounding bridge $h_0$, we develop a nonparametric identification result for the value function of each ITR, which is similar to that by \cite{miao2018confounding} 
 on the average treatment effect. 
 

 
 
 
\begin{theorem}\label{thm: identify D1}
	Let $\calD_1$ be the class of ITRs that map from $(\calL, \calZ)$ to  $\calA$. Under Assumptions \ref{ass: consistency}-\ref{ass: latent exchangeability}, \ref{ass: completeness (1)} and \ref{ass: outcome bridge}, 
 for any $d_1 \in \calD_1$, the value function $V(d_1)$ 
	can be nonparametrically identified by
		\begin{align}\label{outcome identification}
			V(d_1) = \E\left\{h_0(W, d_1(L, Z), L)\right\}. 
		\end{align}
		Then the \textit{restricted} in-class optimal ITR within $\calD_1$,  defined as $d_1^\ast \in \argmax_{d_1 \in \calD_1} V(d_1)$,
 can be almost surely identified by 
 \begin{align}\label{eq: optimal d_1}
 	d_1^\ast(L, Z) 
 	& = 
 	\sign\left[\E\left\{h_0(W, 1, L) \, | \, L, Z\right\} - \E\left\{h_0(W, -1, L) \, | \, L, Z\right\}\right]. 
 \end{align}
\end{theorem}
The proof of Theorem \ref{thm: identify D1} is given in Supplementary Material S2. Theorem \ref{thm: identify D1} indicates that 
the value function 
is identifiable over $\calD_1$ in the presence of unmeasured confounders.
 Due to 
 the use of the outcome confounding bridge  $h_0$, we can only identify the value function over a restricted class of ITRs $\calD_1$ 
 instead of $\calD$. 
 That $W$ are not used as decision variables for the optimal ITR is somewhat reasonable since they have already been used as outcome-inducing confounding proxies. Moreover, in practice, $W$ may be collected after a decision is made, which prevents them from being used in decision making.
 

 The form of $d_1^\ast$ in (\ref{eq: optimal d_1}) shows that the optimal ITR in 
 $\calD_1$ 
 incorporates the effect of treatment-inducing confounding proxies $Z$ on $Y$ unless 
$
\E\left\{h_0(W, 1, L) \, | \, L, Z\right\} - \E\left\{h_0(W, -1, L) \, | \, L, Z\right\}
$
 is independent of $Z$. This is reasonable 
 since $Z$ may contain some useful information 
 of $U$, which can 
 help improve the value function. This is studied by our simulation in Section \ref{sec:simu}. Obviously 
 $V(d_1^\ast) \leq V(d^\ast)$,  
 but due to unmeasured confounding, $V(d_1^\ast)$ is the best we can obtain within $\calD_1$ under the assumptions in Theorem \ref{thm: identify D1}.
 To illustrate how to obtain $d_1^\ast$ via $h_0$, a concrete example is provided in  Supplementary Material S1. 

\subsection{Optimal ITR via Treatment Confounding Bridge} \label{sec:identifytreatment}

In this section, we provide an alternative identification result for the value function, and thus that for a different restricted optimal ITR, without using the outcome confounding bridge as in Section \ref{sec:identifyoutcome}. It requires the following assumptions, which are different from Assumptions \ref{ass: completeness (1)} and \ref{ass: outcome bridge} and were originally used by 
\cite{cui2020semiparametric} to study semiparametric proximal causal inference 
to identify the average treatment effect under unmeasured confounding.



\begin{assumption}[Completeness]\label{ass: completeness treatment}
	\begin{itemize}
				\item[(a)] For any $a \in \calA, l \in \calL$ and measurable function $g$ defined on $\calU$, if  $\E\left\{g(U) \mid W, A = a, L = l\right\} = 0$ almost surely, then $g(U) = 0$ almost surely.
				\item[(b)] For any $a \in \calA$, $l \in \calL$ and measurable function $g$ defined on $\calW$, if  $\E\left\{g(W) \mid Z, A = a, L = l\right\} = 0$ almost surely, then $g(W) = 0$ almost surely.
			\end{itemize}
\end{assumption}

\begin{assumption}[Treatment confounding bridge]\label{ass: treatment bridge}
	There exists a treatment confounding bridge function $q_0$ defined on $(\calZ, \calA, \calL)$ such that 
	\begin{align}\label{eq: treatment bridge}
	\frac{1}{\Pr(A = a \mid W, L)} = \E\left\{q_0(Z, A, L) \mid  W, A = a, L\right\},
	\end{align}
almost surely
\end{assumption}

Assumptions \ref{ass: completeness treatment} and \ref{ass: treatment bridge} play a similar role of 
Assumptions \ref{ass: completeness (1)} and \ref{ass: outcome bridge}. 
In particular, Assumption \ref{ass: treatment bridge} establishes a link between $Z$ and $A$ and \eqref{eq: treatment bridge} 
is also a linear integral equation of the first kind. The existence of such $q_0$ satisfying \eqref{ass: treatment bridge} 
can be guaranteed by Assumption \ref{ass: completeness treatment}~(b) combined with some regularity conditions given in \cite{cui2020semiparametric}. 
Similar to 
Theorem \ref{thm: identify D1}, the value function can be nonparametrically identified, but via the treatment confounding bridge function $q_0$ and over a different class of ITRs. 


\begin{theorem}\label{thm: identify D2}
	Let $\calD_2$ be the class of ITRs that map from $(\calL, \calW)$ to 
	$\calA$. Under Assumptions \ref{ass: consistency}-\ref{ass: latent exchangeability}, \ref{ass: completeness treatment}(a) and \ref{ass: treatment bridge}, 
 for any $d_2\in\calD_2$,
	the value function $V(d_2)$
	can be nonparametrically identified by 
	\begin{align}\label{eq: treatment identification with W}
	V(d_2) = \E\left\{Yq_0(Z, A, L)\mathbb{I}(d_2(L, W) = A)\right\}, 
	\end{align}
	The restricted in-class optimal ITR within $\calD_2$, defined as $
		d^\ast_2 \in \argmax_{d_2 \in \calD_2} V(d_2)$,
	can be 	almost surely identified by 
	\begin{align}\label{eq: optimal d_2}
	d^\ast_2(L, W) 
	& = \sign \left[\E\left\{Y\mathbb{I}(A=1)q_0(Z, 1, L) \mid L, W\right\} - \E\left\{Y\mathbb{I}(A=-1)q_0(Z, -1, L)  \mid L, W\right\}\right]. 
	\end{align}
\end{theorem}


The proof of Theorem \ref{thm: identify D2} is given in Supplementary Material S2.
Similar to the discussion after Theorem \ref{thm: identify D1}, Theorem \ref{thm: identify D2} shows that the value function is identifiable over $\calD_2$ in terms of the treatment confounding bridge $q_0$ despite unmeasured confounding. 
As kindly pointed out by one reviewer, the result in Theorem \ref{thm: identify D2} may be useful to identify a restricted optimal ITR when $W$ should be included as a decision variable but $Z$ should not. For instance, as in the time series example in \cite{miao2018confounding}, if there is no feedback effect, the future exposure may serve as $Z$ and the past outcomes can be $W$, so it is reasonable to include $W$ as a decision variable for the treatment regime but not $Z$.

\subsection{Optimal ITR via Both 
Confounding Bridges} \label{sec:identifyboth}

In Sections \ref{sec:identifyoutcome} and \ref{sec:identifytreatment} above, we have established the identification results for the value function and its corresponding optimal ITR in terms of the outcome confounding bridge and treatment confounding bridge respectively. A natural question is whether it is possible to obtain 
a broader identification result if both confounding bridges coexist. The answer is affirmative. Explicitly, 
if all Assumptions 	\ref{ass: consistency}--\ref{ass: latent exchangeability}, \ref{ass: completeness (1)}~(a),  \ref{ass: outcome bridge}, \ref{ass: completeness treatment}~(a), and \ref{ass: treatment bridge} hold,  
by Theorems \ref{thm: identify D1} and \ref{thm: identify D2}, 
clearly $V(d_4)$ can be identified for any $d_4 \in \calD_1 \cup \calD_2$ by 
\begin{align*}
    V(d_4)  = \mathbb{I}(d_4 \in \calD_1)\E\left\{h_0(W, d_4(L, Z), L)\right\} 
    + \mathbb{I}(d_4 \in \calD_2)\E\left\{Yq_0(Z, A, L)\mathbb{I}(d_4(L, W) = A)\right\}.
\end{align*}
Then the restricted optimal ITR in $\calD_1 \cup \calD_2$  is defined as
\begin{equation}
\label{eq: optimal d_4}
d^\ast_4 \in \argmax_{d_4 \in \calD_1 \cup \calD_2} V(d_4).
\end{equation}
Note that, despite the coexistence of both counfounding bridges, we are still unable to identify the value function $V(d)$ based on observed data over $d \in \tilde \calD$, 
which refers to the class of all ITRs mapping from $(L, W, Z)$ to $\calA$,  since $\E\left\{Y(a) \mid L, W, Z\right\}$ for any $a\in \calA$ is not nonparametrically identifiable due to unmeasured confounding. It is unknown whether there exists a sufficient and necessary condition to identify the conditional average treatment effect given all observed covariates when there exists unmeasured confounding.


To conclude this section, we provide Table \ref{tab: identification} 
that summarizes the identification results we have developed and their required assumptions. More discussions and practical suggestions on these results can be found at the end of Section \ref{sec:policy}.


\begin{table}[!htp]
	\centering
		\caption{A summary of 
		optimal ITR identification results. 
		}
	\label{tab: identification}
	\begin{tabular}{c|c|c}
		\hline\hline
		\textbf{Assumptions} & \textbf{ITR Class} & \textbf{Restricted Optimal ITR}\\
		\hline\hline
		\ref{ass: consistency}-\ref{ass: latent exchangeability}, \ref{ass: completeness (1)}~(a) and \ref{ass: outcome bridge} & $ \calD_1: (\calL, \calZ)\rightarrow \calA $ & $d^\ast_1$ in \eqref{eq: optimal d_1} through $h_0$\\
		\hline
		\ref{ass: consistency}-\ref{ass: latent exchangeability}, \ref{ass: completeness treatment}~(a) and \ref{ass: treatment bridge} & $ \calD_2: (\calL, \calW)\rightarrow \calA $ & $d^\ast_2$ in \eqref{eq: optimal d_2} through $q_0$\\
		\hline
		\ref{ass: consistency}-\ref{ass: latent exchangeability}, \ref{ass: completeness (1)}~(a),  \ref{ass: outcome bridge}, \ref{ass: completeness treatment}~(a) and \ref{ass: treatment bridge} & $ \calD_1 \cup \calD_2 $ & $d^\ast_4$ in \eqref{eq: optimal d_4} through $q_0$ or $h_0$\\
		\hline\hline
	\end{tabular}
\end{table}


\section{Proximal Policy Learning}\label{sec:policy}

In this section, based on the identification results established in Section \ref{sec:identification under U}, we propose several methods to estimate restricted in-class optimal ITRs based on observed 
$n$ independent and identically distributed samples 
$\left\{(L_i, Z_i, W_i, A_i, Y_i): i=1, \ldots, n\right\}.$
In Section \ref{sec:estimatebridge} we first propose the estimation methods for the confounding bridge functions $h_0$ and $q_0$ defined in  Assumptions \ref{ass: outcome bridge} and \ref{ass: treatment bridge} respectively. In Section \ref{sec:ITRlearning}, based on the estimates of $h_0$ and $q_0$, we propose 
several
classification-based methods to estimate the restricted optimal ITRs $d_1^\ast$, $d_2^\ast$, and $d_4^\ast$ defined in Section \ref{sec:identification under U} under their 
corresponding assumptions respectively.
In Section \ref{sec:DRlearning}, under the condition that Assumptions \ref{ass: consistency}-\ref{ass: latent exchangeability}, \ref{ass: completeness (1)}~(a),  \ref{ass: outcome bridge}, \ref{ass: completeness treatment}~(a) and \ref{ass: treatment bridge} hold, we propose an augmented inverse probability weighted (AIPW)-type classification-based method for estimating the optimal ITR in a new class of ITRs. 
For ease of presentation, let Assumptions \ref{ass: completeness (1)}~(b) and \ref{ass: completeness treatment}~(b) always \textit{hold} hereafter so that $h_0$ and $q_0$ can be uniquely identified.


\subsection{Estimation of Confounding Bridge Functions}\label{sec:estimatebridge}


Here we introduce nonparametric estimations of outcome and treatment confounding bridge functions $h_0$ and $q_0$ defined in Assumptions \ref{ass: outcome bridge} and \ref{ass: treatment bridge} respectively. 
\vspace{0.2cm}

\noindent
\textbf{Estimating 
$h_0$:} 
Equation \eqref{eq: outcome bridge} in Assumption  \ref{ass: outcome bridge} is equivalent to 
\begin{align}\label{eq: estimating equation for h}
	\E\left\{Y - h_0(W, A, L) \mid Z, A, L\right\} = 0,
\end{align}
which is known as the instrumental variable model or conditional moment restriction model and has been well studied in econometrics and statistics \citep[e.g.,][]{chamberlain1992efficiency,newey2003instrumental, ai2003efficient,blundell2007semi,chen2007large,chen2012estimation}.

Here we adopt the min-max estimation method by  \cite{dikkala2020minimax} 
to estimate $h_0$ nonparametrically as follows:
  {\small\begin{align}\label{eq: min-max estimation of h}
\hat h_0=\argmin_{h \in \calH} \sup_{f \in \calF}\left[ \frac{1}{n}\sum_{i = 1}^n \left\{Y_i - h(W_i, A_i, L_i)\right\}f(Z_i, A_i, L_i)- \lambda_{1,n} \|f\|^2_{\calF} - \|f\|^2_{2, n} \right]+ \lambda_{2,n}\| h\|^2_{\calH},
 \end{align}
}
where $\lambda_{1, n}>0$ and $\lambda_{2, n}>0$ are 
tuning parameters, $\| \bullet \|_{2,n}$ is the empirical $\ell^2$ norm, i.e., $\| f \|_{2,n} = \sqrt{n^{-1}\sum_{i = 1}^n f^2(Z_i, A_i, L_i)}$, and 
$\calH$ and $\calF$ are some 
functional classes, e.g., 
reproducing kernel Hilbert spaces (RKHS),
with their corresponding norms $\|\bullet\|_\calF$ and $\|\bullet\|_\calH$ respectively. 

The rationale behind  \eqref{eq: min-max estimation of h} is 
the following population version of the min-max optimization problem when 
$\lambda_{1, n}, \lambda_{2, n} \rightarrow 0$ 
as $n \rightarrow \infty$: 
\begin{align}\label{eq: min-max population of h}
\min_{h \in \calH} \sup_{f \in \calF} \left( \E \left[\left\{Y - h(W, A, L)\right\}f(Z, A, L)\right]- \E\left\{f^2(Z, A, L)\right\} \right). 
\end{align}
If 
$
2^{-1}\E\left\{h_0(W, A, L)- h(W, A, L) \mid Z, A, L\right\} \in \calF 
$ 
for every $h \in \calH$, 
then 
the optimization \eqref{eq: min-max population of h} above is equivalent to 
$$
\min_{h \in \calH} \E\left(\left[\E\left\{Y - h(W, A, L) \mid Z, A, L\right\}\right]^2\right).
$$
If we further assume $h_0 \in \calH$, then $h_0$ is the unique global minimizer of \eqref{eq: min-max population of h}. 
Hence the min-max formulation \eqref{eq: min-max estimation of h} is valid.

\vspace{0.2cm}

\noindent
\textbf{Estimating 
$q_0$:} 
Equation \eqref{eq: treatment bridge} in Assumption \ref{ass: treatment bridge} indicates that 
for every $a \in \calA$,
$$
\E\left[\left\{\mathbb{I}(A = a) q_0(Z, A, L) - 1 \right\} \mid W, L\right] = 0.
$$ 
Similar to \eqref{eq: min-max estimation of h}, 
we estimate $q_0$ by 
the following min-max optimization: For each $a \in \calA$, 
\begin{align}\label{eq: min-max estimation of q}
\hat q_0(\bullet, a, \bullet) = \argmin_{q \in \calQ} \left(\sup_{g \in \calG}
\left[\frac{1}{n}\sum_{i = 1}^n \left\{\mathbb{I}(A_i = a) q(Z_i, a, L_i) - 1 \right\}g(W_i, L_i)- \mu_{1,n} \|g\|^2_{\calG} - \|g\|^2_{2, n} \right] + \mu_{2,n}\| q\|^2_{\calQ} \right),
\end{align}
where $\mu_{1, n} > 0$ and $\mu_{2, n} > 0$ are
tuning parameters, 
and 
$\calQ$ and $\calG$ are 
functional classes 
with their corresponding norms $\|\bullet\|_\calQ$ and $\|\bullet\|_\calG$ respectively. 


Generally one may use any functional class for $\calH$ and $\calF$ in \eqref{eq: min-max estimation of h} and for $\calQ$ and $\calG$ in \eqref{eq: min-max estimation of q}. If they are all specified as RKHS, due to the representer theorem, both $\hat h_0$ and $\hat q_0$ will have finite-dimensional representations which lead to fast computations. See Supplementary Material S3 
for details.  


\subsection{Outcome and Treatment Proximal Learning}
\label{sec:ITRlearning}

With estimated $h_0$ and $q_0$, we propose to 
estimate the restricted optimal ITRs $d_1^\ast$, $d_2^\ast$, and $d_4^\ast$ defined in Section \ref{sec:identification under U} using classification-based methods, which are similar to those of  \cite{zhao2012estimating} and \cite{ zhang2012robust} under no unmeasured confounding. 


\vspace{0.2cm}

\noindent
\textbf{Outcome Proximal Learning of $d_1^\ast$:} 
According to \eqref{eq: optimal d_1}, under the assumptions in Theorem \ref{thm: identify D1}, finding $d_1^\ast$ is 
equivalent to minimizing the following 
classification error 
$$
\E\left[\left\{h_0(W, 1, L) - h_0(W, -1, L)\right\}\mathbb{I}\left(d_1(L, Z) \neq 1\right)\right],
$$
over all $d_1 \in \calD_1$. Since each $d_1 \in \calD_1$ can be written as $d_1(L, Z) = \sign\{r_1(L, Z)\}$ for some measurable function $r_1$ defined on $(\calL, \calZ)$, 
we can rewrite $\mathbb{I}(d_1(L, Z) \neq 1) = \mathbb{I}(r_1(L, Z) < 0)$ where  $\sign(0) \triangleq 1$. Then the optimization problem above becomes
$$
\min_{r_1} \, \, \E\left[\left\{h_0(W, 1, L) - h_0(W, -1, L)\right\}\mathbb{I}\left(r_1(L, Z) < 0\right)\right].
$$
Given the observed data and 
$\hat h_0$, we solve its empirical version 
$$
\min_{r_1} \frac{1}{n}\sum_{i = 1}^n\hat \Delta (W_i, L_i)\mathbb{I}\left(r_1(L_i, Z_i) < 0\right), 
$$
where $\hat \Delta (W, L) = \hat h_0(W, 1, L) - \hat h_0(W, -1, L)$, or equivalently 
\begin{align}\label{weighted misclassification}
	& 
	\min_{r_1} \frac{1}{n}\sum_{i = 1}^n \bigg|\hat \Delta (W_i, L_i)\bigg|\mathbb{I}\left(\sign\left(\hat \Delta (W_i, L_i)\right) r_1(L_i, Z_i) < 0 \right). 
\end{align}

The equivalence above motivates us to use a convex surrogate function to replace the indicator function since all weights $|\hat \Delta (W_i, L_i)|$ are non-negative. Similar to \cite{zhao2012estimating} and \cite{zhao2019efficient}, we adopt the hinge loss $\phi(t) = \max(1-t, 0)$ and consider $r_1 \in \calR_1$, a pre-specified class of functions defined on $(\calL, \calZ)$, e.g., a RKHS,  
to obtain the estimated optimal ITR by $\hat d^\ast_1 = \sign (\hat r_1)$ where 
\begin{align}\label{eq: proximal learning 1}
\hat r_1 \in \argmin_{r_1 \in \calR_1} \left\{ \frac{1}{n}\sum_{i = 1}^n \bigg|\hat \Delta (W_i, L_i)\bigg|\phi\left(\sign\left(\hat \Delta (W_i, L_i)\right) r_1(L_i, Z_i)\right) + \rho_{1, n}\| r_1 \|_{\calR_1}^2 \right\}.
\end{align}
Here $\| \bullet \|_{\calR_1}$ is the 
norm of $\calR_1$ 
and $\rho_{1, n} > 0$ is a tuning parameter. 
The optimization in \eqref{eq: proximal learning 1} is convex and thus can be solved efficiently. See Algorithm 1 of Supplementary Material S3. 

\vspace{0.2cm}

\noindent
\textbf{Treatment Proximal Learning of $d_2^\ast$:} Similar to learning $d_1^\ast$, by \eqref{eq: optimal d_2} and with $\hat q_0$, we propose to first find $\hat r_2$, the solution to the following minimization 
\begin{align}\label{eq: proximal learning 2}
\min_{r_2 \in \calR_2} \left\{\frac{1}{n}\sum_{i = 1}^n\bigg|Y_i\hat q_0 (Z_i, A_i,  L_i)\bigg|\phi\left(A_i\sign\left(Y_i\hat q_0 (Z_i, A_i,  L_i)\right) r_2(L_i, W_i)\right) + \rho_{2, n} \| r_2 \|_{\calR_2}^2 \right\},
\end{align}
where $\| \bullet \|_{\calR_2}$ is the 
norm of a pre-specified class of functions $\calR_2$ defined on $(\calL, \calW)$, e.g., a RKHS. 
Then the estimated optimal ITR is obtained by $\hat d^\ast_2 = \sign(\hat r_2)$. 
See details in Algorithm 4 of  Supplementary Material S3.



\vspace{0.2cm}

\noindent
\textbf{Maximum Proximal Learning of $d_4^\ast$:} By \eqref{eq: optimal d_4} and Assumptions \ref{ass: consistency}-\ref{ass: treatment bridge}, 
obtaining $d^\ast_4$ is  equivalent to solving  
\begin{equation}\label{eq: combine learning}
	\max \left\{\max_{d_1 \in \calD_1} V(d_1), \max_{d_2 \in \calD_2} V(d_2)  \right\}.
\end{equation}
Therefore, we combine the learning methods in \eqref{eq: proximal learning 1} and \eqref{eq: proximal learning 2} and use a cross-validation procedure to find the estimated optimal ITR $\hat d^\ast_4$, i.e., the better decision rule between $\hat d^\ast_1$ and $\hat d^\ast_2$. 
See details in Algorithm 5 of  Supplementary Material S3. 

\subsection{Doubly Robust Proximal Learning}\label{sec:DRlearning}


In some applications where $W$ and $Z$ are not observable in future decision makings, one may be interested in ITRs only based on $L$.  
Let $\calD_3$ be the class of all ITRs that map from $\calL$ to $\calA$ and denote $d^\ast_3 \in \argmax_{d_3 \in \calD_3} V(d_3)$.
In this section, we propose an estimation method for $d^\ast_3$. 

Apparently, if all 
Assumptions \ref{ass: consistency}--\ref{ass: treatment bridge} are satisfied, 
the value function $V(d_3)$ can be identified for any $d_3 \in \calD_3$ via either $h_0$ or $q_0$. 
This motivates us to develop a doubly robust estimator for $V(d_3)$ over $\calD_3$, in the sense that the value function estimator is consistent as long as one of $h_0$ and $q_0$ is modelled correctly. 
This can provide a protection against potential model misspecifications of $h_0$ and/or $q_0$. 
The foundation of our method is the efficient influence function of $V(d_3)$ given below. 
\begin{theorem}
	Under Assumptions \ref{ass: consistency}-\ref{ass: treatment bridge} and some regularity conditions given in Theorem 3.1 of \cite{cui2020semiparametric}, the efficient influence function of $V(d_3)$ is
	\begin{align}\label{eq: EIF}
	 \mathbb{I}(A = d_3(L))q_0(Z, A, L)\left\{Y - h_0(W, A, L)\right\} + h_0(W, d_3(L), L) - V(d_3),
	\end{align}
	for any given $d_3 \in \calD_3$.
\end{theorem}
The proof is similar to that of \cite{cui2020semiparametric} and thus omitted.
Define 
\begin{align}\label{eq: doubly robust weight}
	C_{1}(Y, L, W, Z; h_0, q_0) & =\mathbb{I}(A=1)q_0(Z, 1, L)\left\{Y - h_0(W, 1, L) \right\} + h_0(W, 1, L),\\
	\text{and}\quad C_{-1}(Y, L, W, Z; h_0, q_0) & = \mathbb{I}(A=-1)q_0(Z, -1, L)\left\{Y - h_0(W, -1, L)\right\} + h_0(W, -1, L).\nonumber
\end{align}
Based on 
\eqref{eq: EIF}, we can estimate $V(d_3)$ for each $d_3 \in \calD_3$ by
\begin{align}\label{eq: doubly robust value function}
\hat V^{DR}(d_3) & = \frac{1}{n}\sum_{i = 1}^n\left\{C_1(Y_i, L_i, W_i, Z_i; \hat h_0, \hat q_0)\mathbb{I}(d_3(L_i) = 1) \right. \nonumber \\ 
& \left. \quad + C_{-1}(Y_i, L_i, W_i, Z_i; \hat h_0, \hat q_0)\mathbb{I}(d_3(L_i) = -1)\right\}.
\end{align}
It can be shown in Proposition \ref{prop: doubly robust} below that $\hat V^{DR}(d_3)$ enjoys the doubly robust property in the sense that $\hat V^{DR}(d_3)$ is a consistent estimator for $V(d_3)$ for each $d_3 \in \calD_3$ as long as one of 
$q_0$ and $h_0$ is modeled correctly, not necessarily both. 
Following 
similar arguments in Section \ref{sec:ITRlearning}, 
the estimated optimal ITR we propose is $\hat d_3^{DR} = \sign(\hat r_3^{DR})$ and $\hat r_3^{DR}$ is obtained by 
\begin{align}\label{eq: proximal learning 4}
&\hat r_3^{DR} \in  \argmin_{r \in \calR_3} \left[ \frac{1}{n}\sum_{i = 1}^n\left\{ \bigg|C_1(Y_i, L_i, W_i, Z_i; \hat h_0, \hat q_0) \bigg|\phi\left(\sign\left(C_1(Y_i, L_i, W_i, Z_i; \hat h_0, \hat q_0)\right)r(L_i) \right) \right. \right. \nonumber \\
	& \left. \left. + \bigg|C_{-1}(Y_i, L_i, W_i, Z_i; \hat h_0, \hat q_0)\bigg|\phi\left(-\sign \left(C_{-1}(Y_i, L_i, W_i, Z_i; \hat h_0, \hat q_0)\right)r(L_i) \right)\right\} + \rho_{3, n}\|r\|^2_{\calR_3} \right],
\end{align}
where $\calR_3$ is a class of functions defined on $\calL$ with norm $\|\bullet\|_{\calR_3}$ and $\rho_{3, n} > 0$ is a tuning parameter. 



In practice, we can apply the 
cross-fitting technique  \citep{bickel1982adaptive} to 
remove 
the dependence between the nuisance function estimates $\hat h_0$ and $\hat q_0$, and the resulting estimated optimal ITR. 
Thanks to this technique, our proposed learning method does not require restrictive conditions on both nuisance function estimations (e.g., Donsker conditions) to avoid losing efficiency due to nuisance function estimations. 
See \cite{chernozhukov2018double} and \cite{athey2017efficient} for more details.

To implement cross-fitting, we randomly split data into $K$ folds and apply the following procedure: 
first use the $k$-th fold to  
obtain $\hat h_0^{(k)}$ and $\hat q_0^{(k)}$, the estimates of 
$h_0$ and $q_0$ respectively, $k=1,\ldots, K$; then 
for each $k=1,\ldots, K$, compute the decision function by 
solving \eqref{eq: proximal learning 4} based on $\hat h_0^{(k)}$ and $\hat q_0^{(k)}$ using all data except the $k$-th fold;
finally aggregate all $K$ decision rules to obtain our final estimated optimal ITR. 
More details of this algorithm can be found in Supplementary Material S3.
With some abuse of notations, we denote the final decision function by $\hat r_3^{DR}$ and the corresponding estimated optimal ITR by $\hat{d}^{DR}_3=\sign(\hat r_3^{DR})$ . 
To conclude this section, we present
Table \ref{tab:compare} which summarizes each optimal ITR learning and its corresponding assumptions. Note that compared with Table \ref{tab: identification}, we need additional Assumptions \ref{ass: completeness (1)}~(b) and \ref{ass: completeness treatment}~(b) so that $q_0$ and $h_0$ can be uniquely identified 
respectively.


\begin{remark}\label{rmk:tradeoff}
Table \ref{tab:compare} reveals two important trade-offs for proximal ITR learning. The first trade-off is between 
improving the value function and imposing more untestable assumptions. For example, a comparison between $\hat d_4^\ast$ and $\hat d_1^\ast$ shows that 
although we can identify a much larger class of ITRs 
and thus a potentially higher value function for $\hat d_4^\ast$ than $\hat d_1^\ast$, 
learning $\hat d_4^\ast$ requires an additional assumption on 
the existence of a  treatment confounding bridge.  
Conversely, learning $d_1^\ast$ needs fewer assumptions than learning $\hat d_4^\ast$ and is thus more reliable, but its sub-optimality gap to the globally optimal ITR could  be larger compared with $d_4^\ast$. 
The second trade-off is between the estimation robustness and 
the value function improvement. A comparison between $d^\ast_3$ and $d^\ast_4$ shows that to estimate $d^\ast_4$ accurately which belongs to a larger class of ITRs and corresponds to a better value function than $d^\ast_3$, we require both nuisance functions to be estimated consistently. However, if one is willing to consider $\calD_3$, a smaller class of ITRs, one can estimate $d^\ast_3$ consistently as long as one of the two nuisance functions is estimated consistently, but its corresponding value function might be smaller than that of $d^\ast_4$.
\end{remark}


\begin{remark}\label{rmk:practice}
Due to the aforementioned two trade-offs, we suggest practitioners consider a conservative way of choosing the final optimal ITR estimate. 
For example, since learning $d_1^\ast$ or $d_2^\ast$ requires fewer assumptions than learning the other optimal ITRs, the optimal ITR obtained by either the outcome proximal learning or treatment proximal learning is likely to be more trustworthy than those by the other two methods. For example, 
to determine if a subgroup of patients can potentially benefit more from a new treatment than from the standard care, one may 
recommend the new treatment if both of these two optimal ITR estimates agree. In practice, to hopefully achieve conservativeness, robustness and value function improvement simultaneously, one may also recommend treatments for patients when most of the estimated optimal ITRs agree or use the ITR selected by cross-validation.
The ensemble and cross-validation approaches have been applied in our real data application.
More details are in Section \ref{sec: real data} and Supplementary Material S6.
\end{remark}


\begin{table}[H]
	\centering
		\caption{A summary of the proposed proximal learning methods for optimal ITRs. 
		}
	\label{tab:compare}
	\scalebox{0.7}{
	\begin{tabular}{c|c|c|c}
		\hline\hline
		\textbf{Assumptions} & \textbf{ITR Class} & \textbf{Proximal Learning} & \textbf{Estimated Optimal ITR}\\
		\hline\hline
		\ref{ass: consistency}-\ref{ass: latent exchangeability}, \ref{ass: completeness (1)}~(a), \ref{ass: outcome bridge} and \ref{ass: completeness treatment}~(b) 
		& $ \calD_1 $ & Outcome proximal learning \eqref{eq: proximal learning 1} & $\hat d_1^\ast$\\
		\hline
		\ref{ass: consistency}-\ref{ass: latent exchangeability}, \ref{ass: completeness (1)}~(b), \ref{ass: completeness treatment}~(a), and \ref{ass: treatment bridge} & $ \calD_2 $ & Treatment proximal learning \eqref{eq: proximal learning 2}& $\hat d_2^\ast$\\
		\hline
		\ref{ass: consistency}- \ref{ass: treatment bridge}  & $ \calD_3 $ & Doubly robust proximal learning \eqref{eq: proximal learning 4} &  $\hat d_3^{DR}$\\
		\hline
		\ref{ass: consistency}- \ref{ass: treatment bridge}  & $ \calD_1 \cup \calD_2 $ & Maximum proximal learning \eqref{eq: combine learning} & $\hat d_4^\ast$\\
		\hline\hline
	\end{tabular}}
\end{table}

\section{Theoretical Results}\label{sec:theory}

In this section, we 
develop the theoretical properties of our proposed methods, or specifically the finite sample excess risk bound for each estimated optimal ITR. 
For brevity, we only provide the results for 
the doubly robust optimal ITR estimator $\hat d^{DR}_3$, but similar results can be obtained for the other estimators. 
We first show the doubly robust property of 
$\hat V^{DR}(d_3)$ in \eqref{eq: doubly robust value function} 
for any $d_3 \in \calD_3$. 
\begin{proposition}\label{prop: doubly robust}
Under Assumptions \ref{ass: consistency}-\ref{ass: treatment bridge}, if either $\hat h_0$ can consistently estimate $h_0$ in the sup-norm or $\hat q_0$ can consistently estimate $q_0$ in the sup-norm, then $\hat V^{DR}(d_3)$ is a consistent estimator of $V(d_3)$ for any $d_3 \in \calD_3$.
\end{proposition}
The proof is given in Supplementary Material S2.
Next we show Fisher consistency, that is, it is appropriate to replace the indicator function in $\hat V^{DR}$ with the hinge loss as in \eqref{eq: proximal learning 4} to obtain $\hat r_3^{DR}$ and $\hat d_3^{DR} = \sign(\hat r_3^{DR})$ accordingly.
Define the hinge loss based $\phi$-risk by
\begin{align}\label{hinge risk}
	R_\phi(r) = \E\left\{|C_1|\phi(\sign(C_1)r(L)) + |C_{-1}|\phi(\sign(C_{-1})r(L))\right\},
\end{align}
where $C_1$ and $C_{-1}$ denote $C_1(Y, L, W, Z; h_0, q_0)$ and $C_{-1}(Y, L, W, Z; h_0, q_0)$ defined in \eqref{eq: doubly robust weight} respectively for ease of presentation. Let $r^\ast \in \argmin_{r} R_\phi(r)$. 
Then we have the following proposition.
\begin{proposition}\label{prop: finsher consistency}
	Under Assumptions \ref{ass: consistency}-\ref{ass: treatment bridge}, $d_3^\ast(L) = \sign(r^\ast(L))$.
\end{proposition}

The proof 
of Proposition \ref{prop: finsher consistency} is omitted since it can be derived by following similar arguments in the proof of 
Proposition 3.1 of \cite{zhao2019efficient}.
Proposition \ref{prop: finsher consistency}  essentially states that replacing the indicator function by the hinge loss function does not change the goal of finding the optimal ITR. 
We can further link the original value function with the $\phi$-risk as follows. 
\begin{proposition}\label{prop: excess bound}
	Under Assumptions \ref{ass: consistency}-\ref{ass: treatment bridge}, $V(d^\ast_3) - V(d) \leq R_\phi(r) - R_\phi(r^\ast)$ for any $d = \sign(r(L))$.
\end{proposition}
The proof of Proposition \ref{prop: excess bound} is omitted due to the similar reason as that of Proposition \ref{prop: finsher consistency}. 
Proposition \ref{prop: excess bound}
implies that the value function difference between the in-class optimal ITR and any other ITR in $\calD_3$ can be bounded by their $\phi$-risk difference. Therefore the convergence rate of the value function of $\hat d_3^{DR}$ can be bounded by the convergence rate of the $\phi$-risk of our estimated decision function $\hat r_3^{DR}$. 
To establish the finite sample excess risk bound for $\hat d_3^{DR}$, 
we make the following technical assumptions in addition to Assumptions \ref{ass: consistency}-\ref{ass: treatment bridge}.
\begin{assumption}\label{ass: bounded}
	There exists a 
	constant $C_1 > 0$ such that $\max\left\{| Y |, \| h_0 \|_\infty, \| q_0 \|_\infty\right\} \leq C_1$.
\end{assumption}

\begin{assumption}\label{ass: entropy}
	There exist 
	constants $A > 0$ and  $v > 0$ such that $\sup_{Q}N(\calR_3, Q, \varepsilon\|F\|_{Q, 2}) \leq \left(A/\varepsilon\right)^v$ for all $0 < \varepsilon \leq 1$, where $N(\calR_3, Q, \varepsilon\|F\|_{Q, 2})$ denotes the covering number of the space $\calR_3$, $F$ is the envelope function of $\calR_3$,  $\|\bullet\|_{Q, 2}$ denotes the $\ell_2$-norm under some finitely discrete 
	probability measure $Q$ on $(L, \calL)$, and the supremum is taken over all such probability measures. 
\end{assumption}

\begin{assumption}\label{ass: nuisance convergence rate}
	The nuisance function estimators $\hat h^{(k)}_0$ and $\hat q^{(k)}_0$ obtained from the $k$-th fold of the data in the  cross-fitting procedure in Section \ref{sec:DRlearning}
	satisfy that there exist constants $\alpha > 0$ and $\beta > 0$ such that $\left\|h_0(W, a, L) - \hat h^{(k)}_0(W, a, L) \right\|_{P, 2}^2 = O(n^{-2 \alpha})$ and $\left\|q_0(Z, a, L) - \hat q^{(k)}_0(Z, a, L) \right\|_{P, 2}^2 = O(n^{-2 \beta})$ uniformly for all $a \in \calA$ and $1 \leq k \leq K$, 
	where $P$ is the underlying probability measure associated with $(L, W, Z, A, Y)$. In addition, 
	$\max\left\{ \|\hat q_0 \|_\infty, \|\hat h_0 \|_\infty \right\} \leq C_2$ for some  constant $C_2 > 0$.
\end{assumption}
Assumption \ref{ass: bounded} requires that the outcome $Y$ is bounded, 
which is commonly assumed in the literature \citep[e.g.,][]{zhao2012estimating,zhou2017residual}. 
Assumption \ref{ass: bounded} also requires $h_0$ and $q_0$ 
to be both uniformly bounded. Similar conditions on nuisance functions have also been used in ITR learning under no unmeasured confounding \citep[e.g.,][]{zhao2019efficient}. Assumption \ref{ass: entropy} essentially states 
that 
$\calR_3$ is of a certain Vapnik-Chervonenkis class \citep[see Definition 2.1 of ][]{chernozhukov2014gaussian}.
Assumption \ref{ass: nuisance convergence rate} imposes high-level conditions on the convergence rates of the estimated nuisance functions when the cross-fitting technique is applied. Under proper conditions \citep[see,][for details]{dikkala2020minimax}, the nuisance function estimators obtained  via \eqref{eq: min-max estimation of h} and \eqref{eq: min-max estimation of q} can satisfy Assumption \ref{ass: nuisance convergence rate}.

Define the approximation error of $r^\ast$ in terms of the $\phi$-risk by
\begin{align*}
\mathbb{A}(\rho_{3, n}) &= \inf_{r \in \calR_3} \left\{R_\phi(r) + \rho_{3, n} \| r\|_{\calR_3}^2 \right\} - R_{\phi}(r^\ast),
\end{align*}
and consider the irreducible value gap between $d^\ast_3$ and the global optimal ITR $d^\ast$ by
\begin{align*}
\mathbb{G}(d^\ast_3) &= V(d^\ast) - V(d^\ast_3).
\end{align*}
For two positive sequences $\{a_n: n \geq 1\}$ and $\{b_n: n \geq 1\}$, we denote $ a_n \lesssim b_n$ if $\limsup_{n \rightarrow \infty}a_n/b_n < \infty$. 
The finite sample excess risk bound for 
$\hat d^{DR}_3$ is given as follows.
\begin{theorem}\label{thm: regret bound}
	Suppose that Assumptions \ref{ass: consistency}-\ref{ass: treatment bridge} and \ref{ass: bounded}-\ref{ass: nuisance convergence rate} hold. 
	If $\rho_{3, n} \leq 1$, then for any $x > 0$, with probability at least $1 - \exp(-x)$, we have 
	\begin{align}\label{eq: regret bound (II)}
	V(d^\ast) - V(\hat d^{DR}_3) &\lesssim \mathbb{G}(d^\ast_3) + \mathbb{A}(\rho_{3, n}) + \sqrt{v}\rho_{3, n}^{-1/2}n^{-1/2} \nonumber \\
	& + \max\{1, x\}n^{-(\alpha + \beta)}\rho_{3, n}^{-1/2} + \max\{1, x\}\sqrt{v}n^{-1/2-\min\{\alpha, \beta \}}\rho_{3, n}^{-1/2}.
	\end{align}
\end{theorem}
The proof of Theorem \ref{thm: regret bound} is given in Supplementary Material S2. 
Theorem \ref{thm: regret bound} decomposes the finite sample excess risk bound for $\hat d^{DR}_3$ into four components. The first component $\mathbb{G}(d^\ast_3)$ is an irreducible error due to unmeasured confounding. The second component $\mathbb{A}(\rho_{3, n})$  is the approximation error determined by the size of 
$\calR_3$ and tuning parameter $\rho_{3, n}$. The third component $\sqrt{v}\rho_{3, n}^{-1/2}n^{-1/2}$ is the estimation error with respect to 
$\calR_3$ assuming that the nuisance functions are known. 
The last two components,
$\max\{1, x\}n^{-(\alpha + \beta)}\rho_{3, n}^{-1/2}$ and  $\max\{1, x\}\sqrt{v}n^{-1/2-\min\{\alpha, \beta \}}\rho_{3, n}^{-1/2}$, 
are the errors 
resulted from estimating the nuisance functions $h_0$ and $q_0$.
These two errors are negligible if  $\alpha +\beta > \frac{1}{2}$, 
which can be guaranteed by many nonparametric estimators,  
including those defined in \eqref{eq: min-max estimation of h} and \eqref{eq: min-max estimation of q} under certain assumptions.


The finite sample error bounds for other proposed proximal learning methods can be similarly derived and are thus omitted here for brevity. 
Similar to that for $\hat d^{DR}_3$, 
each of their value difference bounds can be decomposed into 
aforementioned four components. The magnitude of the irreducible error due to unmeasured confounding is determined by the size of the corresponding identifiable class of treatment regimes. Obviously the irreducible error related to $d^\ast_4$ is the smallest but at the expense of making additional untestable assumptions. 
The approximation and estimation errors with respect to different pre-specified classes of treatment regimes are similar for all our proposed methods. The errors incurred by estimating nuisance functions 
vary among these methods and they depend on the assumptions imposed on the nuisance function estimation. While the result in Theorem \ref{thm: regret bound} is similar to those from 
\citet{zhao2019efficient},
we consider the ITR estimation problem in the presence of unmeasured confounders while \citet{zhao2019efficient}
considered the scenario of no unmeasured confounding. Therefore, the assumptions and proofs for obtaining our result are different from those in \citet{zhao2019efficient}.

\section{Simulation} \label{sec:simu}


In this section we perform a simulation study to evaluate the numerical performance of our proposed optimal ITR estimators under unmeasured confounding, where 
a state-of-the-art estimator under no unmeasured confounding \citep{ zhao2019efficient} and an \textit{oracle} estimator that can access the unmeasured confounders are compared. Due to the length of presentation, we only report key simulation results in this section. Implementation details and additional numerical results such as statistical inference related to $V(d_3)$ for $d_3 \in \calD_3$ can be found in Supplementary Material S3 -- S5. 

\subsection{Simulated Data Generation}
Here we briefly describe our simulated data generating mechanism, which is motivated by \cite{cui2020semiparametric}. The 
details 
are given in
Supplementary Material S4.
\paragraph{Step 1.} We generate the data $(L,A,Z,W,U)$ by the following steps:
\begin{enumerate}
\item [1.1] 
  The covariates $L$ are generated by $L\sim N([0.25 ,0.25]^\top,\diag\{0.25^2,0.25^2\})$.
  Given $L$, the treatment $A$ is generated by a conditional Bernoulli distribution with
\begin{equation*}
\Pr(A=1\mid L) = \left[ 1+\exp \left\{ [0.125, 0.125] L \right\} \right]^{-1}.
\end{equation*}
As shown in Supplementary Material S4, this is compatible with
\begin{equation} \label{eq: simu prop score}
\Pr(A=1\mid U,L) = [1+\exp \left\{ 0.09375+[0.1875, 0.1875]L -0.25U) \right\}]^{-1},
\end{equation}
where $U$ is generated below.
\item [1.2]
Then we generate $(Z,W,U)$ by the following conditional multivariate normal distribution given
$(A,L)$ with parameters given in Supplementary Material S4:
\begin{equation*}
  (Z,W,U)\mid (A,L) \sim N \left(
    \begin{bmatrix}
      \alpha_0+\alpha_a\mathbb{I}(A=1)+\alpha_l^\top L\\
      \mu_0+\mu_a\mathbb{I}(A=1)+\mu_l^\top L\\
      \kappa_0+\kappa_a\mathbb{I}(A=1)+\kappa_l^\top L\\
    \end{bmatrix},
  \begin{bmatrix}
    \sigma_z^2 & \sigma_{zw} & \sigma_{zu}\\
    \sigma_{zw} & \sigma_w^2 & \sigma_{wu}\\
    \sigma_{zu} & \sigma_{wu} & \sigma_u^2
  \end{bmatrix}
  \right).
\end{equation*}
As shown in Supplementary Material S4, they lead to the following compatible model of $q_0$:
\begin{equation*}
q_0(Z,A,L) = 1+\exp \left\{ A(t_0+t_zZ+t_a\mathbb{I}(A=1)+t_l^\top L) \right\},
\end{equation*}
where the values of $t_0, t_z, t_a$ and $t_l$ are specified in Supplementary Material S4. 

\end{enumerate}

\paragraph{Step 2.} Based on the generated $(L, A, Z, W, U)$, 
we generate $Y$ by adding a random noise from the uniform distribution on $[-1, 1]$ to 
\begin{align*}
\E(Y\mid W,U,A,Z,L) &= \E(Y\mid U,A,Z,L) + \omega\{W-\E(W\mid U,A,Z,L)\}\\
  &=\E(Y\mid U,A,L) + \omega\{W - \E(W\mid U,L)\},
\end{align*}
of which details will be specified  in different scenarios below. 

Next we consider the following outcome confounding bridge function $h_0$, which leads to a linear decision function of $d_1^\ast$. We let 
\begin{align*}
  &h_0(W,A,L) = c_0+c_1\mathbb{I}(A=1)+c_2W+c_3^\top L+\mathbb{I}(A=1)(c_4W+c_5^\top L),\text{ and thus}\\
  &\E(Y\mid W,U,A,Z,L) =c_0+c_1\mathbb{I}(A=1)+ c_3^\top L+\mathbb{I}(A=1)c_5^\top L+\omega W\\ 
  &\qquad\qquad+\left\{c_2+c_4\mathbb{I}(A=1)-\omega \right\}\left\{\mu_0+\mu_l^\top L+\frac{\sigma_{wu}}{\sigma_u^2}(U-\kappa_0-\kappa_l^\top L)\right\},
\end{align*}
where the values of $c_0$--$c_5$ and $\omega$ are specified in Table \ref{tab: simu_linear} and the other 
parameters 
are given in Supplementary Material S4.
Accordingly the global optimal ITR is
\begin{align*}
 d^{*}(L,U) &= \sign [\E(Y\mid L,U,A=1) - \E(Y\mid L,U,A=-1)]\\
 &= \sign\left[c_1+c_5^\top L + c_4 \left\{ \mu_0+\mu_l^\top L+ \frac{\sigma_{wu}}{\sigma_u^2}(U-\kappa_0-\kappa_l^\top L)\right\}\right].
\end{align*}
As shown in Table \ref{tab: simu_linear}, we consider two scenarios.  
In Scenario L1, the global optimal ITR cannot be identified, but the conditional treatment effect $\E(Y\mid L,U,A=1) - \E(Y\mid L,U,A=-1)$ depends on both $U$ and $L$, so taking $W$ or $Z$ into ITRs can potentially improve their value if identifiable.
In Scenario L2, 
the global optimal ITR only depends on $L$, which thus can be identified under our proximal learning framework. 

\begin{table}[h]
    \caption{Simulation scenarios for linear $h_0$.}
	\label{tab: simu_linear}
    \centering
    \scalebox{0.75}{

    \begin{tabular}{c|c|c|c|c|c|c|c}
    \hline\hline
         Scenario & $c_0$ & $c_1$ & $c_2$ & $c_3$ & $c_4$ & $c_5$ & $\omega$ \\
    \hline\hline
         L1 & 2 & 0.5  & 8 & $[0.25, 0.25]^\top $ & 0.25 & $[3,-5]^\top $ & 2 \\
         L2 & 2 & 0.5  & 8 & $[0.25, 0.25]^\top $ & 0    & $[3,-5]^\top $ & 2 \\
    \hline\hline
    \end{tabular}
    }
\end{table}

We also consider a second case where $h_0$ is a nonlinear function of $L$ and $W$. We let
		\begin{align*}
		&h_0(W,A,L) = c_0+c_1(A)+c_2W+c_3(L)+\mathbb{I}(A=1) \left\{ c_4W+c_5(L)+Wc_6(L) \right\}, \text{ and thus}\\
		&\E(Y\mid W,U,A,Z,L) = c_0+c_1(A)+c_3(L)+\mathbb{I}(A=1)c_5(L) + \omega W\\ 
		&\qquad\qquad+\left\{ c_2+c_4\mathbb{I}(A=1)+Ac_6(L)-\omega \right\}\left\{ \mu_0+\mu_l^{\top}L+\frac{\sigma_{wu}}{\sigma_u^2}(U-\kappa_0-\kappa_l^{\top}L) \right\},
		\end{align*}
    where functions $c_1, c_3, c_5, c_6$, and values of $c_0, c_2, c_4$ and $\omega$ are specified in Table \ref{tab: simu_nonlinear} and the other parameters are given in Supplementary Material.
		Accordingly the global optimal ITR is 
		\begin{align*}
	d^*(L,U) &= \sign [\E(Y\mid L,U,A=1) - \E(Y\mid L,U,A=-1)]\\
		&= \sign \left[c_1(1)-c_1(-1) +c_5(L) + [c_4+c_6(L)] \left\{ \mu_0+\mu_l^{\top}L+ \frac{\sigma_{wu}}{\sigma_u^2}(U-\kappa_0-\kappa_l^{\top}L)\right\}\right].
		\end{align*}

		\begin{table}[h]
			\caption{Simulation scenarios for nonlinear $h_0$.}
			\label{tab: simu_nonlinear}
			\centering
    \scalebox{0.75}{
			\begin{tabular}{c|c|c|c|c|c|c|c|c}
				\hline\hline
				Scenario & $c_0$ & $c_1(A)$ & $c_2$ & $c_3(L)$ & $c_4$ & $c_5(L)$ & $c_6(L)$ & $\omega$\\
				\hline\hline
				N1 & 2 & $2.3\mathbb{I}(A=1)$ & 4 & $L^\top  L$ & -2.5 & $|L_1-1|-|L_2+1|$ & $\sin L_1 - 2\cos L_2$ & 2\\
				N2 & 2 & $0.25\mathbb{I}(A=1)$ & 5 & $L^\top  L$ & 0 & $-6L_1 L_2$ & 0 & 2\\
				\hline\hline
			\end{tabular}
    }
		\end{table}
These two scenarios N1 and N2 shown in Table \ref{tab: simu_nonlinear} are analogous to L1 and L2 in Table \ref{tab: simu_linear}.
In each simulation setting described above, we have 100 simulation runs with $n=2,000$ and $5,000$ subjects in each run, and also generate a \textit{noise-free} test dataset of $500,000$ 
used to obtain values of estimated ITRs by \eqref{eq:value fun under no unmeasured confounders} with \eqref{eq: simu prop score}.

\subsection{Estimators and Implementations}

To increase the difficulty of this simulation study, 
we add eight independent variables from the uniform distribution on $[-1, 1]$ in 
$L$ and used the resulting concatenated 10-dimensional $L$ in all ITR learning methods.


We implement all four proposed learning methods in Table \ref{tab:compare} denoted by 
\texttt{d1(L,Z)}, \texttt{d2(L,W)}, \texttt{d3DR(L)} and \texttt{d4} here respectively, 
together with an optimal ITR estimator denoted by \texttt{d1(L)} here, 
which is assumed to only depend on $L$ and obtained by \eqref{eq: proximal learning 1}, as well as 
{the optimal ITR estimator \texttt{d2(L)}, which is assumed to only depend on $L$ and obtained by \eqref{eq: proximal learning 2}.} The latter two are used to compared with \texttt{d3DR(L)}.
We also compare our proposed optimal ITR estimators with that by efficient augmentation and relaxation learning  \citep[EARL,][]{zhao2019efficient} designed under no unmeasured confounding. We denote it by \texttt{dEARL(L)} here and implement it using the R package \texttt{DynTxRegime}. 
To obtain \texttt{dEARL(L)} when $U$ is unobserved, we use all observed variables $(L,W,Z)$ to construct tree-based nonparametric main effect models and propensity score models, and then fit the regime in terms of $L$.
We also create an oracle optimal ITR estimator, denoted by \texttt{NUC}, which is obtained by EARL using $(L,U)$. 
{
Due to the relatviely unsatisfactory computing speed of the R package \texttt{DynTxRegime}, we only obtain \texttt{dEARL(L)} and \texttt{NUC}
	in Scenarios L1 and L2 
  but not in Scenarios N1 and N2.
}
For Scenario L2, we only use $L$ to construct an ITR for \texttt{NUC} since the optimal one only depends on $L$. Note that \texttt{NUC} is unattainable in practice since $U$ is unobserved, but it can be regarded as a benchmark estimator for comparison.

We use the quadratic smoothed hinge loss as surrogate $\phi(\cdot)$ for all proximal learners. The nuisance functions involved in our proposed methods, i.e., the confounding bridge functions $h_0$ and $q_0$, are estimated by \eqref{eq: min-max estimation of h} and \eqref{eq: min-max estimation of q}, respectively, with RKHSs $\calF,\calH,\calG,$ and $\calQ$ all equipped with Gaussian kernels.  
For Scenarios L1 and L2, all decision functions are fitted as linear functions of their corresponding covariates with the $\ell_2$ penalty on coefficients. 
{For Scenarios N1 and N2, $\calR_1,\calR_2,$ and $\calR_3$ are chosen as RKHSs equipped with Gaussian kernels. }
Tuning parameters of the penalties are selected by cross-validation (see Supplementary Material S3 for details). 
For computational acceleration,  all kernels 
are approximated by the Nystr\"om method with $2\lceil\sqrt{n}\rceil$ 
features \citep{yang2012nystrom}.

\subsection{Simulation Results} \label{sec: simu result}


The values of all optimal ITR estimates for $n=5,000$ are illustrated in 
Figures \ref{fig:L_5000} and \ref{fig:N_5000}. The figures for $n=2,000$ {and detailed comparisons with EARL} are given in Supplementary Material S5. 

\begin{figure}[H]
	\captionsetup[subfigure]{aboveskip=-8pt,belowskip=-3pt}
	\centering
	\begin{subfigure}[t]{0.35\textwidth}
		\centering
		\includegraphics[width=\textwidth]{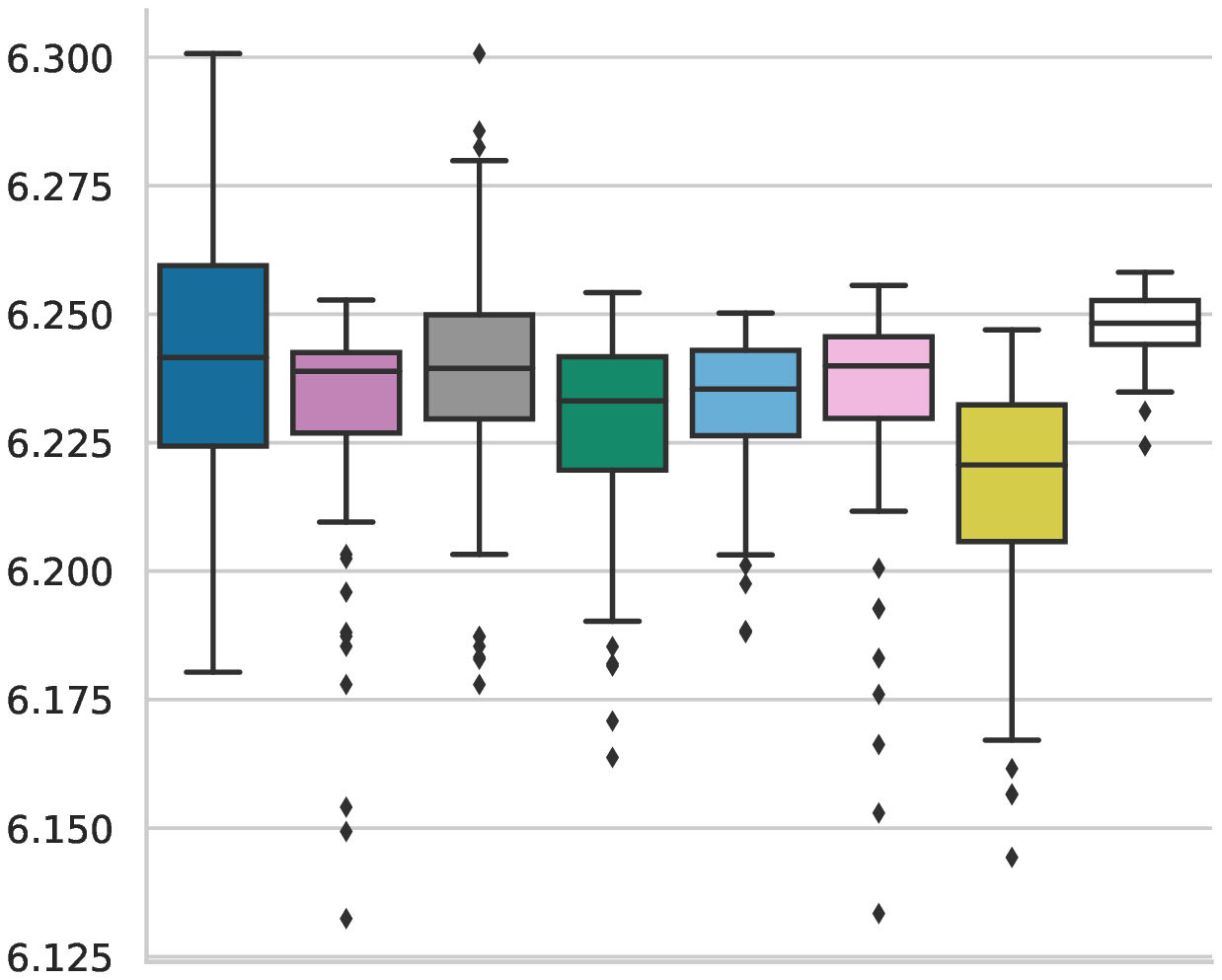}
		\caption{L1}
	\end{subfigure}
	\begin{subfigure}[t]{0.35\textwidth}
		\centering
		\includegraphics[width=\textwidth]{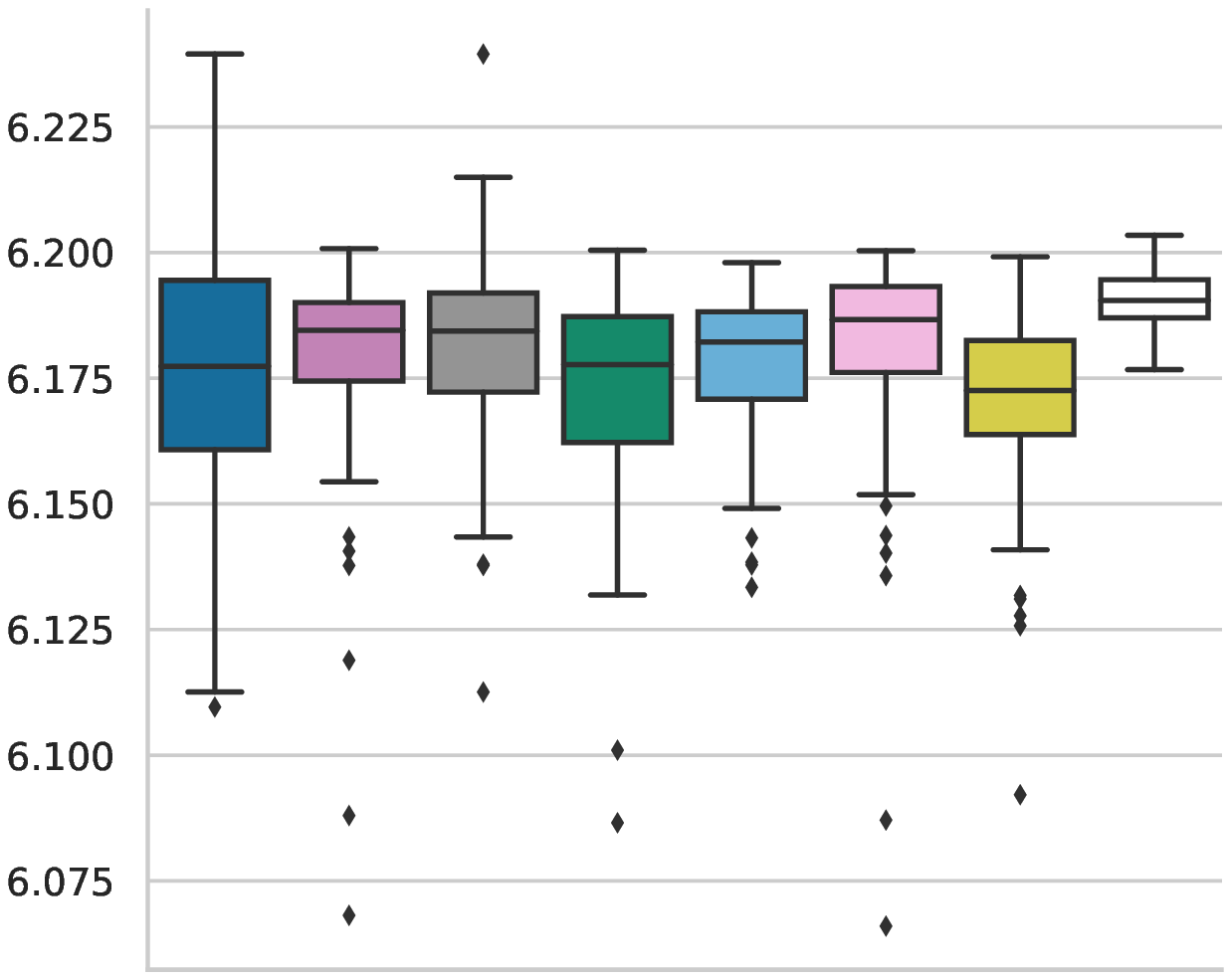}
		\caption{L2}
	\end{subfigure}
	\begin{subfigure}[t]{0.15\textwidth}
		\vspace{-4.5cm}
		\begin{spacing}{0.8}
			{
				\thiscolor{d1(L,Z)}\\
				\thiscolor{d2(L,W)}\\
				\thiscolor{d4}\\
				\thiscolor{d3DR(L)}\\
				\thiscolor{d1(L)}\\
				\thiscolor{d2(L)}\\
				\thiscolor{dEARL(L)}\\
				\thiscolor{NUC}
			}
		\end{spacing}
	\end{subfigure}
	\caption{Boxplots of values for Scenario L1 and L2 with sample size $n=5,000$.}
	\label{fig:L_5000}
\end{figure}

\vspace{-1cm}
\begin{figure}[H]
	\captionsetup[subfigure]{aboveskip=-8pt,belowskip=-3pt}
	\centering
	\begin{subfigure}[t]{0.35\textwidth}
		\centering
		\includegraphics[width=\textwidth]{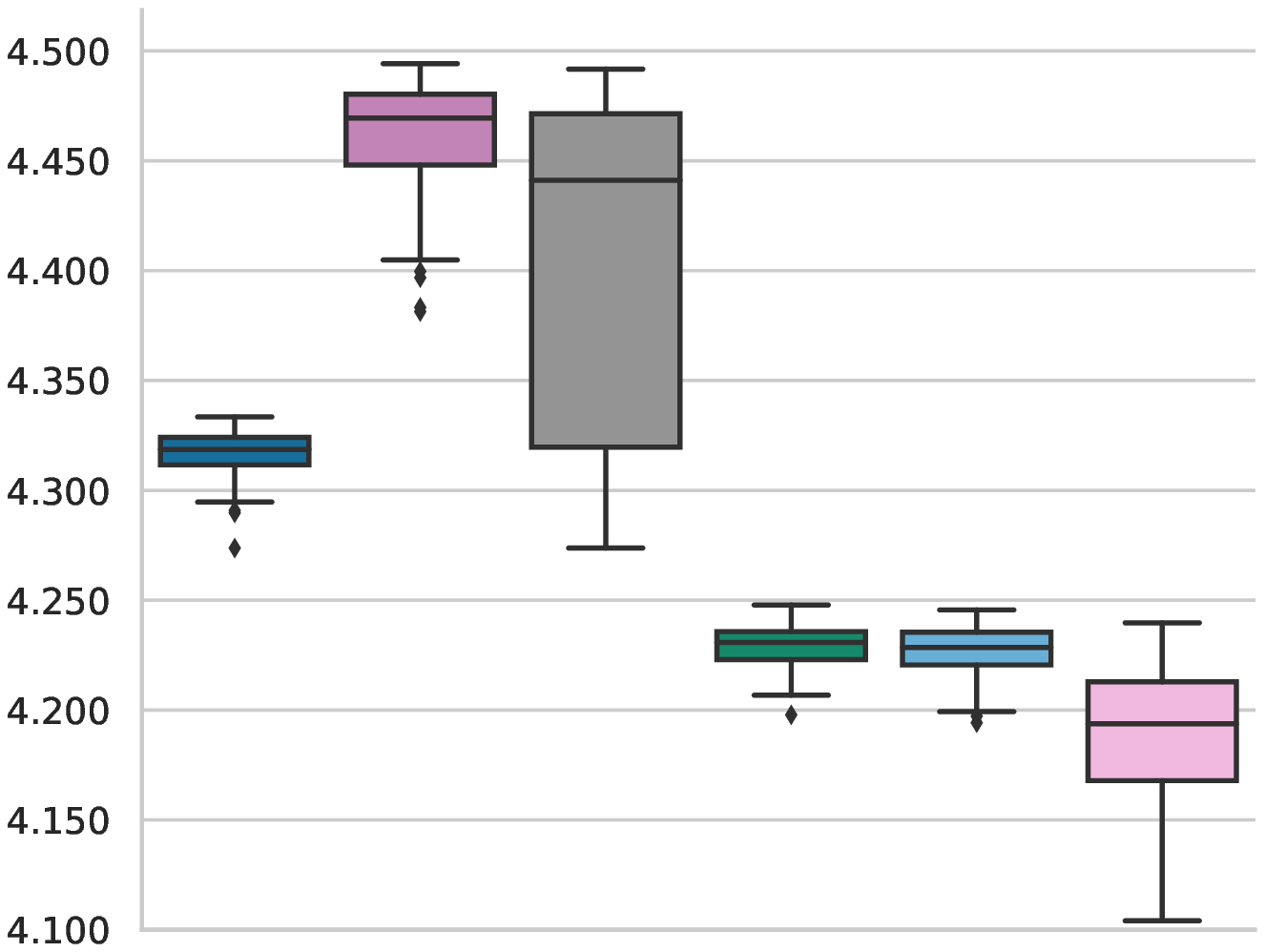}
		\caption{N1}
	\end{subfigure}
	\begin{subfigure}[t]{0.35\textwidth}
		\centering
		\includegraphics[width=\textwidth]{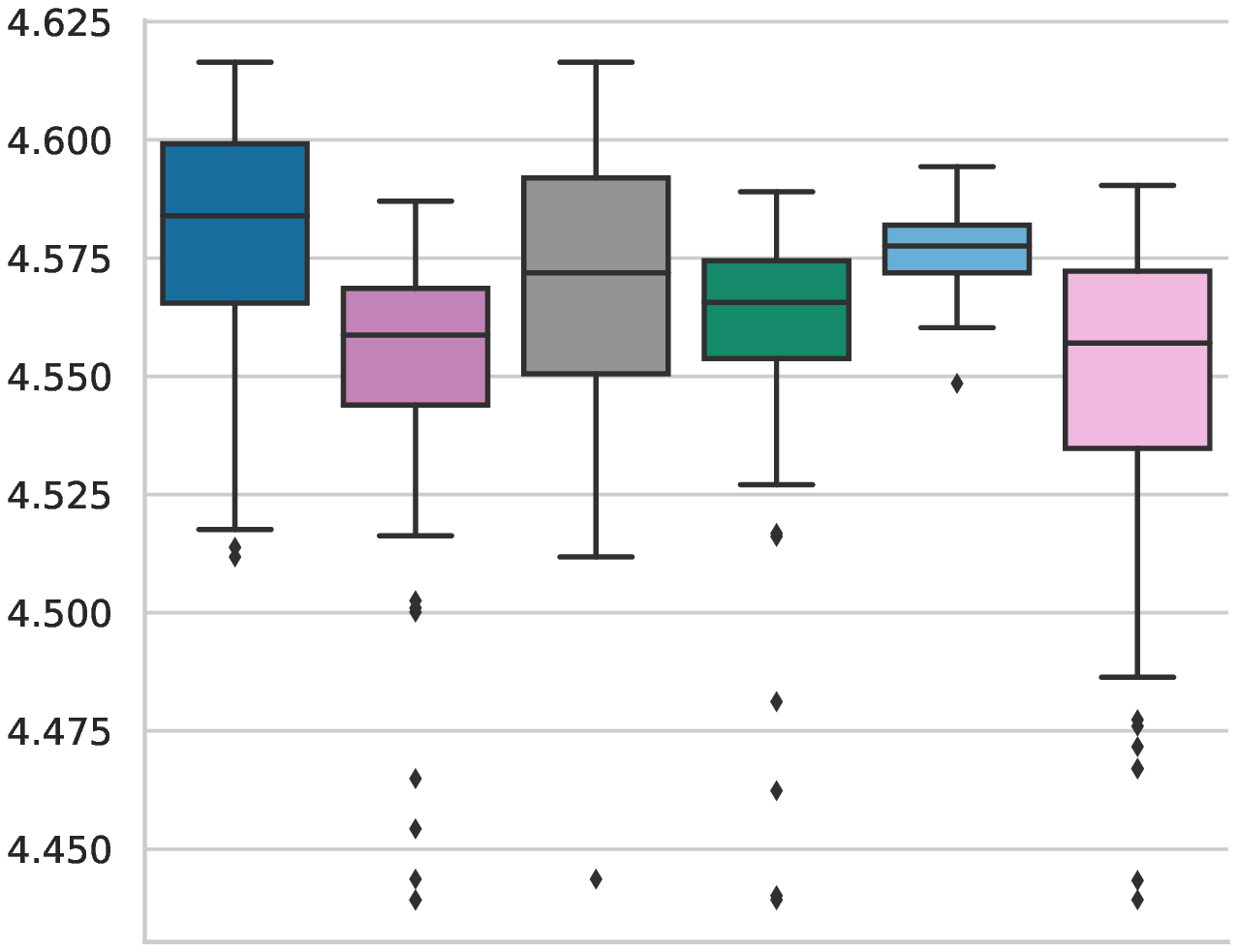}
		\caption{N2}
	\end{subfigure}
	\begin{subfigure}[t]{0.15\textwidth}
		\vspace{-4.5cm}
		\begin{spacing}{0.8}
			{
				\thiscolor{d1(L,Z)}\\
				\thiscolor{d2(L,W)}\\
				\thiscolor{d4}\\
				\thiscolor{d3DR(L)}\\
				\thiscolor{d1(L)}\\
				\thiscolor{d2(L)}
			}
		\end{spacing}
	\end{subfigure}
	\caption{Boxplots of values for Scenario N1 and N2 with sample size $n=5,000$.}
	\label{fig:N_5000}
\end{figure}
\vspace{-0.7cm}



Figure \ref{fig:L_5000} shows that the oracle estimator \texttt{NUC} is the best in both Scenarios L1 and L2. This is not surprising since in Scenario L1 the global optimal ITR cannot be identified by the other methods based on observed data. 
In comparison, since $U$ is unobserved but \texttt{dEARL(L)} is developed under no unmeasured confounding, as expected, all our proposed methods have better values 
than \texttt{dEARL(L)}. 

In Scenarios L1 and N1, \texttt{d1(L,Z)} and \texttt{d2(L,W)} outperform \texttt{d1(L)} and \texttt{d2(L)} respectively, 
which demonstrates the improvement of decision making by including the proxy variables $Z$ and $W$ when the optimal decisions 
depend on unobserved $U$.
For Scenarios L2 and N2, the global optimal ITR can be identified by our proposed methods 
since the difference of outcome bridge only depends on $L$. 
Figure \ref{fig:L_5000} (b) shows that all our proposed methods are comparable to \texttt{NUC} in Scenario L2 but the values of
 \texttt{dEARL(L)} are less satisfactory. 
 Figure \ref{fig:L_5000} also shows that \texttt{d1(L)}, \texttt{d2(L)} and \texttt{d3DR(L)} have similar performances in Scenario L2 
while \texttt{d1(L,Z)}, \texttt{d2(L,W)} and \texttt{d4} lead to values with slightly or substantially higher variations. 
This is consistent with the fact that the global optimal ITR only depends on $L$ in Scenario L2 so adding $Z$ or $W$ into an ITR may only increase its variability. 
{For Scenario N2, similar patterns can be discovered in Figure \ref{fig:N_5000} (b).} The values of the maximum proximal estimator \texttt{d4} mostly lie between those of \texttt{d1(L,Z)} and of \texttt{d2(L,W)}.  
These observations are intuitively reasonable and consistent with the discussion in Section \ref{sec:policy} above.

In all scenarios, the optimal ITR estimators which involve
estimating $q_0$ {typically have relatively longer lower shadows.}
Therefore, it is interesting to study how to improve the 
practical performance of the treatment proximal learning method in the future.

\section{Real Data Application}\label{sec: real data}

In this section, we apply the five proposed proximal learning methods to a dataset from the Study to Understand Prognoses and Preferences for Outcomes and Risks of Treatments 
\citep[SUPPORT,][]{connors1996effectiveness}. SUPPORT examined the effectiveness and safety of the direct measurement of cardiac function by Right Heart Catheterization (RHC) for certain critically ill patients in intensive care units (ICUs). This dataset has been previously analyzed 
for estimating the average treatment effect of using RHC 
\citep[e.g.,][]{lin1998assessing,tan2006distributional,tchetgen2020introduction,cui2020semiparametric}. 

Our objective is to find an optimal ITR on the usage of RHC that maximizes 30-day survival rates of critically ill patients from the day admitted or transferred to ICU. The data include 5,735 patients, of whom 2,184 were measured by RHC in the first 24 hours ($A=1$) and 3,551 were in the control group ($A=-1$). Finally, 3,817 patients survived or censored at day 30 ($Y=1$) and 1,918 died within 30 days ($Y=-1$). For each individual, we consider 71 covariates including demographics, diagnosis, estimated survival probability, comorbidity, vital signs, and physiological status among others. 
See the full list of covariates at \url{https://hbiostat.org/data/repo/rhc.html}. During the first 24 hours in the ICU, ten variables were measured from a blood test for the assessment of the physiological status. Following \citet{tchetgen2020introduction}, among those ten physiological status, we let $Z=$(\texttt{pafi1, paco21}) be treatment-inducing confounding proxies and $W=$(\texttt{ph1, hema1}) be outcome-inducing confounding proxies respectively, where \texttt{pafi1} is the ratio of arterial oxygen partial pressure to fractional inspired oxygen, \texttt{paco21} is the partial pressure of carbon dioxide, \texttt{ph1} is arterial blood pH, \texttt{hema1} is hematocrit.
We apply our proposed methods to obtain \texttt{d1(L,Z)}, \texttt{d2(L,W)}, \texttt{d3DR(L)}, \texttt{d1(L)} and \texttt{d2(L)}
using this dataset with the same configurations of Scenarios L1 and L2 
in Section \ref{sec:simu}.

The coefficient estimates of all covariates are given in Supplementary Material S6. In Table \ref{tab: important coefficients}, we provide a selected number of them with relatively large coefficient estimates in absolute value. 
First, the negative intercepts in Table \ref{tab: important coefficients} indicates that RHC may have a potential negative averaged treatment effect on the 30-day survival rate for critically ill patients, which is consistent with the existing literature \citep[e.g.,][]{tan2006distributional}. Second, the negative coefficients of \texttt{surv2md1} suggest not perform RHC to a patient with a higher survival prediction on day 1. 
In contrast, 
our estimated ITRs tend to suggest trauma patients (\texttt{trauma}), and patients diagnosed with coma (\texttt{cat1\_coma, cat2\_coma}), lung cancer (\texttt{cat1\_lung, cat2\_lung}) or congestive heart failure (\texttt{cat1\_chf}) 
undergo RHC. 
Clinically, RHC is of value when the hemodynamic state of a patient is in question or changing rapidly. It is thus potentially helpful with patients in critical condition whose hemodynamic states are unstable 
\citep{kubiak2019right}. Those findings can be partially supported by \citet{hernandez2019trends} and \citet{tehrani2019standardized},
but require further investigations. 
Our estimated ITRs also imply that patients with upper gastrointestinal bleeding (\texttt{gibledhx}) or autoimmune polyglandular syndrome type 3 (\texttt{aps1}) might be harmed by RHC, which is also 
worthy of future studies. 

\begin{table}[h]
    \caption{Important coefficients of estimated optimal ITRs. 
    \texttt{surv2md1}: 2M model survival prediction at day 1;  
    \texttt{gibledhx}: upper gastrointestinal bleeding;
    \texttt{aps1}: APS III ignoring coma at day 1; 
    \texttt{cat1} and \texttt{cat2}: first and secondary disease category; \texttt{lung}: lung cancer; \texttt{chf}: congestive heart failure. 
    }
	\label{tab: important coefficients}
    \centering
{\scalebox{0.70}{
\begin{tabular}{l|rrrrr}
\hline\hline
Covariate                               & \texttt{d2(L,W)}  & \texttt{d2(L)} & \texttt{d3DR(L)}  & \texttt{d1(L)} & \texttt{d1(L,Z)}  \\
\hline\hline                                                                                                        
\texttt{intercept}                      & -0.204            & -0.258         & -0.722            & -1.215         & -1.222            \\
\texttt{surv2md1}                       & -0.683            & -0.699         & -0.254            & -0.747         & -0.724            \\
\texttt{gibledhx} \hfill(Yes=1/No=0)    & -0.511            & -0.490         & -0.065            & -0.343         & -0.328            \\
\texttt{aps1} \hfill(Yes=1/No=0)        & -0.361            & -0.405         & -0.184            & -0.451         & -0.461            \\
\texttt{trauma} \hfill(Yes=1/No=0)      & 0.862             & 0.925          & 0.020             & 0.622          & 0.533             \\
\texttt{cat1\_coma} \hfill(Yes=1/No=0)  & -0.091            & -0.112         & 0.215             & 2.183          & 2.210             \\
\texttt{cat2\_coma} \hfill(Yes=1/No=0)  & 0.261             & 0.248          & 0.092             & 2.059          & 2.009             \\
\texttt{cat1\_lung} \hfill(Yes=1/No=0)  & -0.025            & -0.098         & 0.062             & 1.740          & 1.711             \\
\texttt{cat2\_lung} \hfill(Yes=1/No=0)  & 0.960             & 1.131          & 0.033             & 1.412          & 1.072             \\
\texttt{cat1\_chf} \hfill(Yes=1/No=0)   & 0.757             & 0.731          & -0.005            & 0.544          & 0.525             \\
\hline\hline
\end{tabular}
}
}
\end{table}
\begin{figure}[h]
	\centering
	\includegraphics[width=0.55\textwidth]{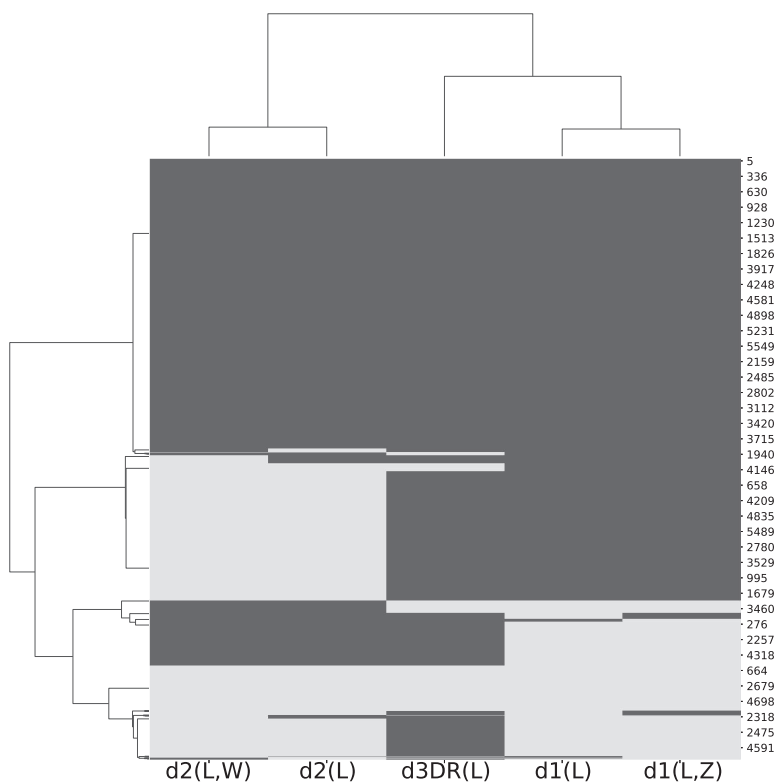}
	\caption{Optimal treatments suggested by the five estimated ITRs. Light portions represent that ITRs suggest the patients to undergo RHC while dark portions represent otherwise.}
	\label{fig:clustermap}
\end{figure}

The treatment recommendations by the five  estimated ITRs for all 5,735 patients, as  illustrated in Figure \ref{fig:clustermap}, show some discrepancies among them. 
The estimated ITR by doubly robust proximal learning, \texttt{d3DR(L)}, only suggests 675 patients to receive RHC, the smallest number among the five. The estimated ITRs by outcome proximal learning, \texttt{d1(L,Z)} and \texttt{d1(L)}, 
both suggest a similar group of around 1,400 patients receive RHC, while those by treatment proximal learning, \texttt{d2(L,W)} and \texttt{d2(L)}, suggest RHC to a similar group of around 2,200 patients.

\begin{figure}[h]
  \centering
  \resizebox{0.6\textwidth}{!}{%
\begin{tikzpicture}
  [
    treenode/.style = {shape=rectangle, rounded corners,
                     draw, align=center,
                     top color=white,
                     font=\ttfamily\normalsize},
    grow                    = right,
    sibling distance        = 4em,
    level distance          = 10em,
    edge from parent/.style = {draw, -latex},
    every node/.style       = {font=\footnotesize},
    sloped
  ]
  \node [treenode] {cat1\_coma\\$n=5735$}
    child { node [treenode] {RHC\checkmark\\$n=436$}
      edge from parent node [below] {Yes} }
    child { node [treenode] {cat2\_mosfm\\$n=5299$}
      child{ node [treenode] {RHC\checkmark\\$n=210$}
        edge from parent node [below] {Yes}}
      child{ node [treenode] {dnr1\\$n=5089$}
        child{ node [treenode] {RHC\checkmark\\$n=510$}
          edge from parent node [below] {Yes}}
        child{ node [treenode] {RHC$\times$\\$n=4579$}
          edge from parent node [above] {No}}
        edge from parent node [above] {No}}
      edge from parent node [above] {No} 
        };
\end{tikzpicture}
  }
  \caption{Decision tree of RHC based on \texttt{d(L,Z)}. 
  \texttt{mosfm}: multiple organ system failure with malignancy. \texttt{dnr1}: ``Do Not Resuscitate'' status on day 1.}
  \label{fig: decision tree}
\end{figure}
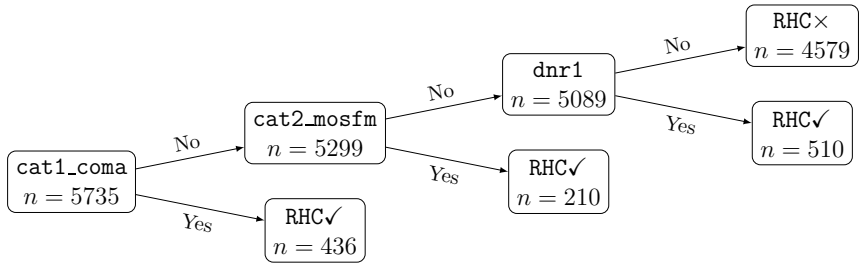

Due to the potentially different recommendations by the five estimated ITRs, we conservatively choose the ITR with the highest 40\% quantile of the estimated values obtained by a 5-fold cross-validation. 
Table \ref{tab: quantiles} lists the 40\% quantiles of estimated values for the five ITRs, of which \texttt{d1(L,Z)} is suggested to be applied. Based on the ITR \texttt{d1(L,Z)} for the 5,735 patients, we construct a decision tree in Figure \ref{fig: decision tree} to illustrate which covariates indicate the usage of RHC. For example, the patients in coma, diagnosed with multiple organ system failure with malignancy or those who agree to ``Do Not Resuscitate'' on the first day of enrollment are recommended to undergo RHC in the first 24 hours. One may also obtain an ensemble ITR by majority voting from the five estimated ITRs. A similar decision tree can be found in Supplementary Material S6.

\begin{table}[h]
    \caption{40\% quantiles of the 5-fold cross-validation values from estimated optimal ITRs.}
	\label{tab: quantiles}
    \centering
{\scalebox{0.70}{
\begin{tabular}{l|rrrrr}
\hline\hline
ITR & \texttt{d2(L,W)}  & \texttt{d2(L)} & \texttt{d3DR(L)}  & \texttt{d1(L)} & \texttt{d1(L,Z)}  \\
\hline\hline
40\% quantile  & 0.3655 & 0.3901 & 0.3735 & 0.4231 & 0.4287 \\
\hline\hline
\end{tabular}
}
}
\end{table}


\section{Discussions}\label{sec:disc}

In this paper, we propose several proximal learning methods to find optimal ITRs under unmeasured confounding. Our methods are built upon the recently developed 
proximal causal inference.  
When the observed covariates can be decomposed into variables that are common causes of the outcome and treatment, namely  outcome-inducing treatment-inducing confounding proxies, we establish 
several identification results on 
a variety of 
classes of ITRs under different assumptions. Based on these results, we propose several classification-based methods to estimate restricted in-class 
optimal ITRs. 
The superior performance of our methods is demonstrated by simulation. 
The real data application above shows the potential of the proposed methods to shed 
light on the 
recommended use of RHC on subgroups of patients, although this requires additional studies for confirmation and validation.

There are several interesting directions for future research. 
First, for the existence of treatment and outcome confounding bridges, if Assumptions \ref{ass: completeness (1)} and \ref{ass: completeness treatment} on completeness hold, then Assumptions \ref{ass: outcome bridge} and \ref{ass: treatment bridge} can be satisfied under some mild regularity conditions. 
As suggested by \cite{cobzaru2022bias}, one can perform a sensitivity analysis on the violation of completeness assumptions, e.g., Assumptions \ref{ass: completeness (1)} and \ref{ass: completeness treatment} under specified structural equation models. Studying this issue in a more general setting would be an interesting future work.
Second, as shown in the simulation study, although 
a flexible nonparametric method is used to estimate $q_0$ to alleviate the issue of the model misspecification,
the finite sample performance of most
optimal ITR estimators that use the estimated $q_0$ is not as good as that of those that use the estimated 
$h_0$. This demonstrates the difficulty in estimating $q_0$, analogous to that in estimating the propensity score in 
average treatment effect estimation under no unmeasured confounding. One possible approach to addressing this limitation is to develop weighted estimators similar to \cite{wong2017kernel} and \cite{athey2018approximate}. Moreover, our proposed methods are developed for a single decision time point. 
It will be interesting to extend them to the longitudinal data 
to estimate the optimal dynamic treatment regimes where individualized decisions are needed at multiple time points. This may be practically useful, e.g., to study the treatment of some chronic disease. 
Lastly, it is also meaningful to study 
the estimation of optimal ITRs for survival outcomes in the presence of unmeasured confounding.


\section*{Acknowledgement}

The authors thank the editor, an associate editor and three referees for their constructive comments and suggestions. Xiaoke Zhang’s research is partially supported by the George Washington University University Facilitating Fund.


\section*{Supplementary Material}
All technical proofs, implementation details and additional numerical results are given in Supplementary Material.

	\section{Example Illustration}\label{supp: concrete example}
	In this section, we provide a concrete example to show how one can obtain restricted in-class optimal ITR via $h_0$ in Section \ref{sec:identifyoutcome}.
	\begin{ex}
		
		Let $W, L, U$ and $Z$ be all  one-dimensional variables, and suppose that 
		\begin{align*}
		&\E\left(Y \mid A, Z, L, U \right) = b_0 + b_1A + b_2U + b_3L + A (b_4U + b_5L),\\
		&\E\left(W \mid A, Z, L, U \right) = a_0 + a_1U + a_2L,
		\end{align*}
		where $b_0$--$b_5$, and $a_0$--$a_2$ are some constants.
		Then we can show that
		\begin{align*}
		&\E\left(Y \mid A, Z, L\right) = b_0 + b_1A + b_2\E\left(U \mid A, Z, L\right) + b_3L + A \, \{b_4\E\left(U \mid A, Z, L\right)  + b_5L\},\\
		&\E\left(W \mid A, Z, L\right) = a_0 + a_1\E\left(U \mid A, Z, L\right)  + a_2L.
		\end{align*}
		If $a_1 \neq 0$, this implies that
		\begin{align*}
		&\E\left(Y \mid A, Z, L\right) = c_0 + c_1A  + c_2\E\left(W \mid A, Z, L\right)+ c_3L + A \, \{c_4\E\left(W \mid A, Z, L\right)  + c_5L\},
		\end{align*}
		where $c_0 = b_0-  (a_0b_2)/a_1$, $c_1 = b1- (a_0b_4)/a_1$, $c_2 = b_2/a_1$ $c_3 = b_3 - (b_2a_2)/a_1$, $c_4 = b_4/a_1$ and $c_5 = b_5 - (a_2b_4)/a_1$. We can also observe that
		$$
		h_0(W, A, L) = c_0 + c_1A + c_2 W + c_3L + A\, (c_4W + c_5 L),
		$$
		and for any $d_1 \in \calD_1$,
		$$
		V(d_1) = c_0 + \E\left\{[c_1d_1(L, Z)  + c_2W + c_3L + d_1(L, Z) (c_4W  + c_5L)\right\}.
		$$
		Then $d_1^\ast$ is 
		\begin{align}\label{eq: example I}
		d_1^\ast(L, Z) & = \sign\left[\E\left\{h_0(W, 1, L) \mid L, Z\right\} - \E\left\{h_0(W, -1, L) \mid L, Z\right\}\right] \nonumber	\\
		&=\sign\left\{c_1 + c_4 \E\left(W \mid L, Z\right) +c_5 L\right\},
		\end{align}
		almost surely.
	\end{ex}
	
	\section{Technical Proofs}\label{sec: technical proof}
	
	\noindent \textbf{Proof of Theorem \ref{thm: identify D1}}: 
	Based on the definition of $V(d)$, we have the following equation for any $d_1 \in \calD_1$:
	\begin{align}
	V(d_1) &= \E\left[Y(1) \mathbb{I}(d_1(L, Z) = 1) + Y(-1) \mathbb{I}(d_1(L, Z) = -1)\right]. 
	\end{align}
	We can first show that
	\begin{align*}
	\E\left[Y(1) \mathbb{I}(d_1(L, Z) = 1) \right] &= \E\left[\E\left[Y(1) \, |\, L, Z\right] \mathbb{I}(d_1(L, Z) = 1) \right] \\[0.1in]
	&= \E\left[\E\left[\E\left[Y(1) \, |\, L, U, Z\right]\, |\, L, Z\right]  \mathbb{I}(d_1(L, Z) = 1) \right] \\[0.1in]
	&= \E\left[\E\left[\E\left[Y \, |\, L, U, A=1\right]\, |\, L, Z\right]  \mathbb{I}(d_1(L, Z) = 1) \right] \\[0.1in]
	&= \E\left[\E\left[\E\left[h_0(W, 1, L) \, |\, L, U\right]\, |\, L, Z\right]  \mathbb{I}(d_1(L, Z) = 1) \right] \\[0.1in]
	&= \E\left[\E\left[\E\left[h_0(W, 1, L) \, |\, L, U, Z\right]\, |\, L, Z\right]  \mathbb{I}(d_1(L, Z) = 1) \right] \\[0.1in]
	&= \E\left[h_0(W, 1, L) \mathbb{I}(d_1(L, Z) = 1) \right],
	\end{align*}
	where the third equality is due to  Assumption \ref{ass: latent exchangeability}, the fourth equality is shown by Theorem 1 of \cite{miao2018identifying} under  Assumptions \ref{ass: completeness (1)}~(a) and \ref{ass: outcome bridge}, and the last but second equality is also due to Assumption \ref{ass: latent exchangeability}.
	
	Similarly, we can show that
	$$
	\E\left[Y(-1) \mathbb{I}(d_1(L, Z) = -1) \right] =\E\left[h_0(W, -1, L) \mathbb{I}(d_1(L, Z) = -1) \right],
	$$
	which concludes the first statement of Theorem \ref{thm: identify D1}. As we can see that
	$$
	V(d_1) = E\left[h_0(W, -1, L) \mathbb{I}(d_1(L, Z) = -1) \right] + E\left[h_0(W, 1, L) \mathbb{I}(d_1(L, Z) = 1) \right],
	$$
	then by minimizing $V(d_1)$ over $\calD_1$ we can show that
	$$
	d_1^\ast(L, Z) = \sign\left(E\left[h_0(W, 1, L) \mid L, Z\right] - E\left[h_0(W, -1, L) \mid L, Z\right] \right),
	$$
	almost surely.

	\vspace{0.2in}
	\noindent 
	\textbf{Proof of Theorem 
		\ref{thm: identify D2}}:
	Based on the definition of $V(d)$, we have the following equation for any $d_2 \in \calD_2$:
	\begin{align}
	V(d_2) &= \E\left[Y(1) \mathbb{I}(d_2(L, W) = 1) + Y(-1) \mathbb{I}(d_2(L, W) = -1)\right]. 
	\end{align}
	Note that $A \independent (Y(a), W) | (U, L)$ for $a \in \calA$. We can first show that
	\begin{align*}
	& \E\left[Y(1) \mathbb{I}(d_2(L, W) = 1) \right] \\[0.1in]
	=& \E\left[\E\left[Y(1) \, |\, L, W\right] \mathbb{I}(d_2(L, W) = 1) \right] \\[0.1in]
	=& \E\left[\E\left[\E\left[Y(1) \, |\, L, U, W\right]\, |\, L, W\right]  \mathbb{I}(d_2(L, W) = 1) \right] \\[0.1in]
	=& \E\left[\E\left[\E\left[Y(1) \, |\, L, U, W, A=1\right]\, |\, L, W\right]  \mathbb{I}(d_2(L, W) = 1) \right], 
	\end{align*}
	where the last equality is due to that $Y(a) \independent A | (U, L, W)$. By the proof of Theorem 2.2 of \cite{cui2020semiparametric}, we have 
	$$
	\E\left[q_0(Z, 1, L) \, |\, U, A =1, L\right] \times \Pr(A =1 \, | \, L, U) = 1.
	$$
	This gives that
	\begin{align*}
	& \E\left[Y(1) \mathbb{I}(d_2(L, W) = 1) \right] \\[0.1in]
	=& \E\left[\E\left[\E\left[Y(1) \, |\, L, U, W, A=1\right]\E\left[q_0(Z, 1, L) \, |\, U, A =1, L\right] \times \Pr(A =1 \, | \, L, U)  \, |\, L, W\right]  \mathbb{I}(d_2(L, W) = 1) \right]\\[0.1in]
	=& \E\left[\E\left[\E\left[Y(1)q_0(Z, 1, L) \, |\, L, U, W, A=1\right]\times \Pr(A =1 \, | \, L, U)  \, |\, L, W\right]  \mathbb{I}(d_2(L, W) = 1) \right]\\[0.1in]
	=& \E\left[\E\left[\E\left[Yq_0(Z, 1, L) \, |\, L, U, W, A=1\right]\times \Pr(A =1 \, | \, L, U, W)  \, |\, L, W\right]  \mathbb{I}(d_2(L, W) = 1) \right]\\[0.1in]
	=& \E\left[\E\left[\E\left[Yq_0(Z, 1, L)\mathbb{I}(A = 1) \, |\, L, U, W \right]\, |\, L, W\right]  \mathbb{I}(d_2(L, W) = 1) \right]\\
	=& \E\left[Yq_0(Z, 1, L)\mathbb{I}(A = 1)  \mathbb{I}(d_2(L, W) = 1) \right]
	\end{align*}
	where the second equality is due to that $Z \independent W | (U, L)$ and $Z \independent Y(a) | (U, L, W, A)$ ensured by Assumption \ref{ass: latent exchangeability},
	and the third equality is based on that $A \independent W | (U, L)$ by Assumption \ref{ass: latent exchangeability} and $Y=Y(A)$ almost surely in Assumption \ref{ass: consistency}.
	Similarly, we can show that
	$$
	\E\left[Y(-1) \mathbb{I}(d_2(L, W) = -1) \right] =\E\left[Y\mathbb{I}(A = -1)q_0(Z, -1, L) \mathbb{I}(d_2(L, W) = -1) \right], 
	$$
	and thus
	\begin{align*}
	V(d_2) &= \E\left[Y\mathbb{I}(A = 1)q_0(Z, 1, L) \mathbb{I}(d_2(L, W) = 1) + Y\mathbb{I}(A = -1)q_0(Z, -1, L) \mathbb{I}(d_2(L, W) = -1)\right]\\[0.1in]
	& =\E\left[Yq_0(Z, A, L) \mathbb{I}(d_2(L, W) = A)\right].
	\end{align*}
	
	Following similar arguments in the proof of Theorem \ref{thm: identify D1}, we can further show that 
	\begin{align}\label{eq: optimal d_3}
	d^\ast_2(L, W) 
	& = \sign \left[\E\left\{Y\mathbb{I}(A=1)q_0(Z, 1, L) \mid L, W \right\} - \E\left\{Y\mathbb{I}(A=-1)q_0(Z, -1, L)  \mid L, W \right\}\right], 
	\end{align}
	almost surely. 

	\vspace{0.2in}
	\noindent 
	\textbf{Proof of Proposition \ref{prop: doubly robust}}: Under some regularity condition, $\hat h_0$ and $\hat q_0$ converge in probability to $\bar h_0$ and $\bar q_0$ respectively in the sup-norm. Suppose $\bar h_0 = h_0$, i.e., $h_0$ is estimated consistently,
	then for any $d \in \calD_3$,
	\begin{align*}
	&\E\left[C_1(Y, L, W, Z, A; h_0, \bar q_0)\mathbb{I}(d(L) = 1) + C_{-1}(Y, L, W, Z; h_0, \bar q_0)\mathbb{I}(d(L) = -1)\right]\\[0.1in]\nonumber
	=&\E\left[\mathbb{I}(A = d(L) = 1)\bar q_0(Z, 1, L)(Y - h_0(W, 1, L)) + \mathbb{I}(A = d(L) = -1)\bar q_0(Z, -1, L)(Y - h_0(W, -1, L)) \right] \\
	+& V(d)\\[0.1in]\nonumber
	=&\E\left[P(A = 1 | Z, L)\mathbb{I}(d(L) = 1)\bar q_0(Z, 1, L)\E\left[Y - h_0(W, 1, L) \, | \, Z, A = 1, L\right] \right] \\[0.1in]\nonumber
	+& \E\left[P(A = -1 | Z, L)\mathbb{I}(d(L) = -1)\bar q_0(Z, -1, L)\E\left[Y - h_0(W, -1, L)\, | \, Z, A = -1, L\right]   \right]+ V(d_3)\\[0.1in]\nonumber
	= & V(d),
	\end{align*}
	where the last equality is due to \eqref{eq: outcome bridge} in Assumption \ref{ass: outcome bridge}. If $\bar q_0 = q_0$, i.e., $q_0$ can be estimated consistently,
	then
	\begin{align}
	&\E\left[C_1(Y, L, W, Z; h_0, \bar q_0)\mathbb{I}(d(L) = 1) + C_{-1}(Y, L, W, Z; h_0, \bar q_0)\mathbb{I}(d(L) = -1)\right]\\[0.1in]\nonumber
	=&\E\left[\mathbb{I}(A = d(L) = 1)q_0(Z, 1, L)Y\right] + \E\left[\mathbb{I}(A = d(L) = -1)q_0(Z, -1, L)Y\right] \\[0.1in]\nonumber
	+& \E\left[\mathbb{I}(d(L) = 1)(1 - \mathbb{I}(A =1)q_0(Z, 1, L))h_0(W, 1, L) \right] \\[0.1in]\nonumber
	+ & \E\left[\mathbb{I}(d(L) = -1)(1 - \mathbb{I}(A =-1)q_0(Z, -1, L))h_0(W, -1, L) \right] \\[0.1in]\nonumber
	=&\E\left[\mathbb{I}(d(L) = 1)Y(1) + \mathbb{I}(d(L) = -1)Y(-1)\right] \\[0.1in]\nonumber
	+& \E\left[\mathbb{I}(d(L) = 1)(1 - \E\left[\mathbb{I}(A =1)q_0(Z, 1, L)\, | \, W, L\right])h_0(W, 1, L) \right] \\[0.1in]\nonumber
	+ & \E\left[\mathbb{I}(d(L) = -1)(1 - \E\left[\mathbb{I}(A =-1)q_0(Z, -1, L)\, | \, W, L\right])h_0(W, -1, L) \right] \\[0.1in]\nonumber
	=&V(d),
	\end{align}
	where the first inequality is due to 
	the proof of Theorem \ref{thm: identify D2} and the last inequality is due to the fact that $\E\left[\mathbb{I}(A = a) q_0(Z, A, L) \, | \, W, L\right] = 1$.
	
	\vspace{0.2in}
	\noindent 
	\textbf{Proof of Theorem \ref{thm: regret bound}:}
	Without loss of generality, we assume that $C_1(Y, L, W, Z; h_0, q_0)$ and $C_{-1}(Y, L, W, Z; h_0, q_0)$ are both positive. For the ease of presentation, we use $C_a$ and $\hat C_a$  to denote $C_a(Y, L, W, Z; h_0, q_0)$ and $C_a(Y, L, W, Z; \hat h_0, \hat q_0)$ respectively for $a \in \left\{-1, 1\right\}$ when there is no confusion.
	As shown in Proposition \ref{prop: doubly robust},  we have
	$$
	V(d_3) = \E\left[C_1\mathbb{I}(d_3(L) = 1) + C_{-1}\mathbb{I}(d_3(L) = -1)\right] \triangleq V^{DR}(d_3),
	$$
	for every $d_3 \in \calD_3$. Then we can have
	\begin{align*}
	V(d^\ast_3) - V(\sign(\hat r_3)) & = V^{DR}(d^\ast_3) - V^{DR}(\sign(\hat r_3)).
	\end{align*}
	As stated in the main text, $r^\ast$ is a  minimizer of $\E \left[C_1 \phi(r(L)) + C_{-1}\phi(-r(L))\right]$ and $d^\ast_3 = \sign(r^\ast)$ by Proposition \ref{prop: finsher consistency}. Define
	\begin{align*}
	r_{\rho_{3, n}} = \arg \min_{r \in \calR_3} \left\{ \E \left[C_1 \phi(r(L)) + C_{-1}\phi(-r(L))\right] + \rho_{3, n}\|r_3\|^2_{\calR_3}  \right\}.
	\end{align*}
	Then the corresponding approximation error is
	\begin{align*}
	\mathbb{A}(\rho_{3, n}) &= \E \left[C_1 \phi(r_{\rho_{3, n}}(L)) + C_{-1}\phi(-r_{\rho_{3, n}}(L))\right] + \rho_{3, n}\|r_{\rho_{3, n}}\|^2_{\calR_3} \\
	& - \E \left[C_1 \phi(r^\ast(L)) + C_{-1}\phi(-r^\ast(L))\right].
	\end{align*}
	For notational convenience, we write $\hat r_3^{DR}$ as $\hat r_3$. By Proposition \ref{prop: excess bound}, we can show that
	\begin{align}
	V^{DR}(d^\ast_3) - V^{DR}(\sign(\hat r_3)) &\leq R_\phi(\hat r_3) - R_\phi(r^\ast)\\
	& = R_\phi\left(\frac{1}{K} \sum_{k = 1}^{K} \hat r_3^{(k)}\right) - R_\phi(r^\ast)\\[0.1in]
	& \leq \frac{1}{K} \sum_{k = 1}^{K} R_\phi( \hat r_3^{(k)}) - R_\phi(r^\ast),
	\end{align}
	where the last inequality is implied by the convexity of $R_\phi(\bullet)$. Thus it suffices  to consider $R_\phi( \hat r_3^{(k)}) - R_\phi(r^\ast)$ for each $k = 1, \cdots K$. Observe that
	\begin{align*}
	& R_\phi( \hat r_3^{(k)}) - R_\phi(r^\ast)\\
	=& \E \left[C_1 \phi(\hat r_{3}^{(k)}) + C_{-1}\phi(-\hat r_3^{(k)})\right] - \E \left[C_1 \phi(r^\ast(L)) + C_{-1}\phi(-r^\ast(L))\right]\\
	\leq & \rho_{3, n} \|r_{\rho_{3, n}} \|^2_{\calR_3} +\E \left[C_1 \phi(r_{\rho_{3, n}}(L)) + C_{-1}\phi(-r_{\rho_{3, n}}(L))\right]  - \E \left[C_1 \phi(r^\ast(L)) + C_{-1}\phi(-r^\ast(L))\right]\\
	+ & \E \left[C_1 \phi(\hat r_3^{(k)}) + C_{-1}\phi(-\hat r_3^{(k)})\right] + \rho_{3, n} \|\hat r^{(k)}_3 \|^2_{\calR_3} - \left\{\rho_{3, n} \|r_{\rho_{3, n}} \|^2_{\calR_3} +\E \left[C_1 \phi(r_{\rho_{3, n}}(L)) + C_{-1}\phi(-r_{\rho_{3, n}}(L))\right]\right\}\\
	= & \mathbb{A}(\rho_{3, n})
	+ \sum_{a \in \calA} \left\{\E \left[C_a \phi(a\hat r_3^{(k)})\right] -\E \left[C_a \phi(ar_{\rho_{3, n}}(L))\right]\right\} + \rho_{3, n} \|\hat r^{(k)}_3 \|^2_{\calR_3} - \rho_{3, n} \|r_{\rho_{3, n}} \|^2_{\calR_3}.
	\end{align*}
	In the following, we apply the empirical process theory to bound the second term on the right hand side of the inequality above. Let 
	\begin{align*}
	\calG_r \triangleq \left\{ \sum_{a \in \calA} \left(C_a \phi(ar) -C_a \phi(ar_{\rho_{3, n}}) \right)+ \rho_{3, n} \|r \|^2_{\calR_3} - \rho_{3, n} \|r_{\rho_{3, n}} \|^2_{\calR_3}  \, \Big| \, \|r\|_{\calR_3} \lesssim \rho_{3, n}^{-\frac{1}{2}}  \right\}.
	\end{align*}
	The reason why we only consider $r$ that satisfies the above norm constraint is motivated by the following argument. By Assumptions \ref{ass: bounded} and \ref{ass: nuisance convergence rate}, $\hat C_a$ is uniformly bounded for every $a \in \calA$. Then according to the optimization property, we can show that
	\begin{align*}
	\sum_{a \in \calA} \E^{(-k)}_{n} \left[\hat C_a \phi(a\hat r_3^{(k)})\right] + \rho_{3, n} \|\hat r^{(k)}_3 \|^2_{\calR_3} \leq  \sum_{a \in \calA}\E^{(-k)}_{n} \left[\hat C_a \phi(0)\right],
	\end{align*}
	which implies that $\rho_{3, n} \|\hat r^{(k)}_3 \|^2_{\calR_3} \lesssim 1$. Based on this, we can further show that for any $g_r \in \calG_r$,
	\begin{align*}
	\|g_r \|_{\infty} \lesssim \rho_{3, n}^{-\frac{1}{2}},
	\end{align*}
	since $\rho_{3, n} \rightarrow 0$ and $\rho_{3, n}\leq 1$. The remaining proof consists of two steps. In the first step, we show
	$$
	\E^{(-k)}_{n}(g_{\hat r_3^{(k)}}) \leq \varepsilon,
	$$
	for some $\varepsilon > 0$ with a high probability. In the second step, we aim to show that
	$$
	\sup_{g_r \in \calG_r} \left|\E^{(-k)}_{n}(g_{r}) - \E(g_{r}) \right| \leq \delta,
	$$ 
	with a high probability for some $\delta$.

	\textbf{Step 1}: This can be shown by first noting that
	\begin{align*}
	&\sum_{a \in \calA} \E^{(-k)}_{n}\left(C_a \phi(a\hat r_3^{(k)}) -C_a \phi(ar_{\rho_{3, n}}) \right)+ \rho_{3, n} \|\hat r_3^{(k)} \|^2_{\calR_3} - \rho_{3, n} \|r_{\rho_{3, n}} \|^2_{\calR_3} \\
	\leq & \sum_{a \in \calA} \E^{(-k)}_{n}\left(\hat C_a \phi(a\hat r_3^{(k)}) -\hat C_a \phi(ar_{\rho_{3, n}}) \right)+ \rho_{3, n} \|\hat r_3^{(k)} \|^2_{\calR_3} - \rho_{3, n} \|r_{\rho_{3, n}} \|^2_{\calR_3}\\
	+ & \sum_{a \in \calA} \E^{(-k)}_{n}\left[\left(C_a- \hat C_a\right) \phi(a\hat r_3^{(k)}) -\left(C_a - \hat C_a\right) \phi(ar_{\rho_{3, n}})\right]\\
	\leq & \sum_{a \in \calA} \E^{(-k)}_{n}\left[\left(C_a- \hat C_a\right) \phi(a\hat r_3^{(k)}) -\left(C_a - \hat C_a\right) \phi(ar_{\rho_{3, n}})\right],
	\end{align*}
	where the last inequality is given by the optimization property. In the following, we bound the last term of the inequality above. It suffices to focus on $\E^{(-k)}_{n}\left[\left(C_a- \hat C_a\right) \phi(a\hat r_3)\right]$ for $a = 1$ while the other terms can be bounded similarly. It can be easily shown that
	\begin{align*}
	&\left|\E^{(-k)}_{n}\left[\left(C_1- \hat C_1\right) \phi(\hat r_3^{(k)})\right]\right|\\
	\leq & \left|\E^{(-k)}_{n}\left[\left(\mathbb{I}(A = 1)q_0(Z, 1, L) - \mathbb{I}(A = 1)\hat q^{(k)}_0(Z, 1, L) \right)\left(\hat h^{(k)}_0(W, 1, L) - h_0(W, 1, L)\right)\phi(\hat r_3^{(k)})\right]\right|\\
	+ & \left|\E^{(-k)}_{n}\left[\left(\mathbb{I}(A = 1)q_0(Z, 1, L) - \mathbb{I}(A = 1)\hat q^{(k)}_0(Z, 1, L) \right)\left(Y - h_0(W, 1, L)\right)\phi(\hat r_3^{(k)})\right]\right|\\
	+ & \left|\E^{(-k)}_{n}\left[\left(\mathbb{I}(A = 1)q_0(Z, 1, L) - 1 \right)\left(\hat h^{(k)}_0(W, 1, L) - h_0(W, 1, L)\right)\phi(\hat r_3^{(k)})\right]\right|\\
	=& (I) + (II) + (III).
	\end{align*}
	For $(II)$, consider 
	$$
	\calG_1 \triangleq \left\{ \left(\mathbb{I}(A = 1)q_0(Z, 1, L) - \mathbb{I}(A = 1)\hat q^{(k)}_0(Z, 1, L) \right)\left(Y - h_0(W, 1, L)\right)\phi(r) \, | \, \|r\|_{\calR_3} \leq \rho_{3, n}^{-\frac{1}{2}} \right\}.
	$$
	By the sample splitting and $\E\left[Y - h_0(W, 1, L) | Z, A=1, L\right] = 0$, we can observe that $\E[g] = 0$ for any $g \in \calG_1$. In addition, the envelop function of $\calG_1$, defined as $G_1$, is $C_0|\hat q^{(k)}_0(Z, 1, L) - q_0(Z, 1, L)||Y-h_0(W, 1, L)|\rho_{3, n}^{-\frac{1}{2}}$. Therefore $\|G_1\|_{2, P} = C_0n^{-\beta}\rho_{3, n}^{-\frac{1}{2}}$ by the error bound condition on $\hat q_0^{(k)}$ given in Assumption \ref{ass: nuisance convergence rate}. By the entropy condition in Assumption \ref{ass: entropy} and Lipschitz property of $\phi$, we can further show that
	$$
	\sup_{Q} N(\calG_1, Q, \varepsilon\|G_1\|_{2, Q}) \lesssim \left(\frac{1}{\varepsilon}\right)^v,
	$$
	which implies that
	$$
	J(1, \calG_1, G_1) \triangleq \int_{0}^1 \sup_{Q} \sqrt{\log N(\calG_1, Q, \varepsilon\|G_1\|_{2, Q})} \lesssim \sqrt{v}.
	$$
	By leveraging the maximal inequality in the empirical process theory, we can show that 
	$$
	\E\sup_{g \in \calG_1}|\E^{(-k)}_{n}g| \lesssim \sqrt{v}n^{-\frac{1}{2}} n^{-\beta}\rho_{3, n}^{-\frac{1}{2}}.
	$$
	Then assuming no measurability issue,  by Talagrand's inequality, we can show with probability $1 - e^{-x}$,
	\begin{align*}
	(II) &\lesssim \E\sup_{g \in \calG_1}|\E^{(-k)}_{n}g| + 2\sqrt{x} \sqrt{\frac{4\sqrt{v}n^{-\frac{1}{2}-\beta}\rho_{3, n}^{-1} + C_0n^{-2\beta}\rho_{3, n}^{-1}}{n}} + \frac{3x\rho_{3, n}^{-\frac{1}{2}}}{n}\\
	&\lesssim \max\{1, x\}\sqrt{v}n^{-\frac{1}{2}} n^{-\beta}\rho_{3, n}^{-\frac{1}{2}}.
	\end{align*}
	Similarly, we can show
	$$
	(III) \lesssim \max\{1, x\}\sqrt{v}n^{-\frac{1}{2}} n^{-\alpha}\rho_{3, n}^{-\frac{1}{2}},
	$$
	with probability at least $1 - e^{-x}$. In addition, we can bound (I) term by Cauchy-Schwarz inequality, i.e., with probability at least $1 - 2e^{-x}$,
	\begin{align*}
	(I) &\leq \left(\E_n^{(-k)}\left[q_0(Z, 1, L) - \hat q^{(k)}_0(Z, 1, L)\right]^2\right)^{\frac{1}{2}} \times \left(\E_n^{(-k)}\left[h_0(W, 1, L) - \hat h^{(k)}_0(W, 1, L)\right]^2\right)^{\frac{1}{2}}\rho_{3, n}^{-\frac{1}{2}}\\
	& \lesssim \max\{1, x\}n^{-(\alpha + \beta)}\rho_{3, n}^{-\frac{1}{2}}.
	\end{align*}
	The last inequality is due to 
	Bernstein's inequality, 
	i.e.,
	$$
	\E_n^{(-k)}\left[q_0(Z, 1, L) - \hat q^{(k)}_0(Z, 1, L)\right]^2 \lesssim n^{-2\beta} + n^{-\frac{1}{2}-\beta}\sqrt{2x} + \frac{x}{3n},
	$$
	and
	$$
	\E_n^{(-k)}\left[h_0(W, 1, L) - \hat h^{(k)}_0(W, 1, L)\right]^2 \lesssim n^{-2\beta} + n^{-\frac{1}{2}-\beta}\sqrt{2x} + \frac{x}{3n},
	$$
	by the uniformly bounded assumption in Assumptions \ref{ass: bounded} and \ref{ass: nuisance convergence rate} and the error bound condition on nuisance function estimation in Assumption \ref{ass: nuisance convergence rate}. Combining the results above together, we can show that with probability at least $1-3e^{-x}$, 
	$$
	\E^{(-k)}_{n}\left[\left(C_a- \hat C_a\right) \phi(a\hat r^{(k)}_3)\right]\lesssim \max\{1, x\}n^{-(\alpha + \beta)}\rho_{3, n}^{-\frac{1}{2}} + \max\{1, x\}\sqrt{v}n^{-\frac{1}{2}-\min\{\alpha, \beta \}}\rho_{3, n}^{-\frac{1}{2}}.
	$$
	Similar result can be obtained by replacing $\hat r^{(k)}_3$ with $r_{\rho_{3, n}}$. So far, we have verified that
	$$
	\E^{(-k)}_{n}(g_{\hat r_3^{(k)}}) \leq \varepsilon,
	$$
	if $\varepsilon \geq C_0 \max\{1, x\}\left(n^{-(\alpha + \beta)}\rho_{3, n}^{-\frac{1}{2}} + n^{-\frac{1}{2}-\min\{\alpha, \beta \}}\rho_{3, n}^{-\frac{1}{2}}\right)$ with probability at least $1 - 12 e^{-x}$.

	\textbf{Step 2}: Again by applying Talagrand's inequality and maximal inequality, we can show that with probability at least $1- e^{-x}$,
	$$
	\sup_{g_r \in \calG_r} \left|\E^{(-k)}_{n}(g_r) - \E(g_r) \right| \lesssim \max\{1, x\} \sqrt{v}\rho_{3, n}^{-\frac{1}{2}}n^{-\frac{1}{2}}.
	$$
	
	Summarizing Step 1 and 2, 
	we can show that with probability $1 - e^{-x}$, 
	\begin{align*}
	& R_\phi( \hat r_3^{(k)}) - R_\phi(r^\ast)\\
	\lesssim &  \mathbb{A}(\rho_{3, n})
	+ \sum_{a \in \calA} \left\{\E \left[C_a \phi(a\hat r_3^{(k)})\right] -\E \left[C_a \phi(ar_{\rho_{3, n}}(L))\right]\right\} + \rho_{3, n} \|\hat r^{(k)}_3 \|^2_{\calR_3} - \rho_{3, n} \|r_{\rho_{3, n}} \|^2_{\calR_3}\\
	\lesssim & \mathbb{A}(\rho_{3, n}) + \sup_{g_r \in \calG_r} \left|\E^{(-k)}_{n}(g_{r}) - \E(g_{r}) \right| + \E^{(-k)}_{n}(g_{\hat r_3^{(k)}})\\
	\lesssim & \max\{1, x\} \sqrt{v}\rho_{3, n}^{-\frac{1}{2}}n^{-\frac{1}{2}} + \max\{1, x\}n^{-(\alpha + \beta)}\rho_{3, n}^{-\frac{1}{2}} + \max\{1, x\}\sqrt{v}n^{-\frac{1}{2}-\min\{\alpha, \beta \}}\rho_{3, n}^{-\frac{1}{2}},
	\end{align*}
	which concludes our proof.
	
	\section{Computation Details} \label{sec: comp details}
	The implementation of the proximal policy learning involves selection of several tuning parameters and the cross-fitting procedure in Section \ref{sec:DRlearning}. Data splitting will be used to implement the cross-fitting. We denote $I_a = \{i: A_i=a, i=1,\dots,n\}$ for the indices of the treatment group ($A=1$) and control group ($A=-1$). Let $I^{(1)}, \dots, I^{(K)}$ denote the indices of the randomly partitioned $K$ folds of the indices $\{1,\dots,n\}$.
	Denote $I^{(-k)} = \{1,\dots,n\}\backslash I^{(k)}$, $k=1,\dots,K$. 
	Let $I_a^{(1)},\dots, I_a^{(K)}$ denote the resulting indices of $I_a$ after the random partition, 
	$a=\pm 1$ and $I_a^{(-k)} = I_a\backslash I_a^{(k)}$, $k=1,\dots,K$.
	
	\subsection{Estimation of Confounding Bridge Functions}
	\paragraph{Estimating $h_0$.}
	We introduce one more tuning parameter in \eqref{eq: min-max estimation of h}. Consider a slightly modified optimization problem
	\begin{align}\label{eq: min-max estimation of h 2}
	\hat h_0=\argmin_{h \in \calH} \left(\sup_{f \in \calF}\left[ \frac{1}{n}\sum_{i = 1}^n \left\{Y_i - h(W_i, A_i, L_i)\right\}f(Z_i, A_i, L_i)- \lambda_1\left( \|f\|^2_{\calF} + \xi \|f\|^2_{2, n}\right) \right]+ \lambda_2\| h\|^2_{\calH} \right),
	\end{align}
	where $\lambda_1 > 0$, $\lambda_2 > 0$ and $\xi>0$ are tuning parameters.
	
	If $\calH$ and $\calF$ are RKHSs equipped with
	kernels $K_{\calH}$ and $K_{\calF}$, 
	and canonical RKHS norms
	$\|\bullet\|_{\calH}$ and $\|\bullet\|_{\calF}$ respectively, we can define the Gram
	matrices $\bK_{\calH,n} = \Big[ K_{\calH} \left( [W_i,A_i,L_i], [W_j,A_j,L_j]
	\right) \Big]_{i,j=1}^n$ and $\bK_{\calF,n} = \Big[ K_{\calF} \left(
	[Z_i,A_i,L_i], [Z_j,A_j,L_j] \right) \Big]_{i,j=1}^n$. 
	Then the solution
	to \eqref{eq: min-max estimation of h 2} has the following closed form:  
	\begin{equation*}
	\hat h_0(w,a,l) = \sum_{i=1}^n \alpha_i K_{\calH}\left( [W_i,A_i,L_i], [w,a,l] \right), \qquad 
	\alpha = \left( \bK_{\calH,n}\bM_h \bK_{\calH,n} + 4 \lambda_{1}\lambda_{2} \bK_{\calH,n} \right)^{\dagger}\bK_{\calH,n}\bM_h Y,
	\end{equation*}
	where $\bM_h = \bK_{\calF,n}^{1/2} (\frac{\xi}{n}K_{\calF,n} + \bI)^{-1} \bK_{\calF,n}^{1/2}$, $Y = [Y_1,\dots,Y_n]^{\top}$, $\alpha=[\alpha_1,\dots,\alpha_n]^{\top}$ and $A^{\dagger}$ is the Moore-Penrose pseudoinverse of $A$. This is a direct application of Propositions 9 and 10 in \citet{dikkala2020minimax}.
	
	In practice, we estimate $h(\bullet,a,\bullet)$ separately for $a=\pm 1$ by
	\begin{align}\label{eq: min-max estimation of h 3}
	\hat h_0(\bullet,a,\bullet)=\argmin_{h \in \calH} \left(\sup_{f \in \calF}\left[ \frac{1}{|I_a|}\sum_{i\in I_a} \left\{Y_i - h(W_i, a, L_i)\right\}f(Z_i, L_i)- \lambda_{1}\left( \|f\|^2_{\calF} + \xi \|f\|^2_{2, n}\right) \right]+ \lambda_{2}\| h\|^2_{\calH} \right),
	\end{align}
	where $\calH$ is RKHS defined on $\calW\times\calL$ and $\calF$ is RKHS defined on $\calZ\times\calL$.
	The Gram matrices $\bK_{\calH,I_a} = \Big[ K_{\calH} \left( [W_i,L_i], [W_j,L_j]
	\right) \Big]_{i,j\in I_a}$ and $\bK_{\calF,I_a} = \Big[ K_{\calF} \left(
	[Z_i,L_i], [Z_j,L_j] \right) \Big]_{i,j\in I_a}$. Similarly, the solution
	to \eqref{eq: min-max estimation of h 3} has the following closed form expression: For each $a=\pm 1$,
	\begin{equation*}
	\hat h_0(w,a,l) = \sum_{i\in I_a}\alpha_i K_{\calH}\left( [W_i,L_i], [w,l] \right), \qquad 
	\alpha = \left( \bK_{\calH,I_a}\bM_{h,I_a} \bK_{\calH,I_a} + 4 \lambda_{1}\lambda_{2} \bK_{\calH,I_a} \right)^{\dagger}\bK_{\calH,I_a}\bM_{h,I_a} Y,
	\end{equation*}
	where $\bM_{h,I_a} = \bK_{\calF,I_a}^{1/2} (\frac{\xi}{|I_a|}\bK_{\calF,I_a} + \bI)^{-1} \bK_{\calF,I_a}^{1/2}$, $Y = [Y_i]_{i\in I_a}^{\top}$, and $\alpha = [\alpha_i]_{i\in I_a}^{\top}$.
	
	In the numerical studies, we equip $\calH$ and $\calF$ with Gaussian kernels $K_{\calH}$ and $K_{\calF}$ respectively.
	The bandwidth of $K_{\calF}$ is selected using median heuristics, e.g., median of pairwise distance \citep{sriperumbudur2009kernel}. 
	The bandwidth of $K_{\calH}$ and tuning parameters $\lambda_1$ and $\lambda_2$ are selected by cross-validation. See details in Algorithm \ref{alg:h0}.
	
	%
	%
	
	\paragraph{Estimating $q_0$.}
	Similarly, we introduce one more tuning parameter in \eqref{eq: min-max estimation of q} and consider optimization problem for each $a=\pm 1$:
	\begin{align}\label{eq: min-max estimation of q 2}
	\hat q_0(\bullet, a, \bullet) = \argmin_{q \in \calQ} \left(\sup_{g \in \calG}
	\left[\frac{1}{n}\sum_{i = 1}^n \left\{\mathbb{I}(A_i = a) q(Z_i, a, L_i) - 1 \right\}g(W_i, L_i)- \mu_1 \left(\|g\|^2_{\calG} + \zeta\|g\|^2_{2, n}\right) \right] + \mu_2\| q\|^2_{\calQ} \right),
	\end{align}
	where $\mu_1 > 0$, $\mu_2 > 0$ and $\zeta>0$ are 
	tuning parameters. 
	
	Similar to the estimation of $h_0$, suppose $\calG$ and $\calQ$ are the RKHSes of the kernels $K_{\calG}$ and $K_{\calQ}$, equipped with the canonical RKHS norms $\|\bullet\|_{\calG}$ and $\|\bullet\|_{\calQ}$. 
	With corresponding Gram matrices $\bK_{\calG,n} = \Big[K_{\calG}\left( [W_i, L_i], [W_j, L_j]\right)\Big]_{i,j=1}^n$ and
	$\bK_{\calQ,n} = \Big[K_{\calQ}\left( [Z_i, L_i], [Z_j, L_j]\right)\Big]_{i,j=1}^n$, the solution to \eqref{eq: min-max estimation of q 2} is
	\begin{equation*}
	\hat q_0(z,a,l) = \sum_{i=1}^n \alpha_i K_{\calQ}\left( [Z_i, L_i], [z,l]\right), \qquad \alpha=\left( \mathring{\bK}_{\calQ,n}^{\top} \bM_q \mathring{\bK}_{\calQ,n} + 4\mu_1\mu_2\bK_{\calQ,n}\right)^{\dagger} \mathring{\bK}_{\calQ,n}^{\top} \bM_q 1_n,
	\end{equation*}
	where $\mathring{\bK}_{\calQ,n} = \diag\{\mathbb{I}(A_i=a)\}_{i=1}^n \bK_{\calQ,n}$, $\bM_q = \bK_{\calG,n}^{1/2} (\frac{\zeta}{n}\bK_{\calG,n} + \bI)^{-1} \bK_{\calG,n}^{1/2}$ and $1_n$ is column vector of $1$'s of length $n$.  See details in Algorithm \ref{alg:q0}.
	
	\paragraph{Selection of tuning parameters.} There are several tuning parameters in the estimation of $h_0$ and $q_0$. We accept the tricks and recommendation defaults by \citet{dikkala2020minimax} and their package \texttt{mliv}. The following parameters will be used to determine $\xi$, $\lambda_1\lambda_2$, $\zeta$ and $\mu_1\mu_2$. In particular, we define following two functions used in our tuning parameter selections.
	\begin{equation}
	\label{eq: varsigma}
	\varsigma(n) = 5/n^{0.4}.
	\end{equation}
	\begin{equation}
	\label{eq: tau}
	\tau(s,n) = \frac{s}{2} \varsigma^4(n).
	\end{equation}
	More details can be found in \citet{dikkala2020minimax} and their package \texttt{mliv}.
	In all our numerical studies, RKHSs $\calF,\calH,\calG, \calQ$ are equipped with Gaussian kernels
	\begin{equation}
	\label{eq:rbf}
	K(x_1,x_2) = \exp\{\gamma\|x_1-x_2\|_2^2\}.
	\end{equation}
	The median heuristic bandwidth parameter $\gamma^{-1} = \text{median}\{\|x_i-x_j\|_2^2\}_{i<j\in I}$ for indices subset $I\subset \{1,\dots,n\}$ \citep{sriperumbudur2009kernel}. Note that we only use median heuristic for $K_\calF$ and  $K_\calG$.
	
	\noindent
	\begin{minipage}{\linewidth}
		\begin{algorithm}[H] \label{alg:h0}
			\SetAlgoLined
			\textbf{Input:} Standardized data $\{(L_i, Z_i, W_i, Y_i)\}_{i\in I_a}$; Set bandwidth of $\calF$ by median heuristic.\\ 
			Repeat for $k=1,\dots,K$:\\
			\Indp 
			Repeat for bandwidth of $\calH$ by letting $\gamma_{\calH}^{-1} \in \{p\text{-quantile of }\{\|[W_i, L_i] - [W_j, L_j]\|_2^2\}_{i<j\in I_a}: p=0.1,\dots,0.9\}$\\
			\Indp
			Repeat for $s$ in a pre-specified collection of positive values:\\
			\Indp
			$\xi^{(-k)} = 1/\varsigma^2(|I_a^{(-k)}|), \lambda_{1}^{(-k)}\lambda_{2}^{(-k)} = \tau(s,|I_a^{(-k)}|), \xi^{(k)} = 1/\varsigma^2(|I_a^{(k)}|)$.\\
			Obtain $\hat h_0^{(-k)} (\bullet,a,\bullet)$ by \eqref{eq: min-max estimation of h 3} with standardized data of indices $I_a^{(-k)}$.\\
			Calculate $r_i = Y_i - \hat h_0^{(-k)}(W_i,a,L_i)$, $i\in I_a^{(k)}$ and the loss $r^T M_{h,I_a^{(k)}} r / |I_a^{(k)}|^2$.\\
			\Indm
			\Indm
			\Indm
			Calculate averaged loss along $k$. Find $s^{\ast}$ and $\gamma_{\calH}^{\ast}$ that minimize averaged loss.\\
			\textbf{Output:} Calculate $\hat h_0(\bullet,a,\bullet)$ by \eqref{eq: min-max estimation of h 3} with $\xi=1/\varsigma^2(|I_a|), \lambda_1\lambda_2 = \tau(s^{\ast}, |I_a|)$ and $\gamma_{\calH}^{\ast}$.
			\caption{Estimating $h_0(\bullet,a,\bullet)$, $a=\pm 1$}
		\end{algorithm}
	\end{minipage}
	
	\noindent
	\begin{minipage}{\linewidth}
		\begin{algorithm}[H] \label{alg:q0}
			\SetAlgoLined
			\textbf{Input:} Standardized data $\{(L_i, Z_i, W_i, A_i)\}_{i=1}^n$; Set bandwidth of $\calG$ by median heuristic.\\
			Repeat for $k=1,\dots,K$:\\
			\Indp
			Repeat for bandwidth of $\calQ$ by letting $\gamma_{\calQ}^{-1} \in \{p\text{-quantile of }\{\|[Z_i, L_i] - [Z_j, L_j]\|_2^2\}_{i<j\in I_a}: p=0.1,\dots,0.9\}$:\\
			\Indp
			Repeat for $s$ in a pre-specified collection of positive values:\\
			\Indp
			Let $n_a^{(k)} = |\{A_i=a:i\in I^{(k)}\}|$, $n_a^{(-k)} = |\{A_i=a:i\in I^{(-k)}\}|$.\\
			$\zeta^{(-k)} = 1/\varsigma^2(n_a^{(-k)}), \mu_1^{(-k)}\mu_2^{(-k)} = \tau(s,n_a^{(-k)}), \zeta^{(k)} = 1/\varsigma^2(n_a^{(k)})$.\\
			Obtain $\hat q_0^{(-k)} (\bullet,a,\bullet)$ by \eqref{eq: min-max estimation of q 2} with standardized data of indices $I^{(-k)}$.\\
			Calculate $r_i = 1 - \hat q_0^{(-k)}(Z_i,a,L_i)$, $i\in I^{(k)}$ and the loss $r^T M_{q,I_{(k)}} r / |I^{(k)}|^2$.\\
			\Indm
			\Indm
			\Indm
			Calculate averaged loss along $k$. Find $s^{\ast}$ and $\gamma_{\calQ}^{\ast}$ that minimize averaged loss.\\
			\textbf{Output:} Calculate $\hat q_0(\bullet,a,\bullet)$ by \eqref{eq: min-max estimation of q 2} with $\zeta = 1/\varsigma^2(\sum_{i=1}^n\mathbb{I}(A_i=a))$, $\mu_1\mu_2=\tau(s^{\ast}, \sum_{i=1}^n\mathbb{I}(A_i=a))$ and $\gamma_{\calQ}^{\ast}$.
			\caption{Estimating $q_0(\bullet,a,\bullet)$, $a=\pm 1$}
		\end{algorithm}
	\end{minipage}
	
	\subsection{Proximal Learning with RKHSs}
	In this subsection, we discuss algorithms of our proposed proximal learning using RKHSs. The algorithm of finding linear decision functions can be similarly obtained. We choose $\calR_i$, $i=1,2$ to be RKHSs equipped with Gaussian kernels in \eqref{eq: proximal learning 1} and \eqref{eq: proximal learning 2}. The bandwidth parameters of those Gaussian kernels are selected by a heuristic approach similar to \citet{damodaran2018fast}. 
	Specifically, for a given dataset $\{(x_i,y_i,w_i)\}_{i=1}^n$ with class label $y_i\in\{-1,1\}$ and sample weight $\{w_i\}_{i=1}^n$ such that $\sum_{i=1}^n w_i=1$ and $w_i>0,i=1,\dots,n$,
	the empirical Hilbert-Schmidt Independence Criterion (HSIC) can be calculated by
	\begin{equation}\label{eq:HSIC}
	\text{HSIC}(\{x_i,y_i,w_i\}_{i=1}^n; K_x,K_y) = n^{-2}\tr (\bK_x\bH\bK_y\bH),
	\end{equation}
	where $\bH = \diag(w) - ww^T$ with $w = (w_1, \cdots, w_n)^T$, $\bK_x$ is the Gram matrix of Gaussian kernel $K_x$ defined in \eqref{eq:rbf} with bandwidth parameter $\gamma$ and 
	$\bK_y$ is the Gram matrix of target kernel $K_y$ defined as 
	\begin{equation*}
	K_y(y_i,y_j) = \frac{\mathbb{I}(y_i=y_j)}{\# c_{y_i}},\quad 1 \leq i, j \leq n, 
	\end{equation*}
	where $\#c_{y_i} = \sum_{j=1}^n \mathbb{I}(y_j=y_i)$ \citep[Section 5.2.2,][]{song2012feature}. 
	Finally, the heuristic optimal bandwidth parameter $\gamma$ is the one that maximizes the HSIC in \eqref{eq:HSIC}.

	In summary, details for implementing \eqref{eq: proximal learning 1} are given in Algorithm \ref{alg:PL1}. Algorithm \ref{alg:PL23} is given for \eqref{eq: proximal learning 2} while Algorithm  \ref{alg:PL4} is given for the maximum proximal learning \eqref{eq: combine learning}. Details for the doubly robust proximal learning  \eqref{eq: proximal learning 4} with  cross-fitting are provided in Algorithm \ref{alg:DR}. 
	Note that cross-validation with cross-fitting requires another random splitting of training data. In Algorithm \ref{alg:DR}, 
	we let $\{I^{(-k,\kappa)}\}_{\kappa=1}^K$ be the resulting indices of the random partition of $I^{(-k)}$ and denote $I^{(-k,-\kappa)} = I^{(-k)}\backslash I^{(-k,\kappa)}$, $k,\kappa=1,\dots,K$.

	\noindent
	\begin{minipage}{\linewidth}
		\begin{algorithm}[H] \label{alg:PL1}
			\SetAlgoLined
			\textbf{Input:} Standardized data $\{L_i, Z_i, W_i, A_i, Y_i\}_{i=1}^n$;\\
			Estimate $\hat h_0$ and calculate $\{\hat\Delta(W_i,L_i)\}_{i=1}^n$;\\
			Select bandwidth parameter $\gamma_{\calR_1}$ of $\calR_1$ by the HSIC heuristic approach.\\
			Repeat for $\rho_{1,n}$ in a pre-specified collection with size $M$:\\
			\Indp
			Repeat for $k=1,\dots,K$:\\
			\Indp
			Estimate $\hat h_0^{(-k)}$ and calculate $\{\hat\Delta(W_i,L_i)\}_{i\in I^{(-k)}}$ with data of indices $I^{(-k)}$;\\
			Find $\hat d_1^{(-k)\ast}$ by \eqref{eq: proximal learning 1} with $\calR_1$ equipped with Gaussian kernel using bandwidth parameter $\gamma_{\calR_1}$ and penalty coefficient $\rho_{1,n}$;\\
			Calculate empirical value $\hat V^{(k)}(\hat d_1^{(-k)\ast})$ with test data of indices $I^{(k)}$ by \eqref{outcome identification}.\\
			\Indm
			Calculate averaged empirical value $K^{-1}\sum_{k=1}^K \hat V^{(k)}(\hat d_1^{(-k)\ast})$.\\
			\Indm
			Find $\rho_{1,n}^{\ast}$ that maximizes the average empirical value over $M$ tuning parameters.\\
			\textbf{Output:} $\hat d_1^*$ by \eqref{eq: proximal learning 1} with $\calR_1$ equipped with Gaussian kernel using bandwidth parameter $\gamma_{\calR_1}$ and penalty coefficient $\rho_{1,n}^{\ast}$.
			\caption{Proximal Learning by \eqref{eq: proximal learning 1}}
		\end{algorithm}
	\end{minipage}

	\noindent
	\begin{minipage}{\linewidth}
		\begin{algorithm}[H] \label{alg:PL23}
			\SetAlgoLined
			\textbf{Input:} Standardized data $\{L_i, Z_i, W_i, A_i, Y_i\}_{i=1}^n$;\\
			Estimate $\hat q_0$ and calculate $\{Y_i\hat q_0(Z_i,A_i,L_i)\}_{i=1}^n$;\\
			Select bandwidth parameter $\gamma_{\calR_2}$ \eqref{eq: proximal learning 2} by the HSIC heuristic approach.\\
			Repeat for $\rho_{2,n}$ in a pre-specified collection with size $M$:\\
			\Indp
			Repeat for $k=1,\dots,K$:\\
			\Indp
			Estimate $\hat q_0^{(-k)}$ and calculate $\{Y_i\hat q_0^{(-k)}(Z_i,A_i,L_i)\}_{i\in I^{(-k)}}$ with data of indices $I^{(-k)}$;\\
			Find $\hat d_2^{(-k)\ast}$ by \eqref{eq: proximal learning 2} with $\calR_2$ equipped with Gaussian kernel using bandwidth parameter $\gamma_{\calR_2}$ and penalty coefficient $\rho_{2,n}$; \\
			Calculate empirical value $\hat V^{(k)}(\hat d_2^{(-k)\ast})$ with test data of indices $I^{(k)}$ by \eqref{eq: treatment identification with W}.\\
			\Indm
			Calculate averaged empirical value $K^{-1}\sum_{k=1}^K \hat V^{(k)}(\hat d_2^{(-k)\ast})$.\\
			\Indm
			Find $\rho_{2,n}^{\ast}$ that maximizes the average empirical value among $M$ tuning parameters.\\
			\textbf{Output:} $\hat d_2^*$ by \eqref{eq: proximal learning 2} with $\calR_2$ equipped with Gaussian kernel using bandwidth parameter $\gamma_{\calR_2}$ and penalty coefficient $\rho_{2,n}^{\ast}$.
			\caption{Proximal Learning by \eqref{eq: proximal learning 2}}
		\end{algorithm}
	\end{minipage}
	
	\noindent
	\begin{minipage}{\linewidth}
		\begin{algorithm}[H] \label{alg:PL4}
			\SetAlgoLined
			\textbf{Input:} Standardized data $\{L_i, Z_i, W_i, A_i, Y_i\}_{i=1}^n$;\\
			Estimate $\hat h_0$ and calculate $\{\hat\Delta(W_i,L_i)\}_{i=1}^n$;\\
			Select bandwidth parameter $\gamma_{\calR_1}$ of $\calR_1$ by the HSIC heuristic approach.\\
			Estimate $\hat q_0$ and calculate $\{Y_i\hat q_0(Z_i,A_i,L_i)\}_{i=1}^n$;\\
			Select bandwidth parameter $\gamma_{\calR_2}$ \eqref{eq: proximal learning 2} by the HSIC heuristic approach.\\
			Do Lines 4-9 in Algorithm \ref{alg:PL1}, Lines 4-9 in Algorithm \ref{alg:PL23}.\\
			\textbf{if} {$\max_{\rho_{1,n}} K^{-1}\sum_{k=1}^K \hat V^{(k)}(\hat d_1^{(-k)\ast})\geq\max_{\rho_{2,n}} K^{-1}\sum_{k=1}^K \hat V^{(k)}(\hat d_2^{(-k)\ast})$}, \textbf{then}\\
			\Indp
			\textbf{Output: $\hat d_1^\ast$} by Lines 10-11 in Algorithm \ref{alg:PL1}\\
			\Indm
			\textbf{else:}\\
			\Indp
			\textbf{Output: $\hat d_2^\ast$} by Lines 10-11 in Algorithm \ref{alg:PL23}\\
			\Indm
			\textbf{end if}\\
			\caption{Proximal Learning by \eqref{eq: combine learning}}
		\end{algorithm}
	\end{minipage}
	\vspace{1cm}

	\noindent
	\begin{minipage}{\linewidth}
		\begin{algorithm}[H] \label{alg:DR}
			\SetAlgoLined
			\textbf{Input:} Standardized data $\{L_i, Z_i, W_i, A_i, Y_i\}_{i=1}^n$;\\
			Estimate $\hat h_0, \hat q_0$ and calculate $\{C_a(Y_i, L_i, W_i,Z_i;\hat h_0,\hat q_0),a=\pm 1\}_{i=1}^n$ by \eqref{eq: doubly robust weight};\\
			Select bandwidth parameter $\gamma_{\calR_3}$ \eqref{eq: proximal learning 4} by the HSIC heuristic approach.\\
			Repeat for $\rho_{3,n}$ in a pre-specified collection with size $M$:\\
			\Indp
			Repeat for $k=1,\dots,K$:\\
			\Indp
			Repeat for $\kappa=1,\dots,K$:\\
			\Indp
			Estimate $\hat h_0^{(-k,\kappa)}, \hat q_0^{(-k,\kappa)}$ with data of indices $I^{(-k,\kappa)}$;\\
			Obtain $\{C_a(Y_i, L_i, W_i,Z_i;\hat h_0^{(-k,\kappa)},\hat q_0^{(-k,\kappa)}), a=\pm 1\}_{i\in I^{(-k,-\kappa)}}$ by \eqref{eq: doubly robust weight} with data of indices $I^{(-k,-\kappa)}$;\\ 
			Find $\hat r_3^{DR,(-k,-\kappa)}$ by \eqref{eq: proximal learning 4} with $\calR_3$ equipped with Gaussian kernel using bandwidth parameter $\gamma_{\calR_3}$ and penalty coefficient $\rho_{3,n}$.\\
			\Indm
			Define aggregated estimator $\hat d_3^{DR,(-k)} = \sign(K^{-1}\sum_{\kappa=1}^K \hat r_3^{DR,(-k,-\kappa)})$.\\
			Calculate empirical value $\hat V^{(k)}(\hat d_3^{DR,(-k)})$ with test data of indices $I^{(k)}$ by \eqref{eq: doubly robust value function}.\\
			\Indm
			Calculate averaged emprical value $K^{-1}\sum_{k=1}^K \hat V^{(k)}(\hat d_3^{DR,(-k)})$.\\
			\Indm
			Find $\rho_{3,n}^{\ast}$ that maximizes averaged empirical value among $M$ tuning parameters.\\
			Repeat for $k=1,\dots,K$:\\
			\Indp
			Estimate $\hat h_0^{(k)}, \hat q_0^{(k)}$ with data of indices $I^{(k)}$;\\
			Obtain $\{C_a(Y_i, L_i, W_i,Z_i;\hat h_0^{(k)},\hat q_0^{(k)}), a=\pm 1\}_{i\in I^{(-k)}}$ by \eqref{eq: doubly robust weight} with data of indices $I^{(-k)}$;\\ 
			Find $\hat r_3^{DR,(-k)}$ by \eqref{eq: proximal learning 4} with $\calR_3$ equipped with Gaussian kernel using bandwidth parameter $\gamma_{\calR_3}$ and penalty coefficient $\rho_{3,n}^{\ast}$.\\
			\Indm
			\textbf{Output:} Aggregated estimator $\hat d_3^{DR} =\sign( K^{-1}\sum_{k=1}^K \hat r_3^{DR,(-k)})$.
			\caption{Proximal Learning for doubly robust ITR \eqref{eq: proximal learning 4}}
		\end{algorithm}
	\end{minipage}
	
	\section{Choices of Parameters for Simulation Data Generation} \label{sec: paras for simu}
	\subsection{Data Generation of $(L,W,A,Z,U)$}
	\paragraph{Step 1.}
	We consider generating data $(L,W,A,Z,U)$ such that for $a=-1,1$,
	\begin{equation*}
	\frac{1}{\Pr(A=a \mid U,L)} = \int q_0(z,a,L) dF(z\mid U,a,L).
	\end{equation*}
	Let
	\begin{equation*}
	q_0(Z,A,L) = 1+\exp \left\{ A(t_0+t_zZ+t_a\mathbb{I}(A=1)+t_lL) \right\}. 
	\end{equation*}
	Therefore, we have that
	\begin{equation*}
	\frac{1}{\Pr(A\mid U,L)} =1+\exp \left\{ A(t_0+t_a\mathbb{I}(A=1)+t_lL) \right\} \int \exp\{At_zz\} dF(z\mid U,A,L)
	\end{equation*}
	Suppose that
	\begin{equation}
	\label{eq:assumpZUAX}
	Z \mid U,A,L \sim N(\theta_0,\theta_a\mathbb{I}(A=1)+\theta_uU+\theta_lL, \sigma_{z|u,a,l}^2), 
	\end{equation}
	so that 
	\begin{equation}
	\label{eq:PrAUX}
	\frac{1}{\Pr(A\mid U,L)} =1+\exp \left\{ A(t_0+t_a\mathbb{I}(A=1)+t_lL) + At_z(\theta_0 + \theta_a\mathbb{I}(A=1)+\theta_uU+\theta_lL) + \frac{t_z^2\sigma_{z|u,a,l}^2}{2}\right\} .
	\end{equation}
	We require that $\Pr(A=1\mid U,L) + \Pr(A=-1\mid U,L)=1$. Since $\Pr(A=a\mid
	U,L)$ is in the form of $\expit$, we only need 
	\begin{align*}
	&t_0+t_a+t_lL+t_z(\theta_0+\theta_a+\theta_uU+\theta_lL)+\frac{t_z^2\sigma_{z|u,a,l}^2}{2}\\
	=& t_0+t_lL+t_z(\theta_0+\theta_uU+\theta_lL)-\frac{t_z^2\sigma_{z|u,a,l}^2}{2},
	\end{align*}
	which implies that
	\begin{equation*}
	t_a = -t_z^2\sigma_{z|u,a,l}^2 - t_z\theta_a.
	\end{equation*}
	Therefore,
	\begin{equation*}
	q_0(Z,A,L) = 1+\exp \left\{ A[t_0+t_zZ + t_lL -t_z^2\sigma_{z|u,a,l}^2\mathbb{I}(A=1) - t_z\theta_a\mathbb{I}(A=1)] \right\}.
	\end{equation*}
	
	\paragraph{Step 2.}
	Let
	\begin{equation*}
	(Z,W,U)\mid A,L \sim N \left(
	\begin{bmatrix}
	\alpha_0+\alpha_a\mathbb{I}(A=1)+\alpha_lL\\
	\mu_0+\mu_a\mathbb{I}(A=1)+\mu_lL\\
	\kappa_0+\kappa_a\mathbb{I}(A=1)+\kappa_lL\\
	\end{bmatrix},
	\begin{bmatrix}
	\sigma_z^2 & \sigma_{zw} & \sigma_{zu}\\
	\sigma_{zw} & \sigma_w^2 & \sigma_{wu}\\
	\sigma_{zu} & \sigma_{wu} & \sigma_u^2
	\end{bmatrix}
	\right)
	\end{equation*}
	Therefore,
	\begin{equation*}
	\E (Z\mid U,A,L) = \alpha_0+\alpha_a+\alpha_lL + \frac{\sigma_{zu}}{\sigma_u^2}(U-\kappa_0 - \kappa_a\mathbb{I}(A=1) - \kappa_lL).
	\end{equation*}
	Compare it with \eqref{eq:assumpZUAX}, we have
	\begin{equation*}
	\theta_0=\alpha_0-\frac{\sigma_{zu}}{\sigma_u^2}\kappa_0,~\theta_a=\alpha_a - \frac{\sigma_{zu}}{\sigma_u^2}\kappa_a, ~\theta_l=\alpha_l-\frac{\sigma_{zu}}{\sigma_u^2}\kappa_l,~\theta_u=\frac{\sigma_{zu}}{\sigma_u^2}.
	\end{equation*}
	In addition, we impose
	\begin{equation*}
	W \indep (A,Z)\mid U,L.
	\end{equation*}
	The independence implies that $W\mid U,A,Z,L$ follows
	\begin{equation*}
	N \left( \mu_0+\mu_a\mathbb{I}(A=1)+\mu_lL + \Sigma_{w(u,l)}\Sigma_{u,z}^{-1}\begin{bmatrix}
	U-\kappa_0-\kappa_a\mathbb{I}(A=1)-\kappa_lL\\
	Z - \alpha_0-\alpha_a\mathbb{I}(A=1)-\alpha_lL
	\end{bmatrix}, \sigma_w^2 - \Sigma_{w(u,l)}\Sigma_{u,z}^{-1}\Sigma_{w(u,l)}^{\top} \right),
	\end{equation*}
	where $\Sigma_{w(u,z)} = \begin{bmatrix}\sigma_{wu}&\sigma_{wz}\end{bmatrix}$,
	$\Sigma_{u,z}=\begin{bmatrix}\sigma_u^2&\sigma_{zu}\\ \sigma_{zu}^2 &
	\sigma_z^2\end{bmatrix}$, such that
	\begin{equation*}
	\E(W\mid U,A,Z,L) = \E (W\mid U,A,L) = \mu_0+\mu_a\mathbb{I}(A=1)+\mu_lL+ \frac{\sigma_{wu}}{\sigma_u^2}\left\{ U-(\kappa_0+\kappa_a\mathbb{I}(A=1)-\kappa_lL) \right\}
	\end{equation*}
	does not depend on $A$ and $Z$. Therefore
	\begin{equation*}
	\frac{\sigma_{wz}\sigma_u^2-\sigma_{wu}\sigma_{zu}}{\sigma_z^2\sigma_u^2 - \sigma_{zu}^2}=0,
	\end{equation*}
	and
	\begin{equation*}
	\mu_a = \frac{\sigma_{wu}}{\sigma_u^2}\kappa_a.
	\end{equation*}
	
	\paragraph{Step 3.1.}
	Notice that $\Pr(A=a \mid U,L) = \Pr(A=a \mid U,W,L)$, their log odds ratio
	with respect to $U$ must be the same, i.e.,
	\begin{align*}
	\underbrace{\log\frac{\Pr(A=1 \mid U=u,L)/\Pr(A=-1 \mid U=u,L)}{\Pr(A=1 \mid U=0,L)/\Pr(A=-1 \mid U=0,L)}}_{\text{logOR1}} \\
	= \underbrace{\log\frac{\Pr(A=1 \mid U=u,W,L)/\Pr(A=-1 \mid U=u,W,L)}{\Pr(A=1 \mid U=0,W,L)/\Pr(A=-1 \mid U=0,W,L)}}_{\text{logOR2}}.
	\end{align*}
	By the expit property of \eqref{eq:PrAUX},
	\begin{equation*}
	\text{logOR1} = -t_z\theta_uu. 
	\end{equation*}
	Moreover, 
	\begin{equation*}
	\Pr(A=a \mid U,W,L) = \frac{\Pr(U\mid A=a,W,L) \Pr(A=a,W,L)}{\Pr(U,W,L)}
	\end{equation*}
	implies that
	\begin{align*}
	&\text{logOR2}\\
	=&\log\frac{\Pr(A=1 \mid U=u,W,L)/\Pr(A=-1 \mid U=u,W,L)}{\Pr(A=1 \mid U=0,W,L)/\Pr(A=-1 \mid U=0,W,L)}\\ 
	=& \log\frac{\Pr(U=u\mid A=1,W,L)/\Pr(U=u\mid A=-1,W,L)}{\Pr(U=0\mid A=1,W,L)/\Pr(U=0\mid A=-1,W,L)}.
	\end{align*}
	Notice that $U\mid A,W,L \sim N\left(\E(U\mid A,W,L),
	\sigma_{u|w,a,l}^2:=\sigma_{u|w,l}^2\right)$ where the variance does not depend on $a=-1$ or $a=1$, so 
	\begin{align*}
	\log \frac{\Pr(U=u\mid A=1,W,L)}{\Pr(U=u\mid A=-1,W,L)} =& -\frac{1}{2\sigma_{u|w,l}^2}\Big[-2u \left\{ \E(U\mid W,A=1,L) - \E(U\mid W,A=-1,L)\right\}\Big.\\ 
	&\quad\qquad\qquad \Big. + \E^2(U|W,A=1,L) -  \E^2(U\mid W,A=-1,L)\Big].
	\end{align*}
	Therefore,
	\begin{align*}
	\text{logOR2}&=\frac{\E(U\mid W,A=1,L) - \E(U\mid W,A=-1,L)}{\sigma_{u|w,l}^2}u = \frac{\kappa_a -\sigma_{wu}\mu_a/\sigma_w^2}{\sigma_{u|w,l}^2}u.
	\end{align*}
	Finally, we have that
	\begin{equation*}
	-t_z\theta_u = \frac{\kappa_a -\sigma_{wu}\mu_a/\sigma_w^2}{\sigma_{u|w,l}^2}.
	\end{equation*}
	
	\paragraph{Step 3.2.} To find the parameters in $q_{0}$, we require the
	constraint 
	\begin{equation*}
	\Pr(A=-1\mid W,L) = \Pr(A=1|W,L)=1.
	\end{equation*}
	Recall that since $\sigma_{u|w,a,l}^2=\sigma_{u|w,l}^2$,  
	\begin{align*}
	\frac{1}{\Pr(A=a\mid W,L)} &= \int \frac{1}{\Pr(A=a\mid U,W,L)} dF(U\mid W,A=a,L)\\
	&=1+\exp \left\{ a[t_0+t_a\mathbb{I}(a=1)+t_lL+t_z(\theta_0+\theta_a+\theta_lL)]+\frac{t_z^2\sigma_{z|u,a,l}^2}{2} \right\}\\
	&\qquad\times \int \exp \left\{ a t_z\theta_u U \right\} dF(U\mid W,A=a,L)\\
	&=1+\exp \left\{ a[t_0+t_a\mathbb{I}(a=1)+t_lL+t_z(\theta_0+\theta_a+\theta_lL)]+\frac{t_z^2\sigma_{z|u,a,l}^2}{2} \right\}\\
	&\qquad\times \exp \left\{ at_z\theta_u\E(U\mid W,A=a,L) +\sigma_{u|w,l}^2 \frac{t_z^2\theta_u^2}{2} \right\}. 
	\end{align*}
	Similarly, we require that
	\begin{align*}
	& t_0+t_a+t_lL+t_z(\theta_0+\theta_a+\theta_lL) + \frac{t_z^2\sigma_{z|u,a,l}^2}{2}\\
	&+ t_z\theta_u \left\{ \E(U\mid W,A=1,L)-\E(U\mid W,A=-1,L) \right\} + \sigma_{u|w,l}^2 \frac{t_z^2\theta_u^2}{2}\\
	=& t_0+t_lL+t_z(\theta_0+\theta_lL) - \frac{t_z^2\sigma_{z|u,a,l}^2}{2}-\sigma_{u|w,l}^2\frac{t_z^2\theta_u^2}{2},
	\end{align*}
	which holds because
	\begin{equation*}
	\E(U\mid W,A=1,L)-\E(U\mid W,A=-1,L) = -t_z\theta_u\sigma_{u|w,l}^2, 
	\end{equation*}
	as shown in Step 3.1 and
	\begin{equation*}
	t_a = -t_z^2\sigma_{z|u,a,l}^2 - t_z\theta_a,
	\end{equation*}
	required in Step 1.
	
	\paragraph{Step 4.}
	Finally, we require that
	\begin{equation*}
	\Pr(A=1\mid U,L) + \Pr(A=-1\mid U,L) = 1.
	\end{equation*}
	Notice that
	\begin{align*}
	&\frac{1}{\Pr(A=a\mid L)}\\
	= & \int \frac{1}{\Pr(A\mid U,L)} dF(U\mid A=a,L)\\
	= & 1+ \exp \left\{ a\left[ t_0 + t_a\mathbb{I}(a=1) + t_lL + t_z(\theta_0+\theta_a\mathbb{I}(a=1)+\theta_lL) \right] + \frac{t_z^2\sigma_{z|u,a,l}^2}{2} \right\}\\
	&\qquad \times\int\exp \left\{ at_z\theta_uU \right\} dF(U\mid A=a,L)\\
	= & 1+ \exp \left\{ a\left[ t_0 + t_a\mathbb{I}(a=1) + t_lL + t_z(\theta_0+\theta_a\mathbb{I}(a=1)+\theta_lL) \right] + \frac{t_z^2\sigma_{z|u,a,l}^2}{2} \right\}\\
	&\qquad \times\exp \left\{ at_z\theta_u\E(U\mid A=a,L) + \sigma_{u|a,l}^2 \frac{t_z^2\theta_u^2}{2} \right\}\\
	= & 1+ \exp \left\{ a\left[ t_0 + t_a\mathbb{I}(a=1) + t_lL + t_z(\theta_0+\theta_a\mathbb{I}(a=1)+\theta_lL) \right] + \frac{t_z^2\sigma_{z|u,a,l}^2}{2} \right\}\\
	&\qquad \times\exp \left\{ at_z\theta_u(\kappa_0+\kappa_a\mathbb{I}(a=1)+\kappa_lL) + \sigma_{u|a,l}^2 \frac{t_z^2\theta_u^2}{2} \right\}.
	\end{align*}
	Thus, $A\mid L$ is generated by
	\begin{align*}
	\frac{1}{\Pr(A=1\mid L)} &= 1+\exp \left\{ t_0+t_a+t_lL+t_z(\theta_0+\theta_a+\theta_lL)+\frac{t_z^2(1-\frac{\sigma_{zu}^2}{\sigma_z^2\sigma_u^2})\sigma_z^2}{2} \right\}\\
	&\qquad\times\exp \left\{ t_z\theta_u(\kappa_0+\kappa_a+\kappa_lL) + \sigma_u^2 \frac{t_z^2\theta_u^2}{2} \right\}.
	\end{align*}
	
	Here we provide a summary of 
	the constraints of data generation:
	\begin{align*}
	t_a = -t_z^2\sigma_{z|u,a,l}^2-t_z\theta_a = -t_z^2(1-\frac{\sigma_{zu}^2}{\sigma_z^2\sigma_u^2})\sigma_z^2 - t_z\theta_a,\\
	\sigma_{wz}\sigma_u^2 - \sigma_{wu}\sigma_{zu}=0,\\
	\mu_a\sigma_u^2 = \sigma_{wu}\kappa_a,\\
	-\theta_ut_z(1-\frac{\sigma_{uw}^2}{\sigma_u^2\sigma_w^2})\sigma_u^2 = -\theta_u\sigma_{u|w,a,l}^2t_z = \kappa_a - \sigma_{wu}\mu_a/\sigma_w^2.
	\end{align*}
	
	
	
	
	
	We used following setting for the generation of $(L,W,A,Z,U)$:
	\begin{itemize}
		\item $\alpha_0=0.25$, $\alpha_a=0.25$, $\alpha_l=[0.25, 0.25]$;
		\item $\mu_0=0.25$, $\mu_a=0.125$, $\mu_l=[0.25, 0.25]$; 
		\item $\kappa_0=0.25$, $\kappa_a=0.25$, $\kappa_l=[0.25, 0.25]$;
		\item
		$\bSigma=\begin{bmatrix}
		1 & 0.25 & 0.5\\
		0.25 & 1 & 0.5\\
		0.5 & 0.5 & 1
		\end{bmatrix}$
	\end{itemize}
	Then the steps above are compatible with the model of $q_0$:
	\begin{equation*}
	q_0(Z,A,L) = 1+\exp \left\{ A(t_0+t_zZ+t_a\mathbb{I}(A=1)+t_lL) \right\},
	\end{equation*}
	where $t_0=0.25$, $t_z=-0.5$, and $t_a=-0.125$. 
	
	\subsection{Data Generation of $Y\mid L,W,A,Z,U$}
	We generate response $Y$ given $L,W,A,Z,U$ such that $Z\indep Y \mid U,A,L$ and Assumption \ref{ass: outcome bridge} holds, i.e., for $a=-1,1$,
	\begin{equation*}
	\E (Y \mid U,a,L) = \int h_0(w,a,L) dF(w \mid U,L),
	\end{equation*}
	which we impose by letting
	\begin{align*}
	\E(Y\mid W,U,A,Z,L) &= \E(Y\mid U,A,Z,L) + \omega\{W-\E(W\mid U,A,Z,L)\}\\
	&=\E(Y\mid U,A,L) + \omega\{W - \E(W\mid U,L)\} \text{ since } W\indep (A,Z)\mid U,L. 
	\numit\label{eq:EYWUAZL}
	\end{align*}
	
	\begin{enumerate}[]
		\item When $\E(Y\mid W,U,A,Z,L)$ is linear of $L$, let
		\begin{equation*}
		h_0(W,A,L) = c_0+c_1\mathbb{I}(A=1)+c_2W+c_3L+\mathbb{I}(A=1)(c_4W+c_5L).
		\end{equation*}
		Then by \eqref{eq:EYWUAZL},
		\begin{align*}
		\E(Y\mid W,U,A,Z,L) &=\int h_0(w,A,L) dF(w\mid U,L) - \omega\E(W\mid U,L) + \omega W\\
		&= c_0+c_1\mathbb{I}(A=1) + c_3L+c_5\mathbb{I}(A=1)L+\omega W\\ &\quad+(c_2+c_4\mathbb{I}(A=1)-\omega)\E(W\mid U,L)\\
		&= c_0+c_1\mathbb{I}(A=1)+ c_3L+c_5\mathbb{I}(A=1)L+\omega W\\ 
		&\quad+(c_2+c_4\mathbb{I}(A=1)-\omega)\left\{\mu_0+\mu_lL+\frac{\sigma_{wu}}{\sigma_u^2}(U-\kappa_0-\kappa_lL)\right\}.
		\end{align*}
		\begin{itemize}
			\item When $c_4=0$, $h_0(W,A=1,L) - h_0(W,A=-1,L)$ only dependes on $L$.
		\end{itemize}
		
		The global optimal ITR is $d^{*}(L,U) = \sign (\E[Y\mid L,U,A=1] - \E[Y\mid L,U,A=-1])$.
		\begin{align*}
		& \E[Y\mid L,U,A=1] - \E[Y\mid L,U,A=-1]\\
		= &\E \left\{ \E[Y\mid W,U,A,Z,L]\mid L,U,A=1 \right\} - \E\left\{ \E[Y\mid W,U,A,Z,L]\mid L,U,A=-1 \right\}\\
		= & c_1+c_5L + \omega \underbrace{\left\{ \E (W\mid L,U,A=1) - \E(W\mid L,U,A=-1)\right\}}_{=0\text{ since } W\indep A\mid L,U} \\
		&+ c_4 \left\{ \mu_0+\mu_lL+ \frac{\sigma_{wu}}{\sigma_u^2}(U-\kappa_0-\kappa_lL)\right\}\\
		= & c_1+c_5L + c_4 \left\{ \mu_0+\mu_lL+ \frac{\sigma_{wu}}{\sigma_u^2}(U-\kappa_0-\kappa_lL)\right\}.
		\end{align*}
		
			The optimal ITR within $\calD_3$ is 
			\begin{align*}
			d_3^*(L) &= \sign(\E[h_0(W,A=1,L)\mid L] - \E[h_0(W,A=-1,L)\mid L])\\
			&= c_1 + c_4 \E(W\mid L) +c_5 L\\
			&= c_1 + c_4 \E[\E(W\mid L,A)\mid L] +c_5 L\\
			&= c_1 + c_4 [(\mu_0+\mu_a+\mu_l L)P(A=1\mid L) + (\mu_0 + \mu_l L)P(A=-1\mid L)] +c_5 L\\
			&= c_1 + c_4 [\mu_0+\mu_l L + \mu_a P(A=1\mid L)] +c_5 L.
			\end{align*}
			
		\item When $\E(Y\mid W,U,A,Z,L)$ is nonlinear of $L$, let
		\begin{equation*}
		h_0(W,A,L) = c_0+c_1(A)+c_2W+c_3(L)+\mathbb{I}(A=1) \left\{ c_4W+c_5(L)+Wc_6(L) \right\}.
		\end{equation*}
		Then by \eqref{eq:EYWUAZL},
		\begin{align*}
		\E(Y\mid W,U,A,Z,L) &=\int h_0(w,A,L) dF(w\mid U,L) - \omega\E(W\mid U,L) + \omega W\\
		&= c_0+c_1(A)+c_3(L)+\mathbb{I}(A=1)c_5(L)\\ 
		&\quad+\left\{ c_2+c_4\mathbb{I}(A=1)+Ac_6(L)-\omega \right\}\E(W\mid U,L) + \omega W\\
		&= c_0+c_1(A)+c_3(L)+\mathbb{I}(A=1)c_5(L) + \omega W\\ 
		&\quad+\left\{ c_2+c_4\mathbb{I}(A=1)+Ac_6(L)-\omega \right\}\left\{ \mu_0+\mu_lL+\frac{\sigma_{wu}}{\sigma_u^2}(U-\kappa_0-\kappa_lL) \right\}.
		\end{align*}
		\begin{itemize}
			\item When $c_4,c_6(\cdot)=0$, $h_0(W,A=1,L) - h_0(W,A=-1,L)$ only dependes on $L$.
		\end{itemize}

		The global optimal ITR is $d^{*}(L,U) = \sign (\E[Y\mid L,U,A=1] - \E[Y\mid L,U,A=-1])$.
		\begin{align*}
		& \E[Y\mid L,U,A=1] - \E[Y\mid L,U,A=-1]\\
		= &\E \left\{ \E[Y\mid W,U,A,Z,L]\mid L,U,A=1 \right\} - \E\left\{ \E[Y\mid W,U,A,Z,L]\mid L,U,A=-1 \right\}\\
		= & c_1(1) - c_1(0)+c_5(L) + \omega \underbrace{\left\{ \E (W\mid L,U,A=1) - \E(W\mid L,U,A=-1)\right\}}_{=0\text{ since } W\indep A\mid L,U} \\
		&+ [c_4+c_6(L)] \left\{ \mu_0+\mu_lL+ \frac{\sigma_{wu}}{\sigma_u^2}(U-\kappa_0-\kappa_lL)\right\}\\
		= & c_1(1)-c_1(0) +c_5(L) + [c_4+c_6(L)] \left\{ \mu_0+\mu_lL+ \frac{\sigma_{wu}}{\sigma_u^2}(U-\kappa_0-\kappa_lL)\right\}.
		\end{align*}
		
			The optimal ITR within $\calD_3$ is 
			\begin{align*}
			d_3^*(L) &= \sign(\E[h_0(W,A=1,L)\mid L] - \E[h_0(W,A=-1,L)\mid L])\\
			&= c_1(1)-c_1(-1) + c_5(L) + (c_4+c_6(L)) \E(W\mid L)\\
			&= c_1(1)-c_1(-1) + c_5(L) + (c_4+c_6(L))\E[\E(W\mid L,A)\mid L])\\
			&= c_1(1)-c_1(-1) + c_5(L) + (c_4+c_6(L))\left\{\mu_0+\mu_l L + \mu_a P(A=1\mid L)\right\}.
			\end{align*}

		
	\end{enumerate}
	
	\section{Additional Simulation Results}\label{sec: additional simu result}
	In addition to the results in Section 6.2, for L1 and L2, we also compare our proposed optimal ITR estimators
	with \texttt{dEARL(L,W)}, \texttt{dEARL(L,Z)} and \texttt{dEARL(L,Z,W)}. Similar to \texttt{dEARL(L)}, we use
	all observed variables $(L,W,Z)$ to construct tree-based nonparametric main effect models and propensity score models,
	then fit the regime in terms of $(L,W)$, $(L,Z)$ and $(L,Z,W)$, respectively.

\begin{figure}[H]
		\centering
		\begin{subfigure}[t]{0.4\textwidth}
			\centering
			\includegraphics[width=\textwidth]{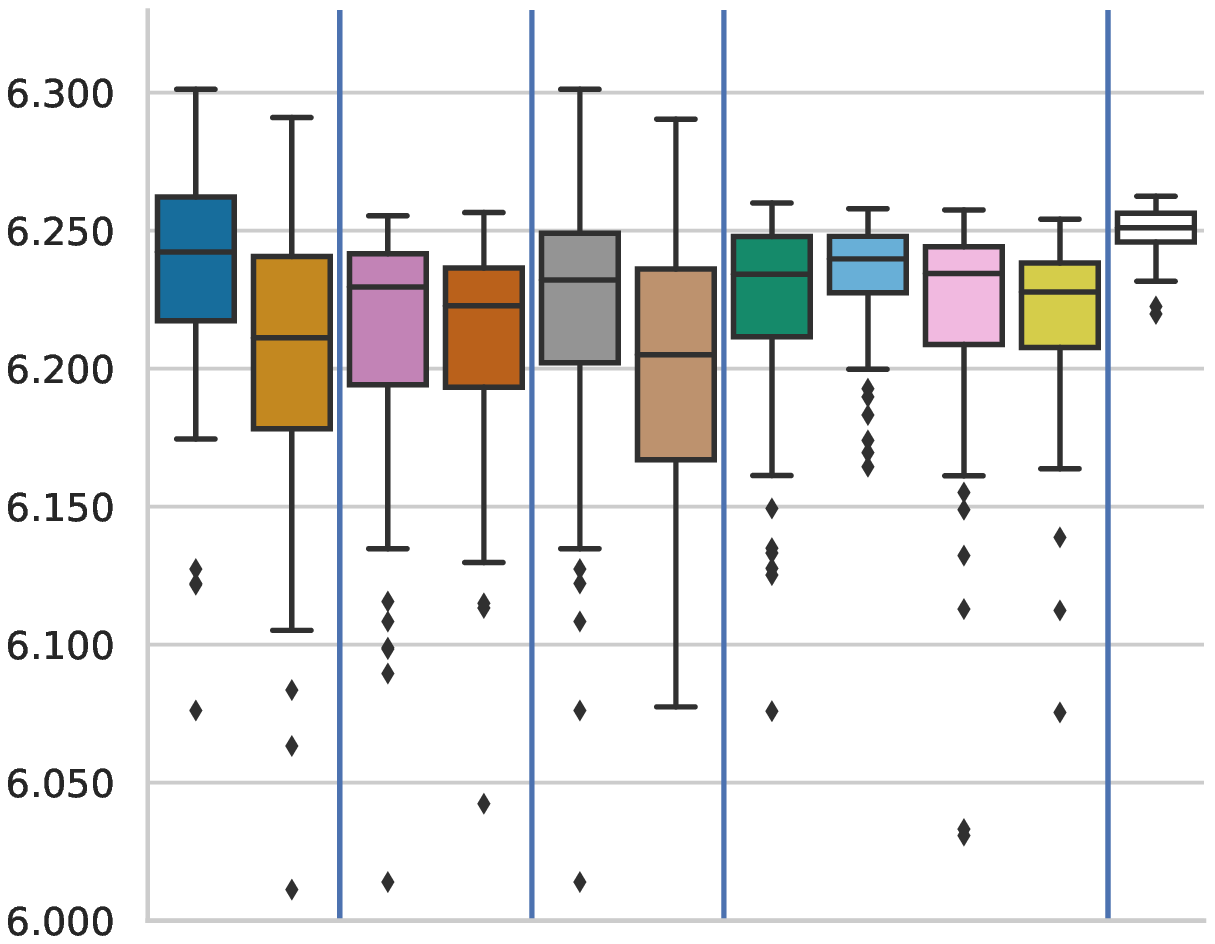}
			\caption{L1}
		\end{subfigure}
		\begin{subfigure}[t]{0.4\textwidth}
			\centering
			\includegraphics[width=\textwidth]{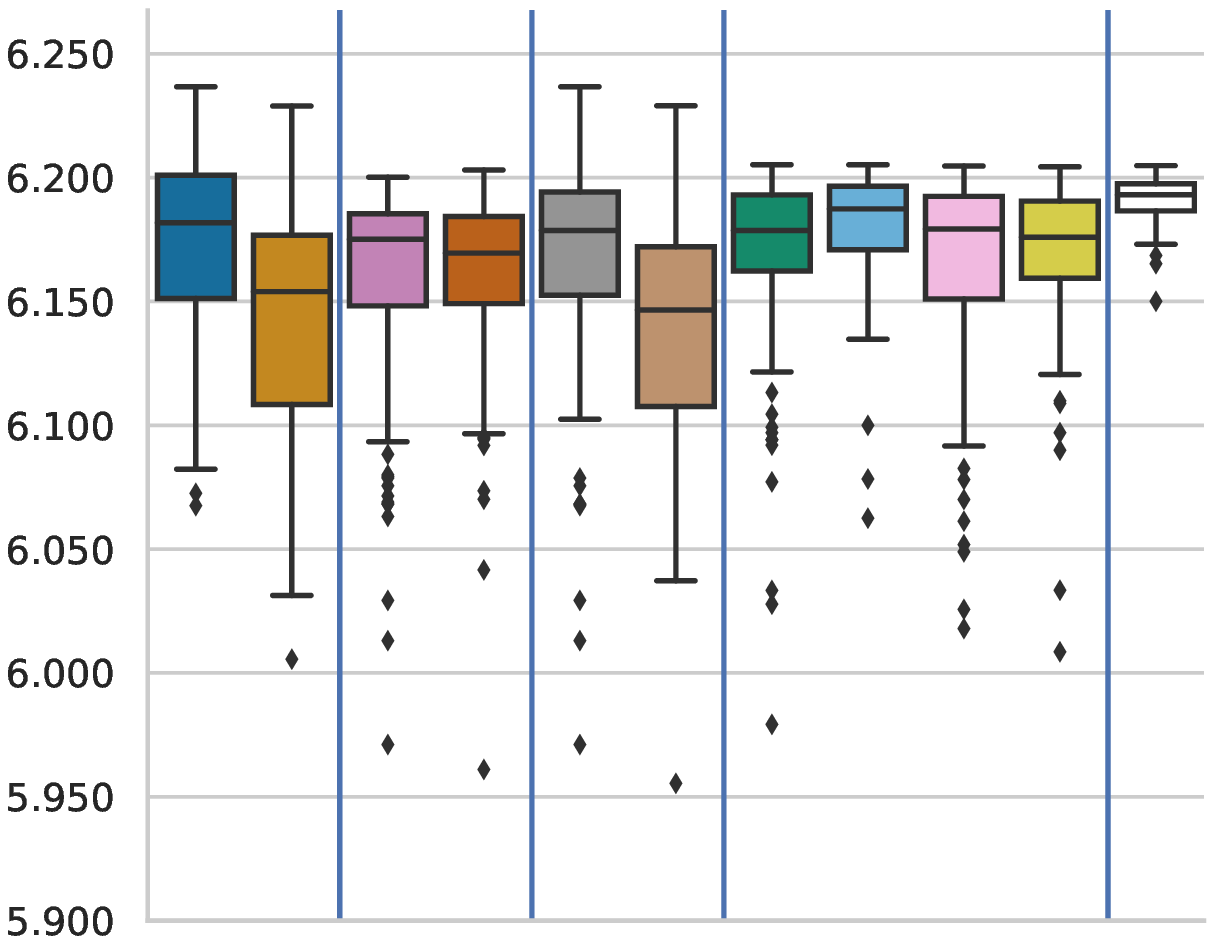}
			\caption{L2}
		\end{subfigure}
		\begin{subfigure}[t]{0.18\textwidth}
			\vspace{-5cm}
			\begin{spacing}{0.8}
				{\small
					\thiscolor{d1(L,Z)}\\
					\thiscolor{dEARL(L,Z)}\\
					\thiscolor{d2(L,W)}\\
					\thiscolor{dEARL(L,W)}\\
					\thiscolor{d4}\\
					\thiscolor{dEARL(L,W,Z)}\\
					\thiscolor{d3DR(L)}\\
					\thiscolor{d1(L)}\\
					\thiscolor{d2(L)}\\
					\thiscolor{dEARL(L)}\\
					\thiscolor{NUC}
				}
			\end{spacing}
		\end{subfigure}
		\caption{Boxplots of values for Scenarios L1 and L2 with sample size $n=2,000$ without adding noisy variables.}
		\label{fig:L_2000}
	\end{figure}

\begin{figure}[H]
		\centering
		\begin{subfigure}[t]{0.4\textwidth}
			\centering
			\includegraphics[width=\textwidth]{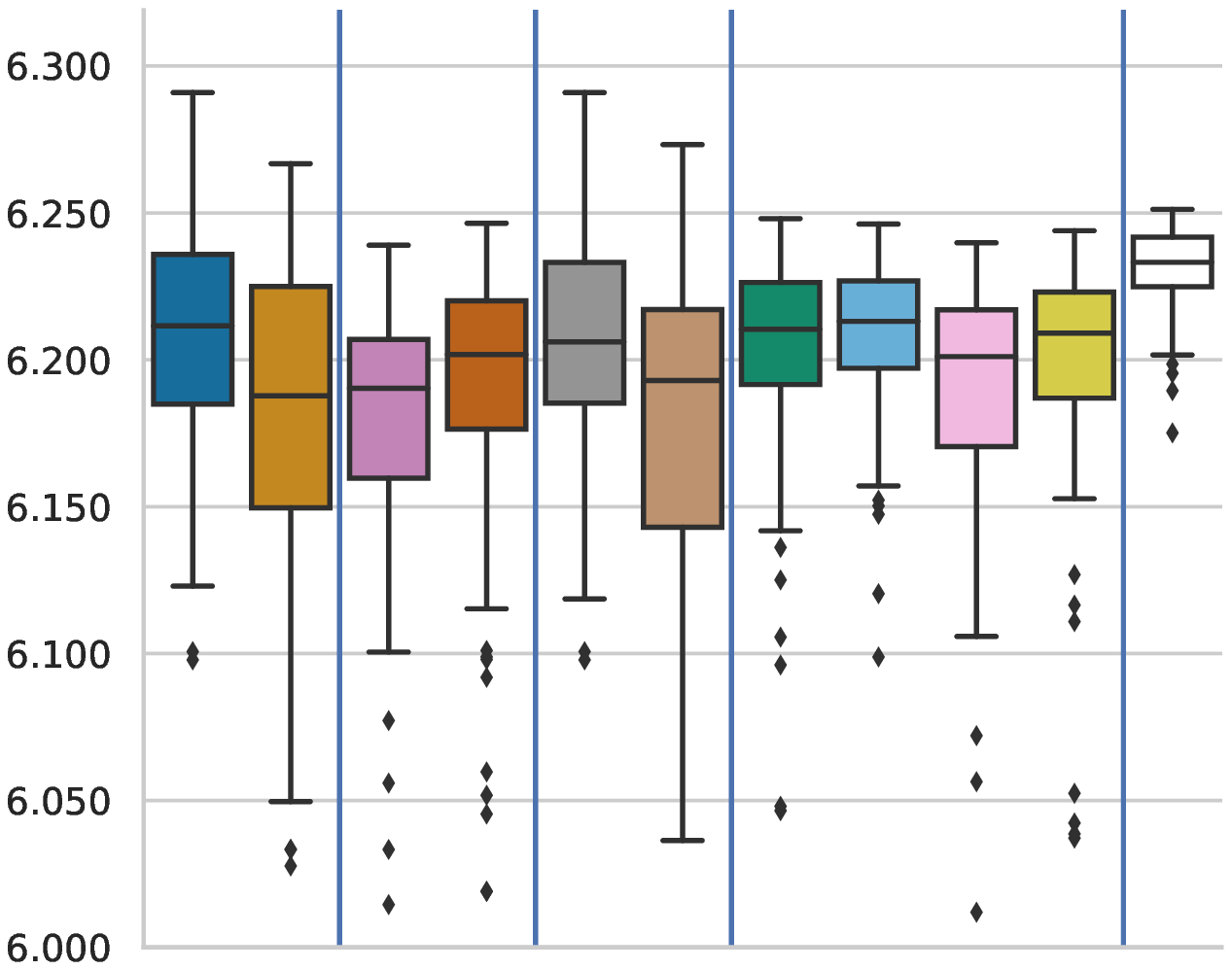}
			\caption{L1}
		\end{subfigure}
		\begin{subfigure}[t]{0.4\textwidth}
			\centering
			\includegraphics[width=\textwidth]{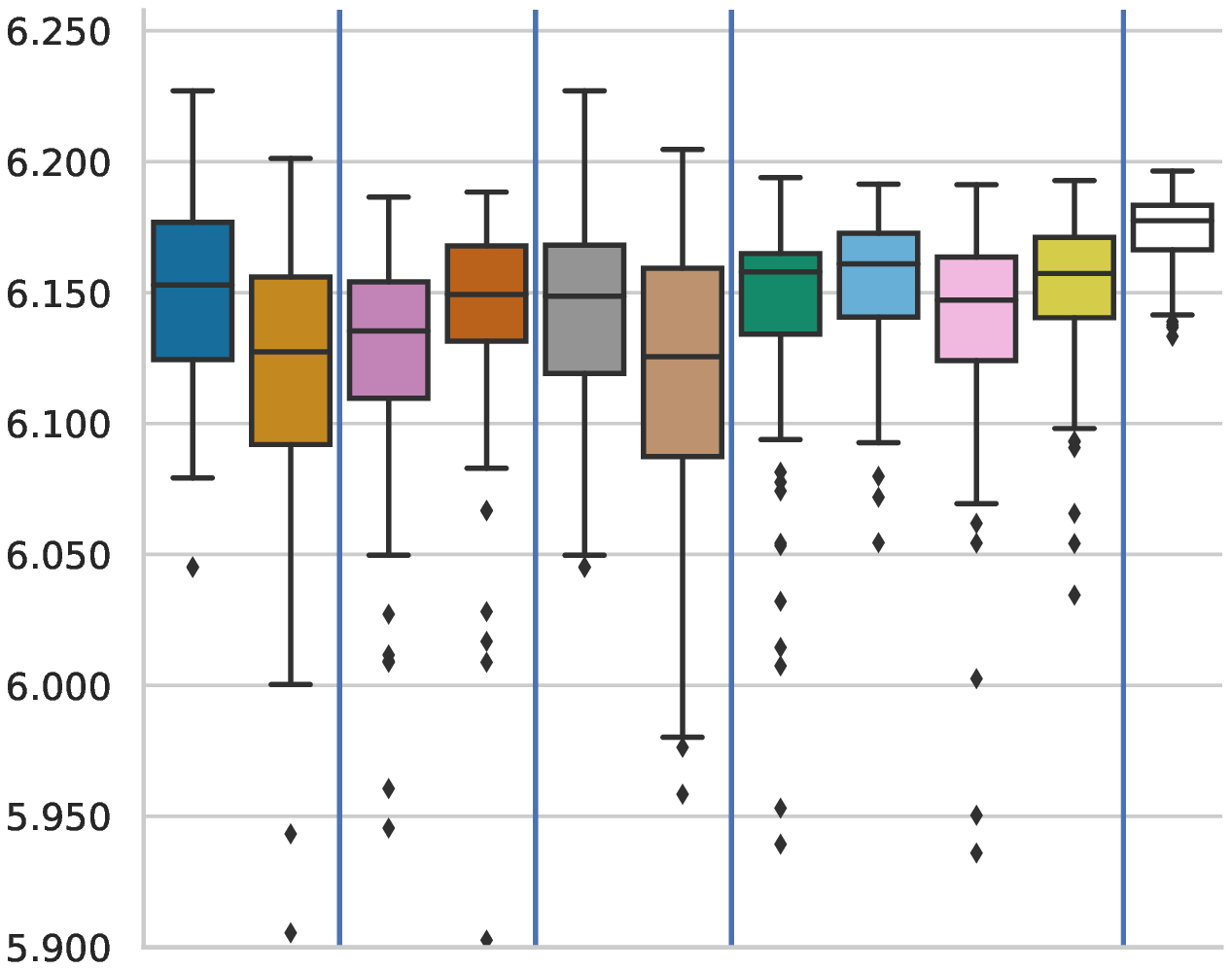}
			\caption{L2}
		\end{subfigure}
		\begin{subfigure}[t]{0.18\textwidth}
			\vspace{-5cm}
			\begin{spacing}{0.8}
				{\small
					\thiscolor{d1(L,Z)}\\
					\thiscolor{dEARL(L,Z)}\\
					\thiscolor{d2(L,W)}\\
					\thiscolor{dEARL(L,W)}\\
					\thiscolor{d4}\\
					\thiscolor{dEARL(L,W,Z)}\\
					\thiscolor{d3DR(L)}\\
					\thiscolor{d1(L)}\\
					\thiscolor{d2(L)}\\
					\thiscolor{dEARL(L)}\\
					\thiscolor{NUC}
				}
			\end{spacing}
		\end{subfigure}
		\caption{Boxplots of values for Scenarios L1 and L2 with sample size $n=2,000$ with noisy variables.}
		\label{fig:L_2000_noisy}
	\end{figure}

Figures \ref{fig:L_2000} -- \ref{fig:L_5000} correspond to scenarios L1 and L2 with sample sizes $n=2,000$ (with and without noisy variables) and $n=5,000$. 
For $n=2,000$ without noisy variables (Figure \ref{fig:L_2000}), the dimension of $L$ is $2$. To increase the difficulty, for $n=2,000$ or $5,000$ with noisy variables (Figures \ref{fig:L_2000_noisy} and \ref{fig:L_5000}), we added $8$ independent noisy variables from the uniform distribution on $[-1,1]$ into $L$ and used the resulting concatenated 10-dimensional $L$ in all ITR learning method.
Generally, our methods perform better than \text{EARL} in all considered policy classes such as $(L,W)$, $(L,Z)$, $(L,Z,W)$ or $L$, when $n=2,000$ and $5,000$. However, for $n=2,000$, \texttt{d2(L,W)} only improves marginally over \texttt{dEARL(L,W)} in Figure \ref{fig:L_2000} and is no better than  \texttt{dEARL(L,W)} in Figure \ref{fig:L_2000_noisy}. This demonstrates the difficulty of estimating $q_0$ for small sample.

In addition, under scenarios L1 and L2, Figures \ref{fig:L_2000_pess} -- \ref{fig:L_5000_pess} show the performance of \texttt{dPESS}, which is obtained by choosing the ITR with the highest 40\% quantile of estimated values from 5-fold cross-validations. 
In general, the values obtained by this selection procedure, which we called it \texttt{dPESS}, are as high as those by \texttt{d4}. We remark that in our previous submission, the performance of \texttt{d4} should be the best theoretically if all assumptions are met.
For Scenario L2, where the true ITR is a function of $L$ only, the performance of \texttt{dPESS} is still comparable to the best ITR that only depends on $L$ (i.e., \texttt{d3DR(L),d1(L)} and \texttt{d2(L)}). 

	Figure \ref{fig:N_2000} refers to our N1 and N2 scenarios with 
	sample size $n = 2,000$. 
	Similar interpretations can be obtained as those in the main text.

	In the following, we also study the statistical inference related to $V(d_3)$ for $d_3 \in \calD_3$. Table \ref{tab:EIF} refers to the inference results of $V(d_3)$ at $d_3=d_3^*$ for all simulation scenarios. 
	The point estimates of $V(d_3^*)$ were calculated by \eqref{eq: doubly robust value function} in the main text. The 95\% confidence intervals of $V(d_3)$ were constructed
	according to Theorem 4.1. For each scenario L1, L2, N1, N2 and sample size $n=2,000$, $5000$, we ran 1,000 simulations
	to obtain the mean squared error (MSE), averaged length of CIs (Avg. CI Length) and coverage rate. Results shown in Table \ref{tab:EIF} indicate satisfactory performances of using doubly robust estimator in evaluating $V(d_3)$.

	\begin{figure}[H]
		\centering
		\begin{subfigure}[t]{0.4\textwidth}
			\centering
			\includegraphics[width=\textwidth]{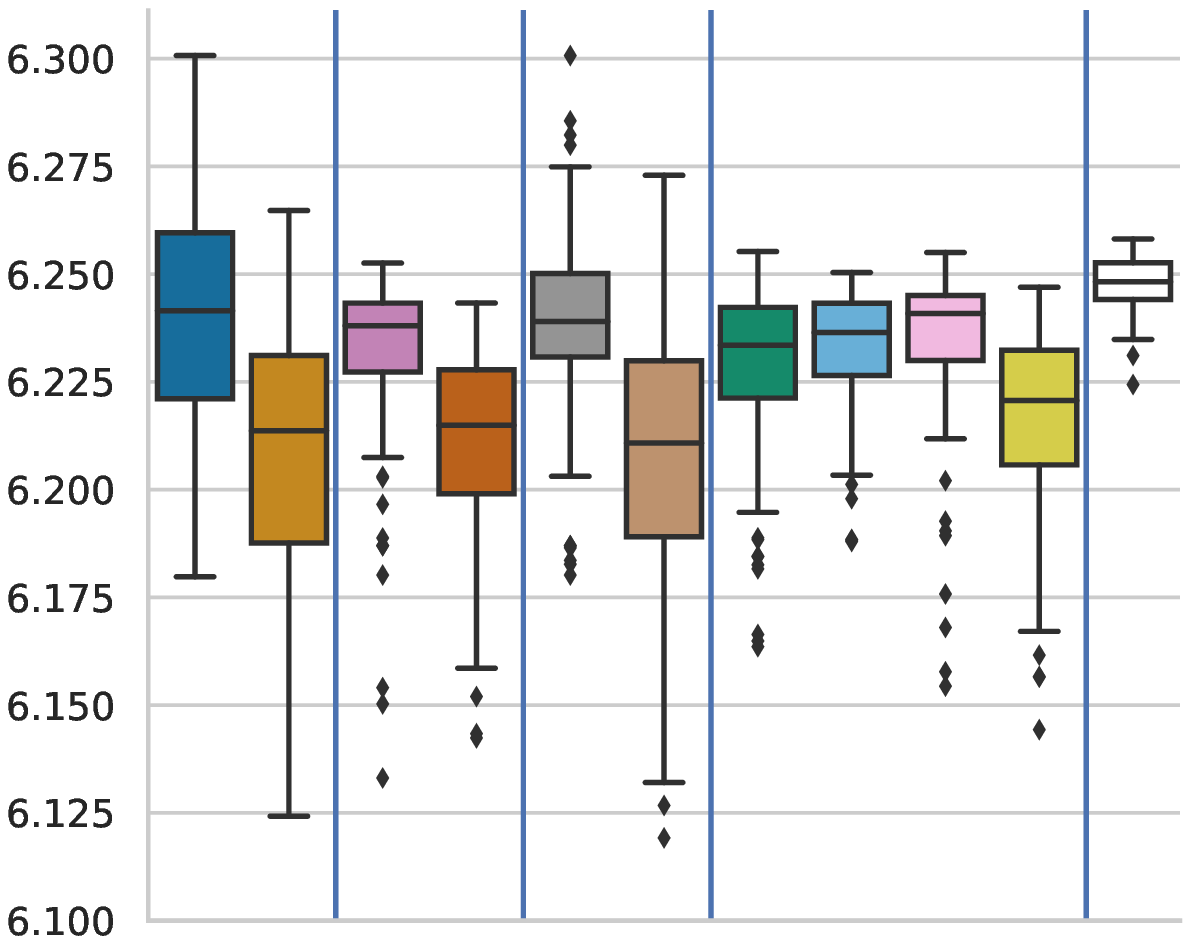}
			\caption{L1}
		\end{subfigure}
		\begin{subfigure}[t]{0.4\textwidth}
			\centering
			\includegraphics[width=\textwidth]{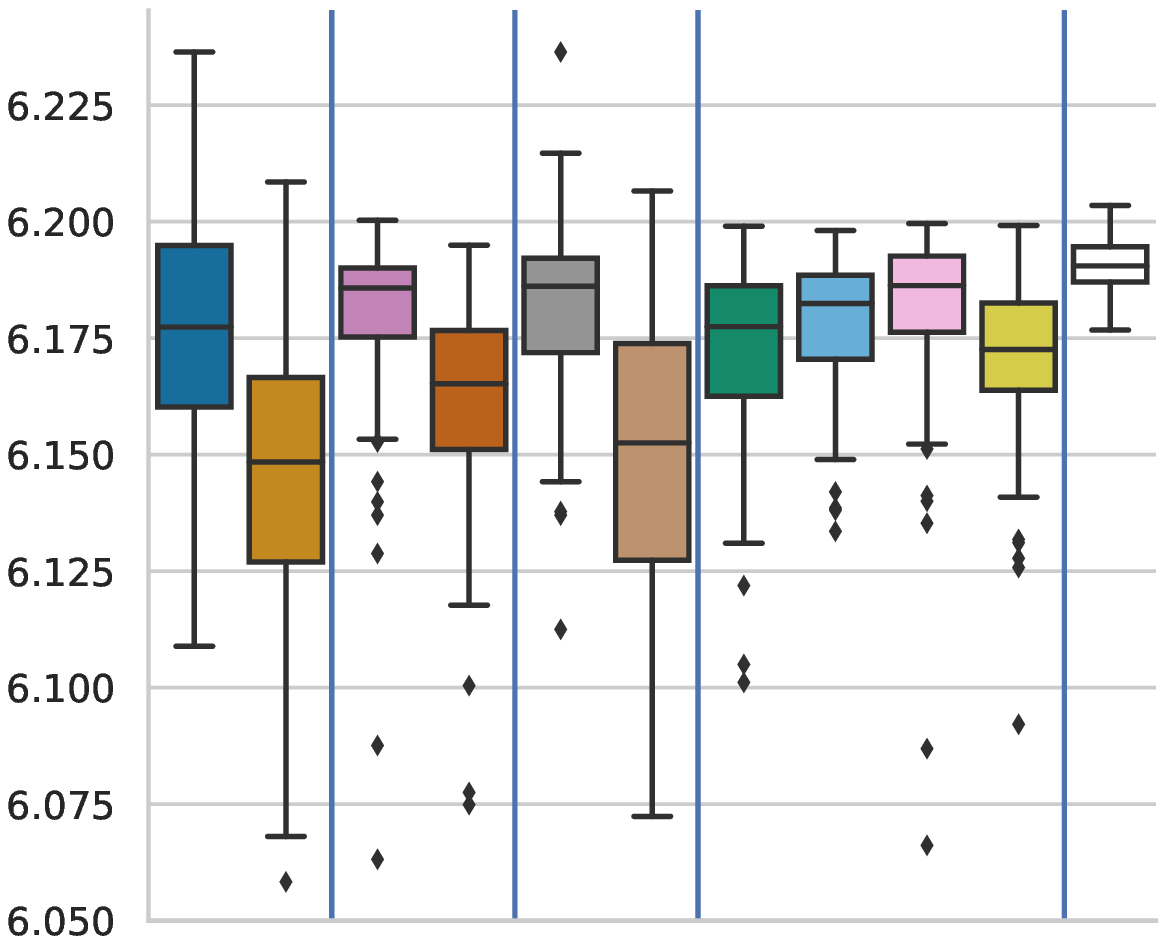}
			\caption{L2}
		\end{subfigure}
		\begin{subfigure}[t]{0.18\textwidth}
			\vspace{-5cm}
			\begin{spacing}{0.8}
				{\small
					\thiscolor{d1(L,Z)}\\
					\thiscolor{dEARL(L,Z)}\\
					\thiscolor{d2(L,W)}\\
					\thiscolor{dEARL(L,W)}\\
					\thiscolor{d4}\\
					\thiscolor{dEARL(L,W,Z)}\\
					\thiscolor{d3DR(L)}\\
					\thiscolor{d1(L)}\\
					\thiscolor{d2(L)}\\
					\thiscolor{dEARL(L)}\\
					\thiscolor{NUC}
				}
			\end{spacing}
		\end{subfigure}
		\caption{Boxplots of values for Scenarios L1 and L2 with sample size $n=5,000$ with noisy variables.}
		\label{fig:L_5000}
	\end{figure}
	
	\begin{figure}[H]
		\centering
		\begin{subfigure}[t]{0.4\textwidth}
			\centering
			\includegraphics[width=\textwidth]{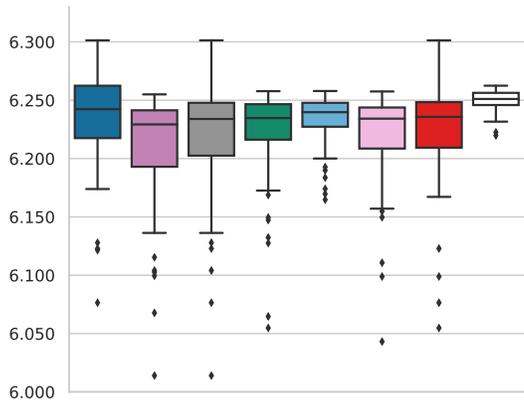}
			\caption{L1}
		\end{subfigure}
		\begin{subfigure}[t]{0.4\textwidth}
			\centering
			\includegraphics[width=\textwidth]{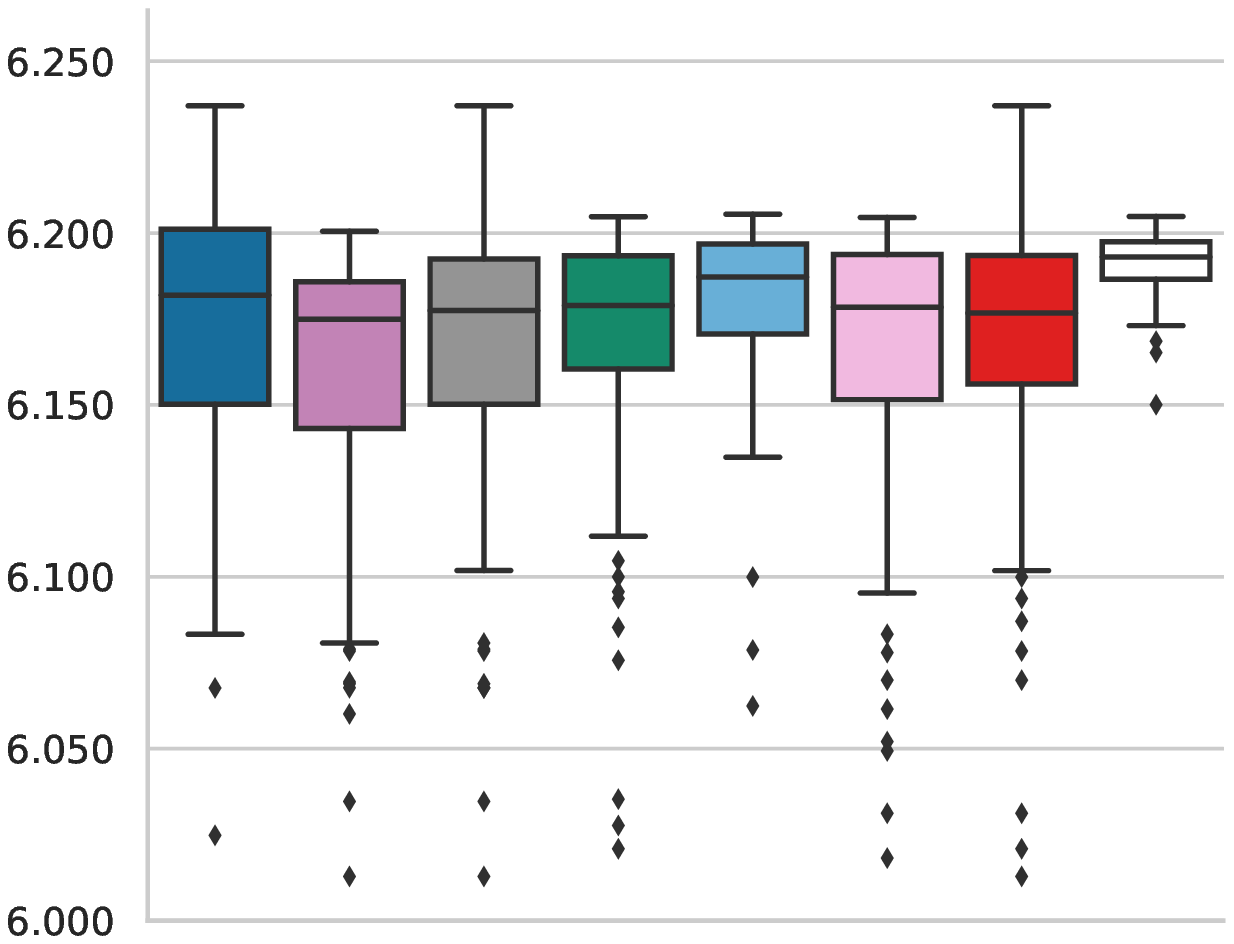}
			\caption{L2}
		\end{subfigure}
		\begin{subfigure}[t]{0.18\textwidth}
			\vspace{-5cm}
			\begin{spacing}{0.8}
				{\small
					\thiscolor{d1(L,Z)}\\
					\thiscolor{d2(L,W)}\\
					\thiscolor{d4}\\
					\thiscolor{d3DR(L)}\\
					\thiscolor{d1(L)}\\
					\thiscolor{d2(L)}\\
					\thiscolor{dPESS}\\
					\thiscolor{NUC}
				}
			\end{spacing}
		\end{subfigure}
		\caption{Boxplots of values for evaluating \texttt{dPESS} under Scenarios L1 and L2 with $n=2,000$ without adding noisy variables.}
		\label{fig:L_2000_pess}
	\end{figure}

	\begin{figure}[H]
		\centering
		\begin{subfigure}[t]{0.4\textwidth}
			\centering
			\includegraphics[width=\textwidth]{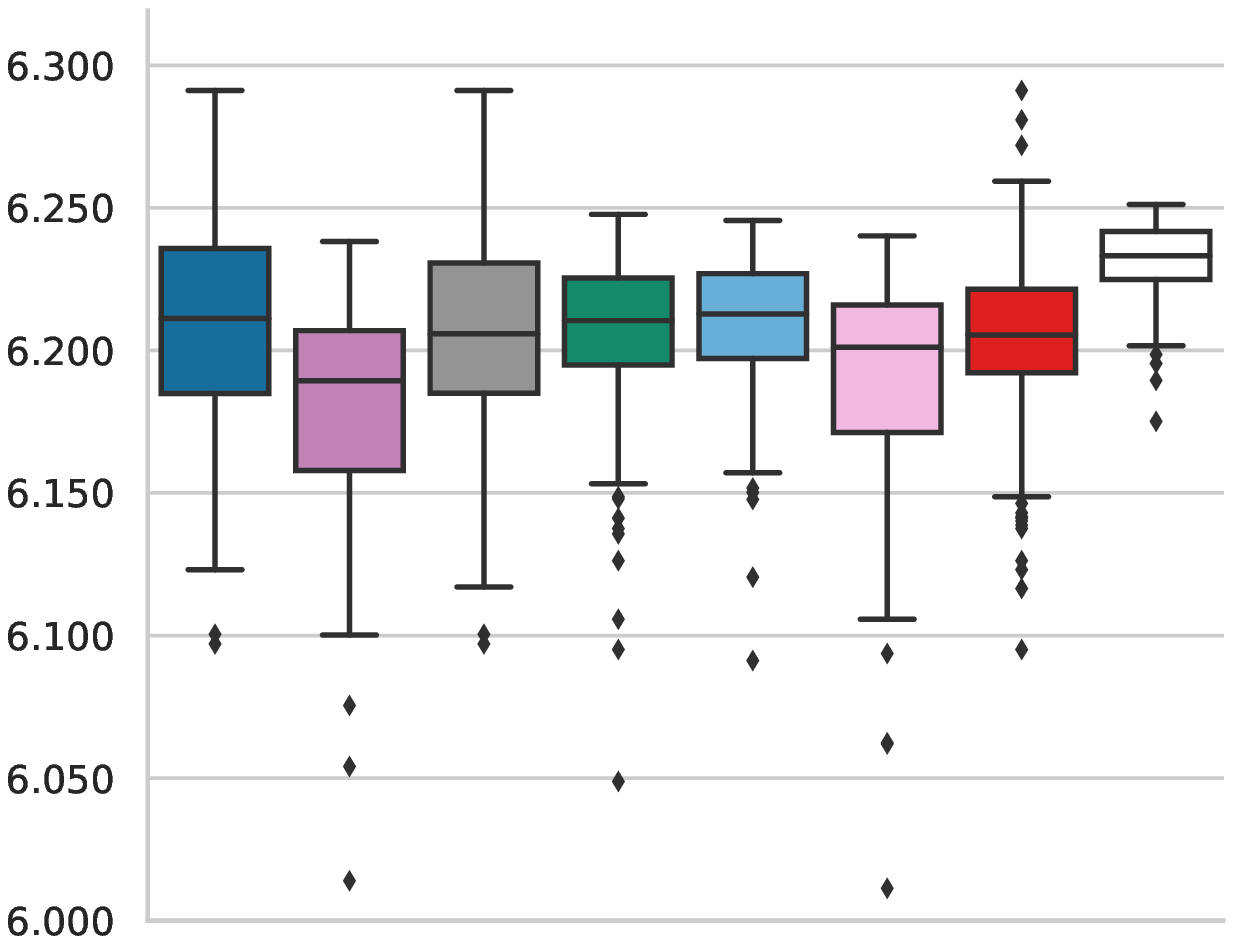}
			\caption{L1}
		\end{subfigure}
		\begin{subfigure}[t]{0.4\textwidth}
			\centering
			\includegraphics[width=\textwidth]{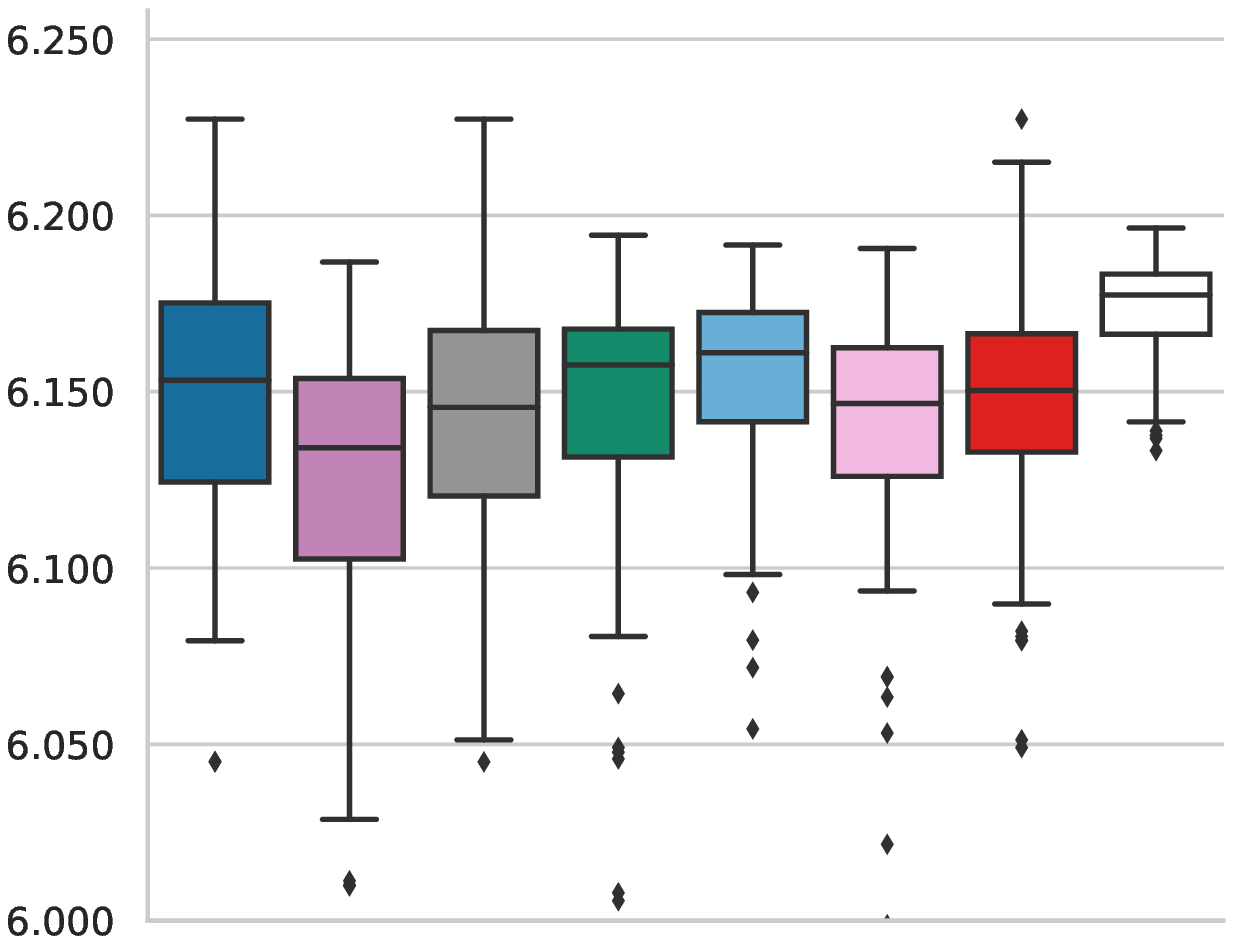}
			\caption{L2}
		\end{subfigure}
		\begin{subfigure}[t]{0.18\textwidth}
			\vspace{-5cm}
			\begin{spacing}{0.8}
				{\small
					\thiscolor{d1(L,Z)}\\
					\thiscolor{d2(L,W)}\\
					\thiscolor{d4}\\
					\thiscolor{d3DR(L)}\\
					\thiscolor{d1(L)}\\
					\thiscolor{d2(L)}\\
					\thiscolor{dPESS}\\
					\thiscolor{NUC}
				}
			\end{spacing}
		\end{subfigure}
		\caption{Boxplots of values for evaluating \texttt{dPESS} under Scenarios L1 and L2 with $n=2,000$ with noisy variables.}
		\label{fig:L_2000_noisy_pess}
	\end{figure}

	\begin{figure}[H]
		\centering
		\begin{subfigure}[t]{0.4\textwidth}
			\centering
			\includegraphics[width=\textwidth]{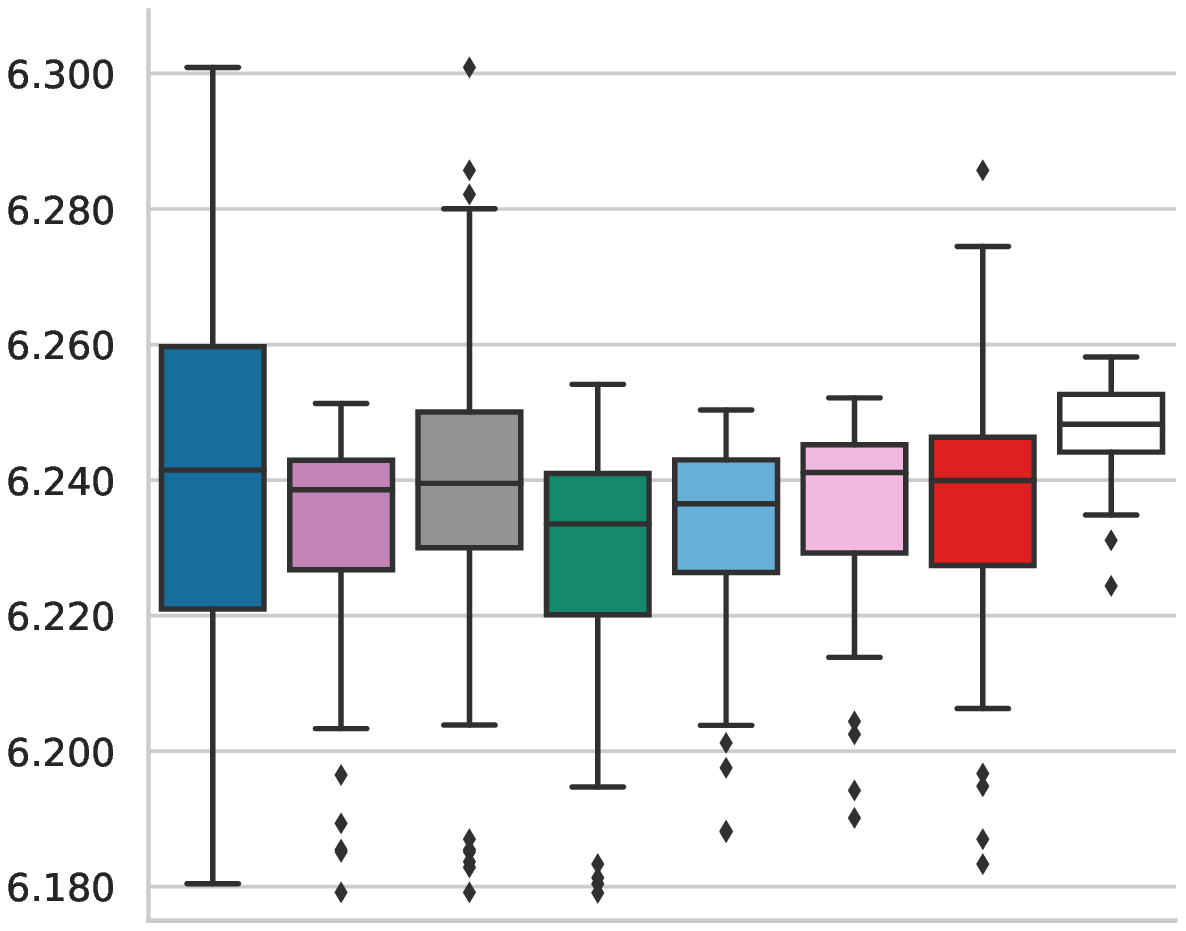}
			\caption{L1}
		\end{subfigure}
		\begin{subfigure}[t]{0.4\textwidth}
			\centering
			\includegraphics[width=\textwidth]{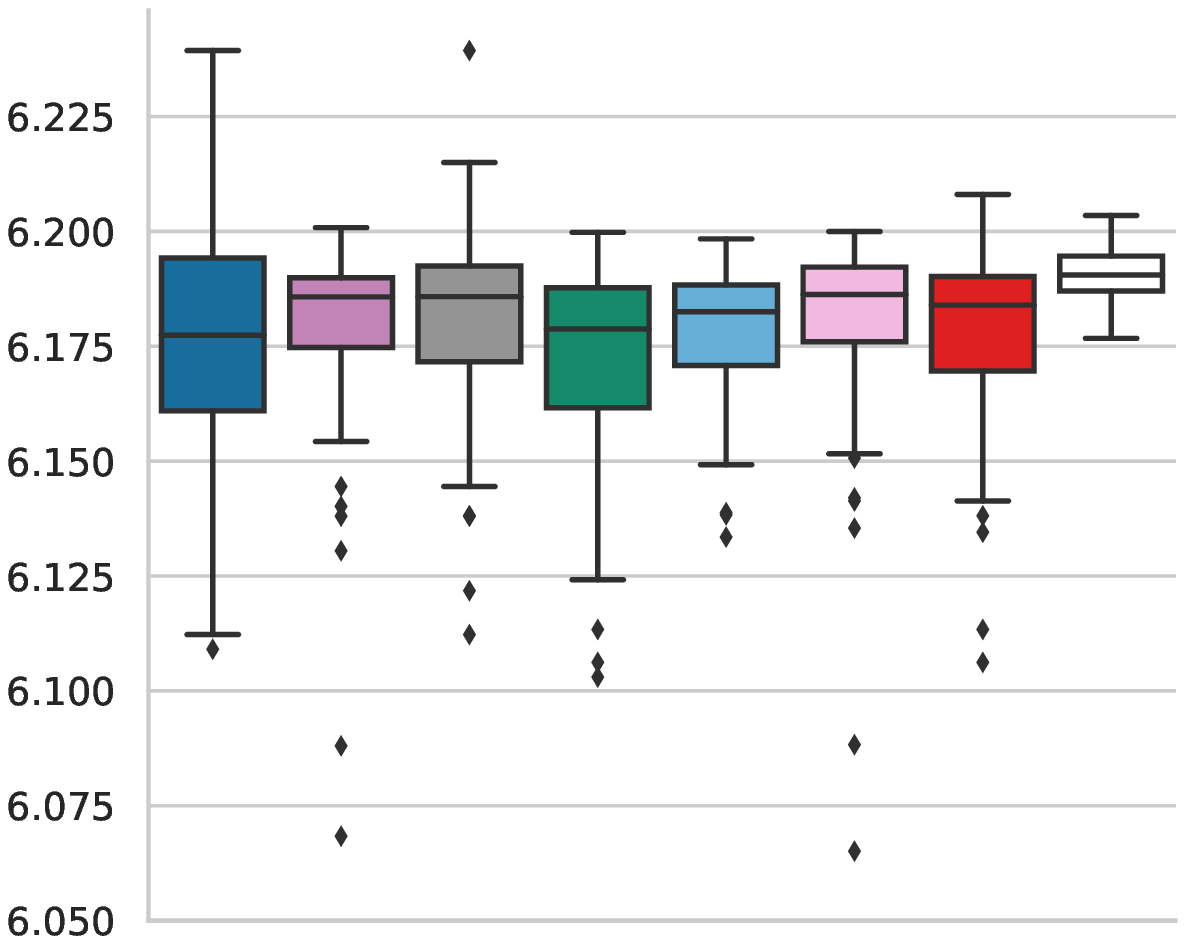}
			\caption{L2}
		\end{subfigure}
		\begin{subfigure}[t]{0.18\textwidth}
			\vspace{-5cm}
			\begin{spacing}{0.8}
				{\small
					\thiscolor{d1(L,Z)}\\
					\thiscolor{d2(L,W)}\\
					\thiscolor{d4}\\
					\thiscolor{d3DR(L)}\\
					\thiscolor{d1(L)}\\
					\thiscolor{d2(L)}\\
					\thiscolor{dPESS}\\
					\thiscolor{NUC}
				}
			\end{spacing}
		\end{subfigure}
		\caption{Boxplots of values for evaluating \texttt{dPESS} under Scenarios L1 and L2 with $n=5,000$ with noisy variables.}
		\label{fig:L_5000_pess}
	\end{figure}
	\begin{figure}[H]
		\centering
		\begin{subfigure}[t]{0.4\textwidth}
			\centering
			\includegraphics[width=\textwidth]{Nonlinear_LU_2000.eps}
			\caption{N1}
		\end{subfigure}
		\begin{subfigure}[t]{0.4\textwidth}
			\centering
			\includegraphics[width=\textwidth]{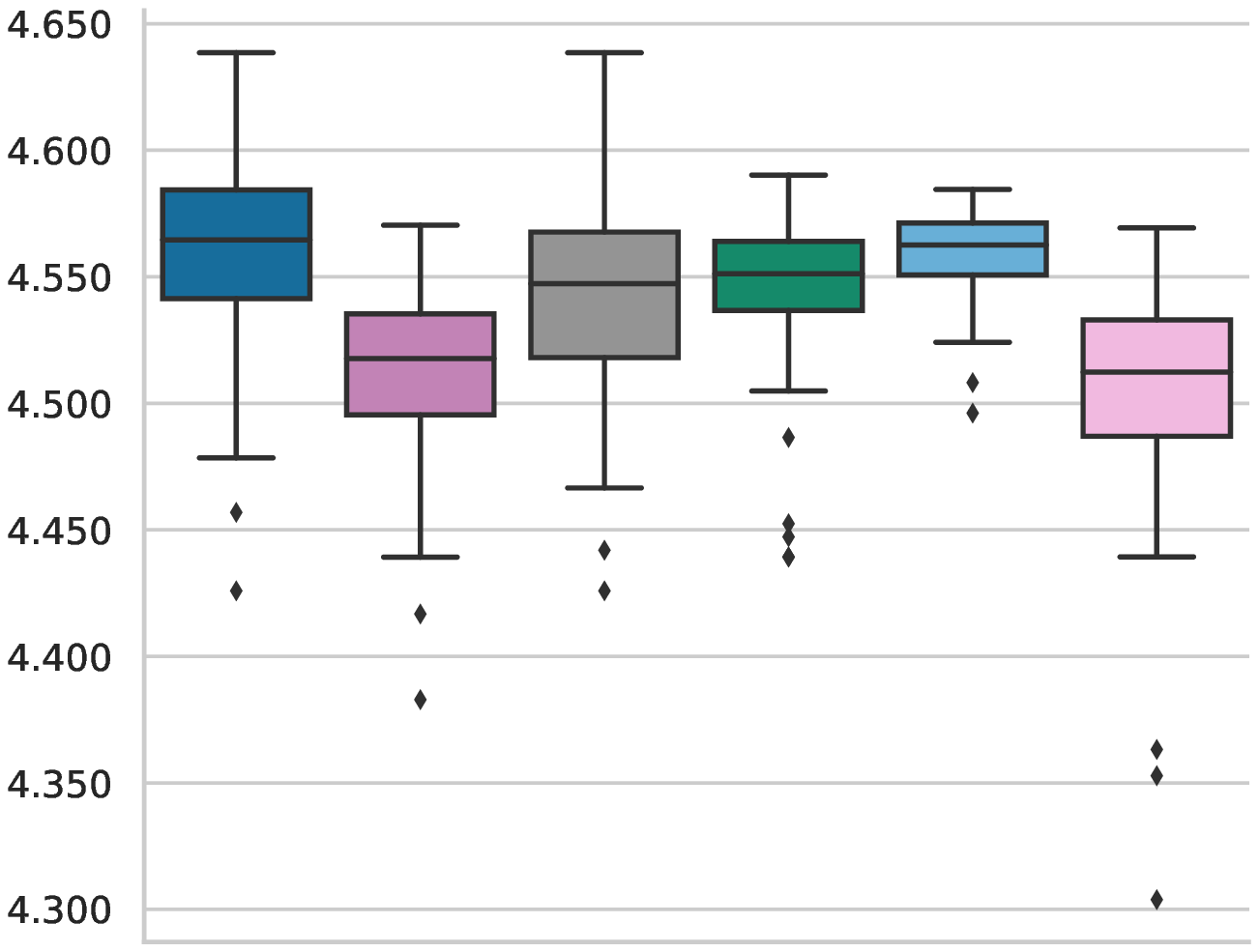}
			\caption{N2}
		\end{subfigure}
		\begin{subfigure}[t]{0.18\textwidth}
			\vspace{-4.5cm}
			\begin{spacing}{0.8}
				{
					\thiscolor{d1(L,Z)}\\
					\thiscolor{d2(L,W)}\\
					\thiscolor{d4}\\
					\thiscolor{d3DR(L)}\\
					\thiscolor{d1(L)}\\
					\thiscolor{d2(L)}
				}
			\end{spacing}
		\end{subfigure}
		\caption{Boxplots of values for Scenario N1 and N2 with sample size $n=2,000$.}
		\label{fig:N_2000}
	\end{figure}

	\begin{table}[h]
		\caption{Inferences of efficient influence function at $d_3^*$.}
		\label{tab:EIF}
		\centering
		\begin{tabular}{cc|cccc}
			\hline
			\hline
Scenario & $n$ & $V(d_3^*)$ & MSE & Avg. CI Length & Coverage Rate\\
			\hline
			\hline
L1 & 2,000 & 6.258 & 0.0191 & 0.628 & 97.6\%\\
L1 & 5,000 & 6.258 & 0.0108 & 0.425 & 96.2\%\\
L2 & 2,000 & 6.203 & 0.0183 & 0.615 & 98.3\%\\
L2 & 5,000 & 6.203 & 0.0104 & 0.416 & 96.1\%\\
N1 & 2,000 & 4.244 & 0.0067 & 0.332 & 95.3\%\\
N1 & 5,000 & 4.244 & 0.0031 & 0.218 & 95.8\%\\
N2 & 2,000 & 4.602 & 0.0081 & 0.378 & 96.3\%\\
N2 & 5,000 & 4.602 & 0.0043 & 0.247 & 94.0\%\\
			\hline
		\end{tabular}
	\end{table}

	\section{Additional Results of Real Data Application} \label{appendix:data}
	
\begin{figure}[H]
		\centering
		\includegraphics[width=0.8\textwidth]{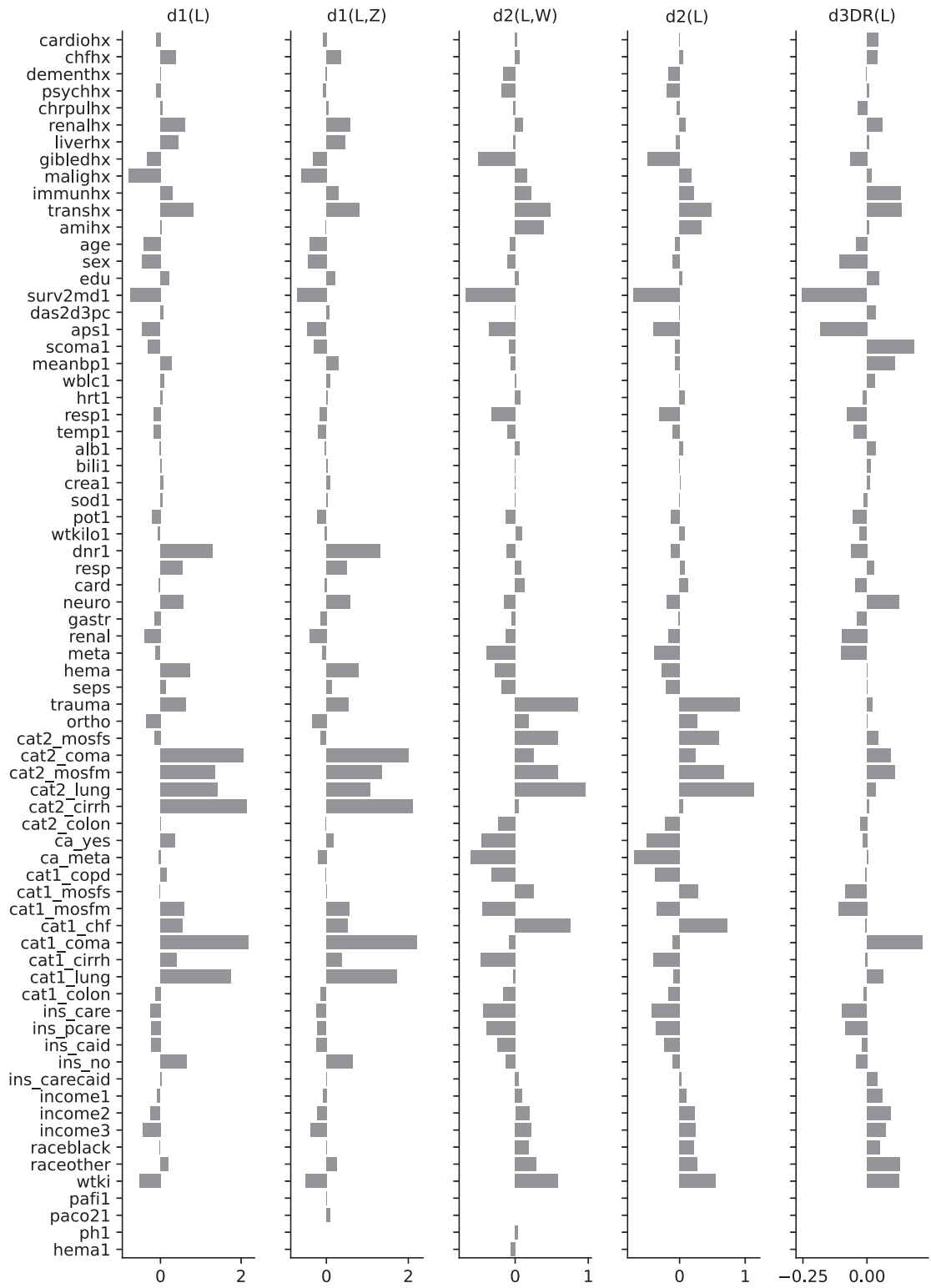}
		\caption{Sizes of all coefficients of the five optimal linear ITR estimates.}
		\label{fig:rhc_coef}
	\end{figure}

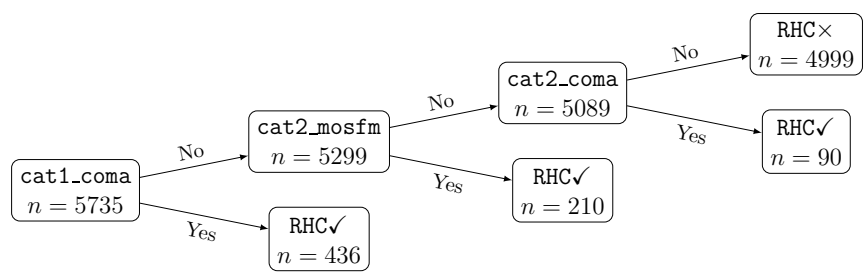
\begin{figure}[h]
  \centering
  \resizebox{0.6\textwidth}{!}{%
\begin{tikzpicture}
  [
    treenode/.style = {shape=rectangle, rounded corners,
                     draw, align=center,
                     top color=white,
                     font=\ttfamily\normalsize},
    grow                    = right,
    sibling distance        = 4em,
    level distance          = 10em,
    edge from parent/.style = {draw, -latex},
    every node/.style       = {font=\footnotesize},
    sloped
  ]
  \node [treenode] {cat1\_coma\\$n=5735$}
    child { node [treenode] {RHC\checkmark\\$n=436$}
      edge from parent node [below] {Yes} }
    child { node [treenode] {cat2\_mosfm\\$n=5299$}
      child{ node [treenode] {RHC\checkmark\\$n=210$}
        edge from parent node [below] {Yes}}
      child{ node [treenode] {cat2\_coma\\$n=5089$}
        child{ node [treenode] {RHC\checkmark\\$n=90$}
          edge from parent node [below] {Yes}}
        child{ node [treenode] {RHC$\times$\\$n=4999$}
          edge from parent node [above] {No}}
        edge from parent node [above] {No}}
      edge from parent node [above] {No}
        };
\end{tikzpicture}
  }
  \caption{Decision tree of RHC based on the ensemble ITR, which recommends a patient to undergo RHC if at least four of the estimated ITRs all recommend it.
  \texttt{mosfm}: multiple organ system failure with malignancy.}
  \label{fig: decision tree majority voting}
\end{figure}

	All coefficients of the five optimal linear ITR estimates in comparison are given in Table \ref{tb:all_coef} and illustrated in Figure \ref{fig:rhc_coef}. The decision tree of ensemble ITR by majority voting is given in Figure \ref{fig: decision tree majority voting}.

{\scriptsize
	\begin{longtable}{l|rrrrr}
		\caption{All coefficients of the five optimal linear ITR estimates.}\label{tb:all_coef}
		\\ 
		\hline\hline
		Covariate   & \texttt{d1(L)} & \texttt{d1(L,Z)} & \texttt{d2(L,W)} & \texttt{d2(L)}  & \texttt{d3DR(L)} \\
		\hline\hline
		intercept    & -1.215 & -1.222  & -0.204  & -0.258 & -0.722  \\
		cardiohx     & -0.109 & -0.085  & 0.033   & 0.007  & 0.045   \\
		chfhx        & 0.391  & 0.348   & 0.059   & 0.055  & 0.041   \\
		dementhx     & 0.001  & -0.021  & -0.171  & -0.167 & -0.003  \\
		psychhx      & -0.112 & -0.079  & -0.188  & -0.200 & 0.007   \\
		chrpulhx     & 0.046  & 0.061   & -0.023  & -0.042 & -0.038  \\
		renalhx      & 0.612  & 0.579   & 0.111   & 0.096  & 0.061   \\
		liverhx      & 0.447  & 0.451   & -0.025  & -0.056 & 0.008   \\
		gibledhx     & -0.343 & -0.328  & -0.511  & -0.490 & -0.065  \\
		malighx      & -0.797 & -0.622  & 0.170   & 0.181  & 0.017   \\
		immunhx      & 0.296  & 0.296   & 0.217   & 0.226  & 0.132   \\
		transhx      & 0.827  & 0.802   & 0.486   & 0.488  & 0.136   \\
		amihx        & 0.031  & -0.019  & 0.388   & 0.333  & 0.008   \\
		age          & -0.421 & -0.413  & -0.079  & -0.065 & -0.043  \\
		sex          & -0.459 & -0.455  & -0.112  & -0.105 & -0.109  \\
		edu          & 0.209  & 0.211   & 0.050   & 0.039  & 0.046   \\
		surv2md1     & -0.747 & -0.724  & -0.683  & -0.699 & -0.254  \\
		das2d3pc     & 0.070  & 0.069   & -0.005  & 0.002  & 0.033   \\
		aps1         & -0.451 & -0.461  & -0.361  & -0.405 & -0.184  \\
		scoma1       & -0.309 & -0.307  & -0.082  & -0.073 & 0.184   \\
		meanbp1      & 0.288  & 0.286   & -0.057  & -0.064 & 0.109   \\
		wblc1        & 0.093  & 0.099   & 0.013   & 0.007  & 0.031   \\
		hrt1         & 0.042  & 0.040   & 0.076   & 0.081  & -0.016  \\
		resp1        & -0.172 & -0.166  & -0.327  & -0.310 & -0.078  \\
		temp1        & -0.180 & -0.204  & -0.108  & -0.111 & -0.051  \\
		alb1         & -0.032 & -0.040  & 0.064   & 0.061  & 0.033   \\
		bili1        & 0.022  & 0.022   & 0.007   & 0.008  & 0.015   \\
		crea1        & 0.081  & 0.093   & 0.001   & 0.013  & 0.012   \\
		sod1         & 0.051  & 0.035   & 0.010   & 0.008  & -0.015  \\
		pot1         & -0.209 & -0.221  & -0.130  & -0.130 & -0.055  \\
		wtkilo1      & -0.059 & -0.046  & 0.091   & 0.083  & -0.028  \\
		dnr1         & 1.294  & 1.315   & -0.117  & -0.131 & -0.062  \\
		resp         & 0.543  & 0.508   & 0.083   & 0.076  & 0.027   \\
		card         & -0.037 & -0.049  & 0.131   & 0.131  & -0.045  \\
		neuro        & 0.570  & 0.588   & -0.156  & -0.191 & 0.126   \\
		gastr        & -0.147 & -0.144  & -0.052  & -0.022 & -0.038  \\
		renal        & -0.403 & -0.414  & -0.135  & -0.167 & -0.098  \\
		meta         & -0.122 & -0.100  & -0.389  & -0.391 & -0.100  \\
		hema         & 0.733  & 0.786   & -0.278  & -0.275 & 0.003   \\
		seps         & 0.139  & 0.138   & -0.185  & -0.210 & 0.002   \\
		trauma       & 0.622  & 0.533   & 0.862   & 0.925  & 0.020   \\
		ortho        & -0.353 & -0.349  & 0.187   & 0.275  & 0.000   \\
		cat2\_mosfs   & -0.160 & -0.140  & 0.588   & 0.604  & 0.042   \\
		cat2\_coma    & 2.059  & 2.009   & 0.261   & 0.248  & 0.092   \\
		cat2\_mosfm   & 1.363  & 1.360   & 0.586   & 0.674  & 0.108   \\
		cat2\_lung    & 1.412  & 1.072   & 0.960   & 1.131  & 0.033   \\
		cat2\_cirrh   & 2.134  & 2.101   & 0.047   & 0.061  & 0.007   \\
		cat2\_colon   & -0.005 & -0.013  & -0.239  & -0.223 & -0.028  \\
		ca\_yes       & 0.355  & 0.182   & -0.458  & -0.498 & -0.018  \\
		ca\_meta      & -0.053 & -0.205  & -0.608  & -0.698 & 0.004   \\
		cat1\_copd    & 0.155  & -0.013  & -0.320  & -0.374 & -0.006  \\
		cat1\_mosfs   & -0.017 & -0.007  & 0.254   & 0.284  & -0.084  \\
		cat1\_mosfm   & 0.586  & 0.568   & -0.447  & -0.355 & -0.112  \\
		cat1\_chf     & 0.544  & 0.525   & 0.757   & 0.731  & -0.005  \\
		cat1\_coma    & 2.183  & 2.210   & -0.091  & -0.112 & 0.215   \\
		cat1\_cirrh   & 0.398  & 0.386   & -0.468  & -0.399 & -0.008  \\
		cat1\_lung    & 1.740  & 1.711   & -0.025  & -0.098 & 0.062   \\
		cat1\_colon   & -0.129 & -0.138  & -0.171  & -0.176 & -0.013  \\
		ins\_care     & -0.255 & -0.236  & -0.439  & -0.421 & -0.097  \\
		ins\_pcare    & -0.242 & -0.224  & -0.391  & -0.359 & -0.085  \\
		ins\_caid     & -0.242 & -0.252  & -0.247  & -0.229 & -0.021  \\
		ins\_no       & 0.647  & 0.644   & -0.127  & -0.111 & -0.043  \\
		ins\_carecaid & 0.036  & 0.006   & 0.048   & 0.028  & 0.041   \\
		income1      & -0.078 & -0.089  & 0.094   & 0.110  & 0.061   \\
		income2      & -0.246 & -0.234  & 0.194   & 0.238  & 0.093   \\
		income3      & -0.442 & -0.390  & 0.224   & 0.251  & 0.072   \\
		raceblack    & -0.018 & 0.009   & 0.192   & 0.226  & 0.051   \\
		raceother    & 0.206  & 0.251   & 0.292   & 0.277  & 0.130   \\
		wtki         & -0.526 & -0.518  & 0.586   & 0.546  & 0.125   \\
		pafi1        & -      & -0.011  & -       & -      & -       \\
		paco21       & -      & 0.090   & -       & -      & -       \\
		ph1          & -      & -       & 0.037   & -      & -       \\
		hema1        & -      & -       & -0.061  & -      & -      \\
		\hline\hline
	\end{longtable}
}	

\begingroup
\setstretch{0.9}

\bibliographystyle{abbrvnat}
\bibliography{reference}

\endgroup

\end{document}